%% file: cfvirgo.tex
\documentclass[useAMS,usenatbib]{mn2e}

\usepackage{graphicx}
\usepackage{lscape}
\usepackage{amssymb}

\newcommand{\Sub}{_{\mathrm{sub}}}

\newcommand{\Max}{_{\mathrm{max}}}
\newcommand{\Min}{_{\mathrm{min}}}

\newcommand{\Ew}{_{\mathrm{ew}}}

\newcommand{\Sun}{_{\sun}}
\newcommand{\Break}{_{\mathrm{break}}}
\newcommand{\damp}{_{\mathrm{damp}}}

\newcommand{\degree}{^o}

\newcommand{\K}{\,\textrm{K}}
\newcommand{\KeV}{\,\textrm{keV}}
\newcommand{\Kpc}{\,\textrm{kpc}}
\newcommand{\Mpc}{\,\textrm{Mpc}}
\newcommand{\PC}{\,\textrm{pc}}

\newcommand{\Myr}{\,\textrm{Myr}}
\newcommand{\Gyr}{\,\textrm{Gyr}}
\newcommand{\Kms}{\,\textrm{km}\,\textrm{s}^{-1}}

\newcommand{\gccm}{\,\textrm{g}\,\textrm{cm}^{-3}}
\newcommand{\Presunit}{\,\textrm{erg}\,\textrm{cm}^{-3}}

\title[Cold fronts in Virgo]{Gas sloshing, cold front formation, and metal redistribution:  the Virgo cluster as a quantitative test case}

\author[E. Roediger et al.]%
{E. Roediger$^{1}$\thanks{E-mail: e.roediger@jacobs-university.de},
M. Br\"uggen$^{1}$,
A. Simionescu$^{2}$,
H. B\"ohringer$^{3}$,\newauthor
E. Churazov$^{4,5}$ and
W. R. Forman$^{6}$\\
$^{1}$Jacobs University Bremen, PO Box 750 561, 28725 Bremen, Germany\\
$^{2}$KIPAC, Stanford University, 452 Lomita Mall, Stanford, CA 94305, USA\\
$^{3}$MPE, Gie§enbachstr. 1, 85748 Garching, Germany\\
$^{4}$MPA, Karl Schwarzschild Str. 1, 85748 Garching, Germany\\
$^{5}$Space Research Institute (IKI), Profsoyuznaya 84/32, Moscow 117810, Russia\\
$^{6}$Harvard-Smithsonian Center for Astrophysics, 60 Garden St., Cambridge, MA 02138, USA}

\begin{document}

\date{Accepted 1988 December 15. Received 1988 December 14; in original form 1988 October 11}

\pagerange{\pageref{firstpage}--\pageref{lastpage}} \pubyear{2009}

\maketitle

\label{firstpage}

\begin{abstract}
We perform hydrodynamical simulations of minor-merger induced gas sloshing and the subsequent formation of cold fronts. Using the Virgo cluster as a test case, we show for the first time that the sloshing scenario can reproduce the radii and the contrasts in X-ray brightness, projected temperature, and metallicity across the observed cold fronts quantitatively. 
The comparison suggests a third cold front 20 kpc NW of the Virgo core.

We identify several new features typical for sloshing cold fronts: an alternating distribution of cool, metal enriched X-ray brightness excess regions and warm brightness deficit regions of reduced metallicity; a constant or radially decreasing temperature accompanied by a plateau in metallicity inside the cold fronts; a warm rim outside the cold fronts; and a large-scale brightness asymmetry. We can trace these new features not only in Virgo, but also in other clusters exhibiting sloshing cold fronts. 

By comparing synthetic and real observations, we estimate that the original minor merger event took place 
about 1.5 Gyr ago 
when a subcluster of 2--4 $\times 10^{13}M\Sun$ passed the Virgo core at 100 to 400 kpc distance, where a smaller mass corresponds to a smaller pericentre distance, and vice versa.
From our inferred merger geometry, we derive the current location of the disturbing subcluster to be about 1-2 Mpc E of the Virgo core. A possible candidate is M60.

Additionally, we quantify the metal redistribution by sloshing and discuss its importance. 

We verify that the subcluster required to produce the observed cold fronts could be completely ram pressure stripped before reaching the Virgo centre, and discuss the conditions required for this to be achieved.

Finally, we demonstrate that the bow shock of a fast galaxy passing the Virgo cluster at $\sim 400\Kpc$ distance also causes sloshing and leads to very similar cold front structures. 
The responsible galaxy would be located about 2 Mpc north of the Virgo centre. 
A possible candidate is M85.
\end{abstract}


%\setcounter{tocdepth}{4}
%\tableofcontents

%***************************************************
%         MAIN TEXT
%***************************************************
%iiiiiiiiiiiiiiiiiiii
\input{intro}

\input{method}

\input{compare}

\input{metals}

\input{gas}

\input{shock}

\input{discussion}

\input{summary}
%iiiiiiiiiiiiiiii

%***********************
\section*{Acknowledgments}
We acknowledge the support of the Priority Programme "Witnesses of Cosmic History'' of the DFG (German Research Foundation) and the supercomputing  grants NIC 3229 and 3711 at the John-Neumann Institut at the Forschungszentrum J\"ulich. AS acknowledges the support provided by NASA through Einstein Postdoctoral Fellowship grant number PF9-00070 awarded by the Chandra X-ray Center, which is operated by the Smithsonian Astrophysical Observatory for NASA under contract NAS8-03060.
We thank  John ZuHone and Ryan Johnson for helpful discussions and the anonymous referee for the useful suggestions.
The results presented were produced using the FLASH code, a product  of the DOE ASC/Alliances-funded Center for Astrophysical Thermonuclear Flashes  at the University of Chicago. This research has made use of the NASA/IPAC Extragalactic Database (NED) which is operated by the Jet Propulsion Laboratory, California Institute of Technology, under contract with the National Aeronautics and Space Administration.

%*******************************************************************
%*************** R E F E R E N C E S *******************************
%*******************************************************************
%
\bibliographystyle{mn2e}
\bibliography{library}

%iiiiiiiiiiii
\appendix
\input{appendix}

\input{fiducial}

\input{details}

\bsp

\label{lastpage}

\end{document}

%% file: intro.tex
%****************
\section{Introduction}

During the last decade, high-resolution X-ray observations have revealed a wealth of structure in the intra-cluster medium (ICM) of galaxy clusters, among them cold fronts. The first ones were noticed as sharp discontinuities in X-ray brightness on one side of  the cores of  A2142 (\citealt{Markevitch2000}), A3667 (\citealt{ Vikhlinin2001}), and the bullet cluster 1E 0657-56 (\citealt{Markevitch2002}). Temperature measurements revealed that these brightness edges were not the expected bow shocks but cold fronts (CFs), where the brighter and denser side is also the cooler one. For a shock, the temperature jump across the brightness edge would be opposite. Several such CFs associated with merging clusters have been found (see review by \citealt{Markevitch2007}), hence they are called merger CFs. They have been interpreted as the contact discontinuity between the gaseous atmospheres of two different clusters. 
More recently, also shocks associated with the merging have been discovered in some of these clusters (see \citealt{Markevitch2010shocks} for a review). 

Additionally, a second class of CFs, named sloshing CFs after their most likely origin, have been detected. Here, the CFs form arcs around the cool cores of apparently relaxed clusters. This type of CFs is the subject of this paper. For clarity, we will first introduce the sloshing scenario (Sect.~\ref{sec:intro_scenario}) which is commonly believed to explain this type of CFs. Section~\ref{sec:intro_obs} summarises the observational characteristics of sloshing CFs known so far. Targeting the CFs in the Virgo cluster, we show that the sloshing scenario can explain the observational characteristics not only qualitatively, but also quantitatively. Furthermore, we obtain insights in the correct way of interpreting CF observations, and we uncover a number of new observational characteristics, which can also be transferred to other clusters.

%******
\subsection{Gas sloshing scenario} \label{sec:intro_scenario}
Originally proposed by \citet{Markevitch2001} (see also review by \citealt{Markevitch2007}), the idea of the sloshing scenario is the following: A gas-free subcluster falls through the main galaxy cluster, which  initially has a near-hydrostatic ICM distribution. 
During the pericentre passage, the interaction slightly offsets the gas in the main cluster core, but does not disrupt the main cluster core. 
The offset mechanism is described in detail in \citet{Ascasibar2006} (AM06 hereafter): while the subcluster approaches the main cluster, the core of the main cluster, i.e.~both, dark matter (DM) peak and cool gas peak, move towards the subcluster. When the subcluster passes the main core, it gravitationally pulls the main cluster core along. However, the ICM velocity field surrounding the main cluster core opposes this pull. While the DM peak can still move towards the receding subcluster, the gas peak is held back.
After the subcluster has passed the central region and moves away, the offset ICM falls back towards the main cluster DM peak and starts to slosh inside the main potential well. Generally, the highest frequencies of sloshing occur at small radii. Thus, gas at a given radius moving into one direction will frequently encounter gas from larger radii still moving in the opposite direction. Such opposing flows lead to the formation of density, temperature, and consequently X-ray brightness discontinuities. The formation of a similar discontinuity occurs at the leading side of a cool cloud moving through a hotter ambient medium (\citealt{Heinz2003}). As the central cool gas moves outwards into regions of lower pressure, adiabatic expansion enhances the temperature contrast across the fronts.  Thus, sloshing CFs are contact discontinuities between gases of different entropy, originating from different cluster radii. Usually, the subcluster passes the main cluster core at some distance, it transfers angular momentum to the ICM, and the sloshing takes on a spiral-like appearance.

AM06 have performed hydrodynamical SPH+$N$-body simulations confirming this scenario. Clearly, the gravity of the subcluster does not only influence the main cluster's ICM, but also its DM distribution. However, the interaction with the subcluster induces only a slow motion of  the central DM peak  w.r.t.~the overall cluster potential, while in the meantime the ICM sloshes inside the slowly moving central potential well. These simulations also demonstrated that the sloshing reproduces the morphology of observed CFs, where less massive subclusters cause weaker signatures.
 
AM06 also tested the impact of a subcluster containing a gaseous atmosphere of its own. They showed that the gas-gas interaction tends to leave clear observable signatures like a tail of ram-pressure stripped gas
and an overall much more disturbed appearance.
 As the clusters with observed sloshing CFs do not show such signatures, the scenario favours gas-free subclusters. 
\citet{Johnson2010} presented observations of the double cluster Abell 1644, where the two subclusters A1644N and A1644S appear to be merging. While both subclusters contain an ICM of their own, the core of A1644S exhibits a characteristic spiral-shaped CF. However, in contrast to the clusters with relaxed appearance and sloshing CFs, both subclusters of A1644 show a disturbed large-scale morphology.

The simulations of \citet{ZuHone2010} are similar to the ones of AM06, but concentrate on the heating efficiency of the sloshing process.

As an alternative to the scenario described above, the oscillation of the central galaxy or central dark matter peak of the cluster has been proposed (\citealt{Lufkin1995,Fabian2001,Tittley2005}). The moving central galaxy could experience ram pressure stripping by the surrounding ICM, and the contact discontinuity between the cooler central and warmer ambient gas could form the CFs. However, it is unclear whether this scenario would produce multiple fronts, which arise naturally in the sloshing scenario. 

\citet{Birnboim2010} propose that at least some CFs are caused by merging of shocks. In this case, the CFs should form quasi-spherical rings around the cluster centre, and could be found at large radii out to 1 Mpc. We show that sloshing CFs are accompanied by a characteristic large-scale asymmetry, which would be absent in this scenario.

%******
\subsection{Observed characteristics of sloshing CFs} \label{sec:intro_obs} 
Most sloshing CFs manifest themselves as brightness edges forming arcs around the cool cores of clusters that otherwise appear relaxed and show no obvious signatures of recent merging (\citealt{Markevitch2007}, \citealt{Owers2009hifid}). Often, pairs or even triplets of edges appear on more or less opposite sides of the cluster core. 
Their morphology depends mainly on the angle between our line-of-sight (LOS) and the orbital plane of the subcluster: If the interaction is seen face-on, the edges form a spiral-like structure. If the LOS is parallel to the orbital plane, we see arcs on alternating sides of the cluster core. The brightness and temperature contrast across the edges is more modest compared to the merging CFs. Typical temperature contrasts range between 1.2 and 2.5.

This type of CF is reported to be ubiquitous (\citealt{Markevitch2003,Markevitch2007}, \citealt{Ghizzardi2010}), but  high-resolution observations are available for only 13 clusters:  RX J1720.1+2638 (\citealt{Mazzotta2001,Mazzotta2008,Owers2008phd,Owers2009hifid}), MS1455.0+2232 (\citealt{Mazzotta2008,Owers2009hifid}), A2142 (\citealt{Markevitch2000,Owers2009hifid}), A496 (\citealt{Tanaka2006,Dupke2007}), 2A0335+096 (\citealt{Mazzotta2003,Sanders2009_2a}), A2029 (\citealt{Clarke2004,Million2009sample}), A2204  (\citealt{Sanders2005a2204,Sanders2009a2204}), Ophiuchus (\citealt{Million2009oph}),  A1795 (\citealt{Markevitch2001,Bourdin2008}), Perseus (\citealt{Churazov2003,Sanders2005perseus}), Centaurus (\citealt{Fabian2005centaurus,Sanders2006centaurus}), A1644 (\citealt{Johnson2010}).  Our work will concentrate on the Virgo cluster (\citealt{Simionescu2010}, S10 hereafter).

In addition to the edges in brightness and temperature, temperature maps show spiral- or arc-shaped cool regions inside the actual fronts. In X-ray brightness residual maps, an excess  corresponding spatially to the structure in the temperature map is found (A2029, A2204, 2A0335+096, Perseus). For some clusters (RX J1720.1+2638, MS1455.0+2232, A496, Ophiuchus, 2A0335+096, A2204, Centaurus), even a corresponding structure in metallicity maps is seen.  As X-ray brightness maps can be obtained more easily than temperature or metallicity maps, a spiral-shaped excess wrapped around the cluster core is often regarded as an indicator for ongoing sloshing.

Figure~\ref{fig:obs_excess} shows two versions of a residual X-ray brightness map for Virgo: 
%FFFFFFF
\begin{figure}
\begin{center}
\includegraphics[trim=0 0 0 50,clip,width=0.4\textwidth,angle=0]{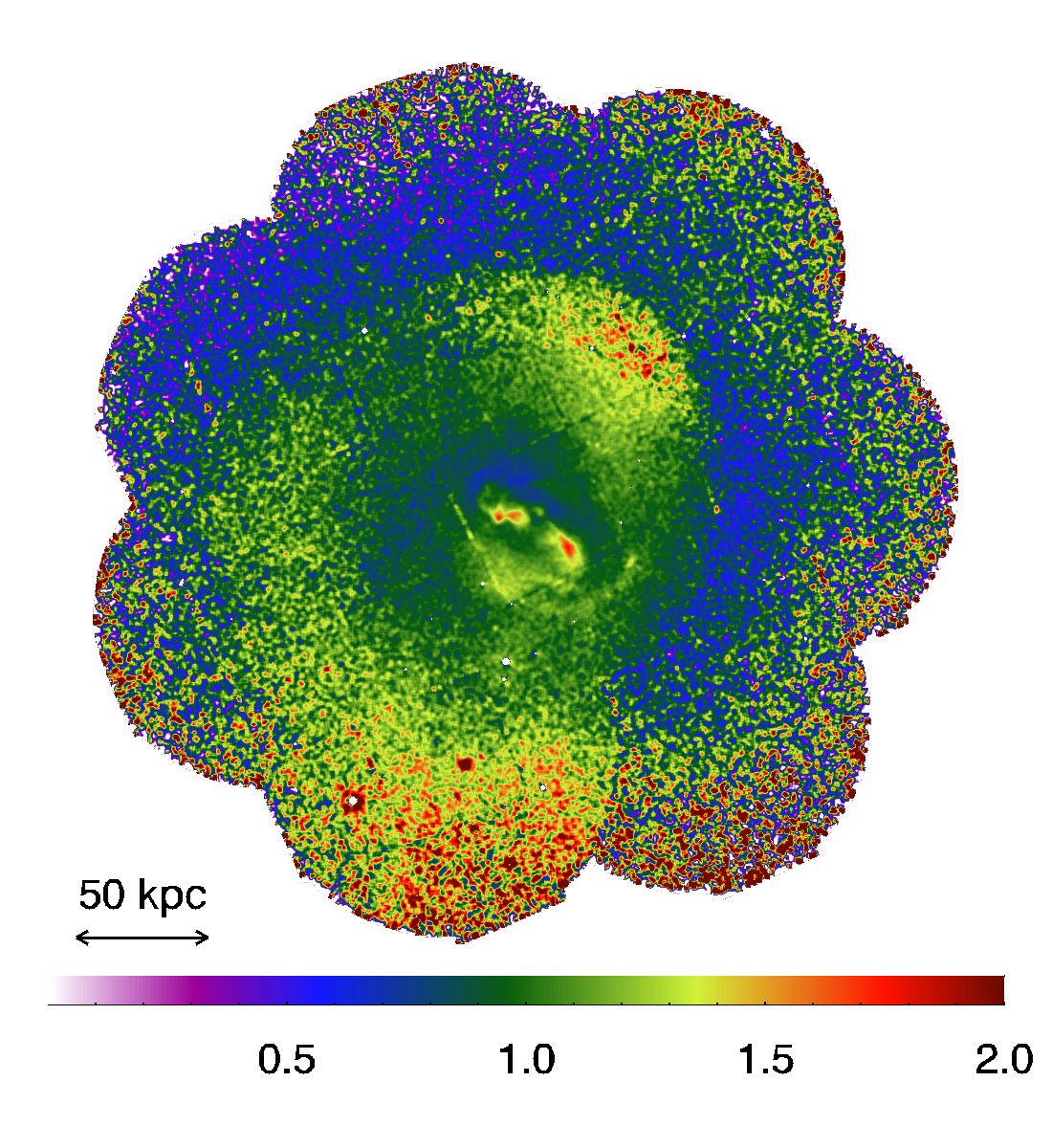}
\includegraphics[trim=0 200 0 50,clip,width=0.4\textwidth,angle=0]{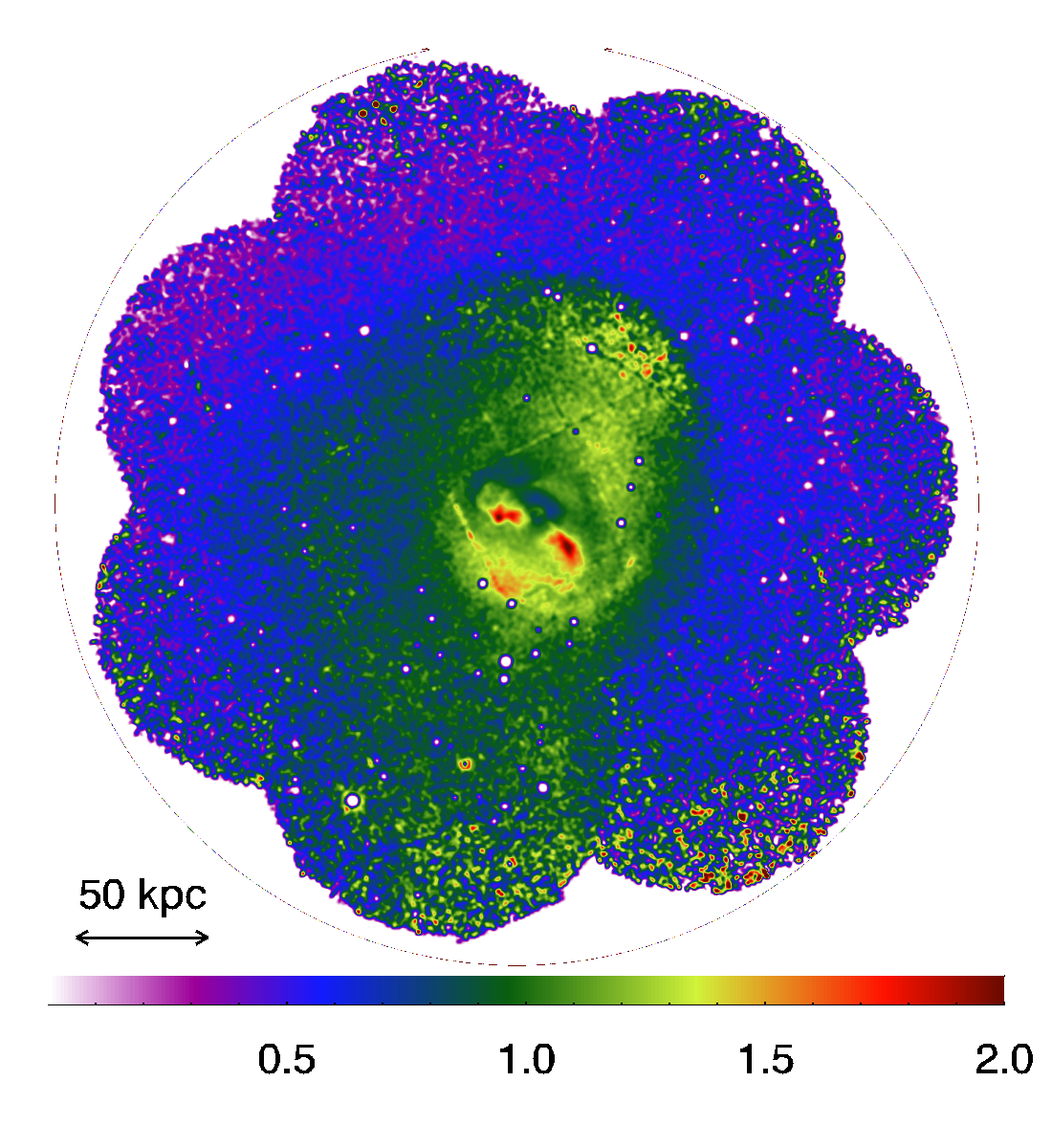}
\caption{Observed residual maps (data divided by average or model) of the Virgo cluster. The top panel shows residuals w.r.t.~the azimuthally averaged brightness. The bottom panel shows residuals w.r.t.~a $\beta$-profile fit to the observed X-ray brightness (same as in \citealt{Simionescu2010}, but with colour scale consistent to this paper).}
\label{fig:obs_excess}
\end{center}
\end{figure}
%FFFFFFF
The top panel displays the brightness residuals w.r.t. the azimuthally averaged brightness. The bottom panel shows residuals w.r.t. the  best-fitting radially symmetric $\beta$-model. This map is the same as Fig.~2 in S10, but in a colour-scale  used consistently throughout this paper. Both residual maps show a sharp brightness edge about 90 kpc  NW of the core.  The edge about 33 kpc towards the SE is somewhat more subtle. The Virgo observations are described in more detail in Sect.~\ref{sec:compare} along with the comparison to the simulation results.

%******
\subsection{Aims and outline} \label{sec:intro_aims}
Previous work has shown that the sloshing scenario (Sect.~\ref{sec:intro_scenario}) can in principle explain the known observed characteristics  (Sect.~\ref{sec:intro_obs})  of the resulting CFs. 
The quantitative characteristics of the CFs in a cluster (radius, contrast of X-ray brightness, temperature, metallicity across the edges, morphology in maps) depend on several parameters:
\begin{itemize}
\item host cluster potential and ICM distribution;
\item characteristics of subcluster: mass, size, pericentre of orbit, velocity along orbit;
\item age: time since the pericentre passage of the subcluster;
\item angle of our line-of-sight (LOS) w.r.t.~the orbit and orbital plane of the subcluster.
\end{itemize}

In this work, we concentrate on the Virgo cluster and focus on the following aims:
\begin{itemize}
\item Test the sloshing scenario quantitatively. We show that we can find a reasonable set of parameters to reproduce the observations well (Sect.~\ref{sec:compare}). We note that there is a degeneracy between subcluster mass and pericentre distance/velocity along orbit, so that only a combination of these can be constrained.
\item Refine our understanding of observable signatures of sloshing CFs (Sect.~\ref{sec:newfeatures}).
\item Study the ability of sloshing to redistribute metals throughout the cluster core (Sect.~\ref{sec:metals}). 
\item Test if the basic assumption of the scenario, i.e., that the subcluster is gas-free (see Sect.~\ref{sec:intro_scenario}), can be true (Sect.~\ref{sec:gas}). Once we have identified a suitable subcluster, we show that a gaseous atmosphere could be ram-pressure stripped off this subcluster before it reaches the cluster core.  However, the outcome of this test is not unique, and depends on the details of the subcluster's atmosphere. 
\item Test an alternative scenario: \citet{Churazov2003} propose that the passage of a shock through a cluster can cause the formation of CFs and demonstrated this with an idealised plane shock. We study the more realistic situation of the bow shock of a galaxy and its ability to trigger sloshing and CF formation (Sect.~\ref{sec:bowshock}).
\end{itemize}

To this end, we run hydrodynamical simulations of an idealised minor merger of a (gas-free) subcluster with a main cluster (see Sect.~\ref{sec:method}), which induces sloshing of the ICM of the main cluster. We perform a parameter study, varying the discussed parameters, and present a detailed comparison to the observations (Sects.~\ref{sec:compare}, \ref{sec:newfeatures}).

%% file: method.tex
%*************
\section{Method}  \label{sec:method}
%*************
%
In a set of idealised merger simulations,  we model the interaction between a massive galaxy cluster ("main cluster") and a smaller subcluster ("disturber").  We concentrate on minor mergers, where the disturber is significantly less massive than the main cluster. A major merger would  cause additional structure, e.g., destroy the cool core, which is not the case for sloshing CF clusters. In the Virgo cluster, only in the inner few kpc the cooling time is shorter than 2 Gyr. Hence, in the outer regions of interest for our analysis, the cooling time is long and we neglect radiative cooling in our simulations. Finally, we will only model the first core passage of the subcluster and the subsequent gas sloshing.

%*************
\subsection{Code}

Our simulations are run using the FLASH code (version 3.2, \citealt{Dubey2009}).  FLASH is a modular block-structured AMR code, parallelised using the Message Passing Interface (MPI) library. It solves the Riemann problem on a Cartesian grid using the Piecewise-Parabolic Method (PPM). The simulations are performed in 3D and all boundaries are reflecting. We use a simulation grid of size $3\times 3.5\times 3\Mpc^3$. 
In most simulations, we resolve the inner 32 kpc with $\Delta x=1\Kpc$, the inner $64\Kpc$ with $\Delta x=2\Kpc$, and the inner 128 kpc with $\Delta x=4\Kpc$. At this moderate resolution, one simulation can be completed in about 100 CPUh. In Appendix~\ref{sec:resolution}, we present a high-resolution run of our fiducial case which resolves the inner 8 kpc with $\Delta x=0.5\Kpc$ and the inner 128 kpc with $\Delta x=1\Kpc$, and a very high-resolution run with an again twice as good resolution.
None of our results depends on resolution.

%*******************
\subsection{Rigid potential approximation} \label{sec:method_PRA}
The gas dynamics is described by the hydrodynamical equations. Additionally, the ICM is subject to the gravitational acceleration due to the main cluster and the subcluster. In order to sample the parameter space of the subcluster mass, scale radius, pericentre distance, and apocentre distance with a sufficient grid of simulations, we use the computationally much cheaper rigid potential approximation for both, the main and the subcluster.

Our simulations are run in the rest frame of the main cluster. The orbit of the subcluster through the main cluster is assumed to be that of a test particle falling through the main cluster, given an initial position and velocity. 

As the rest frame of the main cluster is no inertial frame, the ICM is subject to a pseudo-acceleration due to the attraction of the main cluster core towards the approaching subcluster. The simplest approximation to account for this is to assume that the main cluster responds to the gravity of the subcluster as a whole, like a rigid body. Therefore, we calculate the inertial acceleration felt by the main cluster centre due to the subcluster and add this pseudo-acceleration to the entire ICM. This procedure was also used by \citet{ZuHone2010}. It is a reasonable approximation for the central region of the main cluster, but it is inaccurate for its outer parts, where it will lead to unrealistic flows. For small pericentre distances, during pericentre passage of the subcluster the inertial acceleration can be large and even produce supersonic motions in the outer parts of the main cluster. Hence, we apply the pseudo-acceleration only to the central region of the main cluster inside a characteristic radius, $R\damp$, and damp it outside this radius exponentially with a length scale, $L\damp$. Roediger \& ZuHone (in prep.) perform a direct comparison between sloshing simulations using this rigid potential approximation and full hydro+Nbody simulations for a more massive cluster resembling A 2029. They show that for $R\damp\gtrsim 2a\Sub$, where $a\Sub$ is the scale radius of the subcluster,  and $L\damp \approx 0.5 R\damp$ the rigid potential simulations reproduce the full hydro+Nbody simulations well. The orientation of the cold spiral structure is reproduced accurately. The evolution of the size of the CF structure lags behind in the rigid potential approximation by about 200 Myr. When this lag is taken into account, the size of the CF structure is reproduced well. Due to this artefact, we will over-estimate the age of the CFs derived from the rigid potential simulations by about 200 Myr. For a given size of the CF structure, the density contrast across the  CFs is reproduced well. The asymmetry in density out to at least 500 kpc is reproduced qualitatively.  The temperature at the inner side of each CF is also reproduced well. The temperature just outside the CFs is too high due to enhanced compressional heating from unrealistic gas motions outside the central region. The damping of the inertial frame correction explained above can reduce the over-estimate of the temperature on the outer side of the CFs somewhat, but cannot prevent this artefact completely.  

In the case of Virgo, we use $R\damp = 500\Kpc$ and $L\damp =300\Kpc$. Apart from lowering the temperature outside the CFs, this choice does not influence the resulting CF structure.

%*******************
\subsection{Model setup}

The Virgo cluster ICM is modelled by a spherically symmetric density and temperature profile, chosen to fit observational data.  Assuming  hydrostatic equilibrium, we calculate the gravitational acceleration due to the underlying main cluster potential as a function of radius and thus the underlying cluster potential and mass profile. (see Sect.~\ref{sec:maincluster}.)
When the cluster is allowed to evolve for 2 Gyr in isolation, no sloshing or other modification of the ICM distribution is observed.

Once we know the main cluster's potential, we calculate the orbit of a test mass moving through this potential.
The subcluster will move along one of the orbits discussed in Sect.~\ref{sec:method_orbits}. 
Initially, the subcluster is placed at its orbit with a position and velocity such that it passes the pericentre after 1 Gyr. For the "slow", "medium" and "fast" orbits (see Sect.~\ref{sec:method_orbits}) this corresponds to an initial separation of approximately 1 Mpc,  1.4 Mpc, and 1.6 Mpc, respectively.
 During the course of the simulation, the subcluster potential is shifted along its orbit through the main cluster (see Fig.~\ref{fig:orbits}). The time normalisation is chosen such that pericentre passage happens at $t=0$.

We stop the simulations at $t\approx 1.7 \Gyr$ after core passage when a good match with the observed CFs in Virgo is reached. This moment is well before the second core passage of the subcluster.

%*********************
\subsection{Cluster models}

%**********
\subsubsection{Main cluster -- Virgo} \label{sec:maincluster}
%
%FFFF
\begin{figure}
\begin{center}
\includegraphics[width=0.4\textwidth]{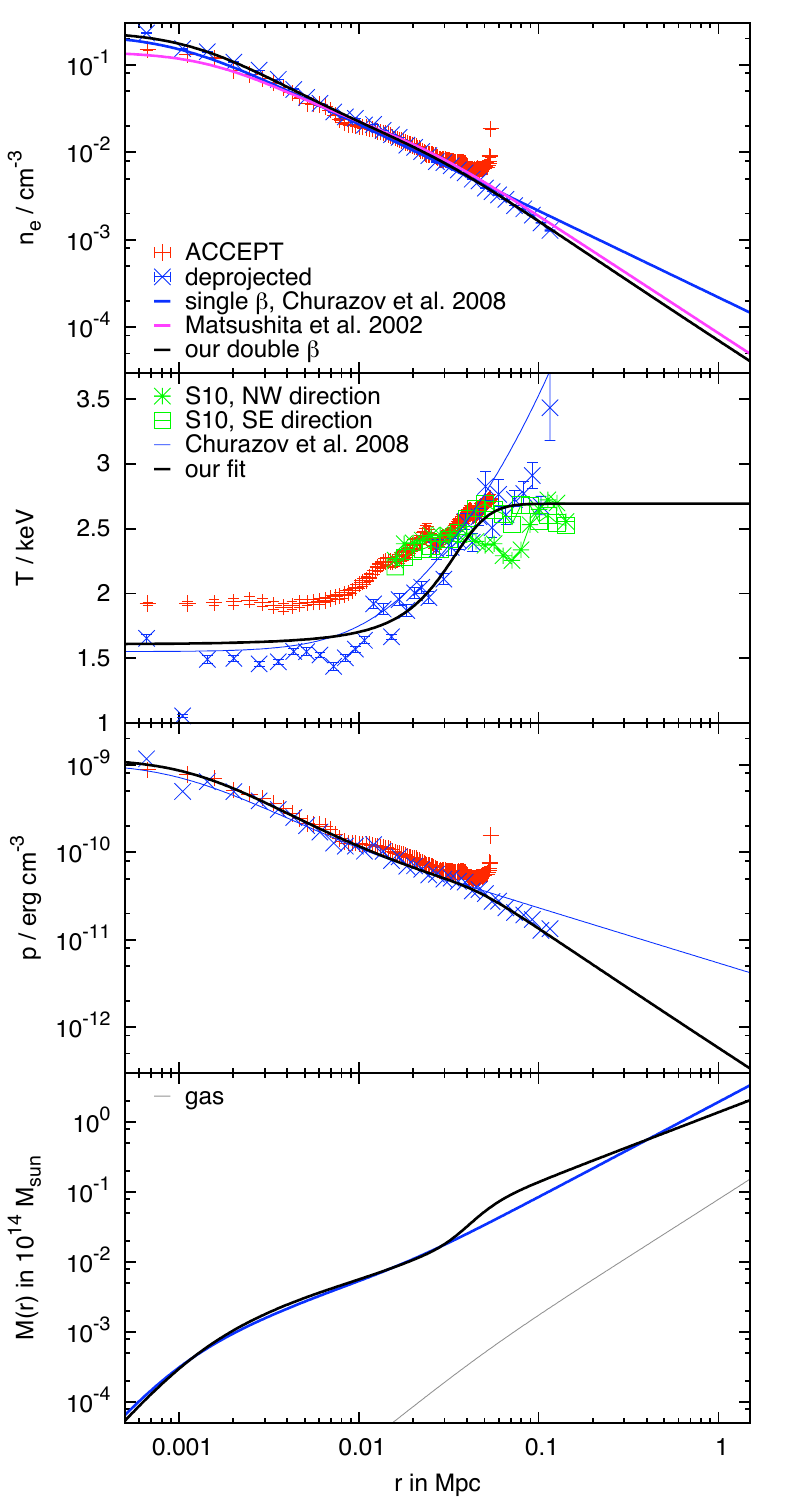}
\caption{Overview of the Virgo cluster model, comparison with observations:  Panels show electron density, temperature, pressure,  and cumulative mass. For details see Sect.~\ref{sec:maincluster}.}
\label{fig:iniprofs}
\end{center}
\end{figure}
%FFFFFFFF
%
Figure~\ref{fig:iniprofs} summarises the characteristics of our Virgo cluster model. The first panel shows observational data for the electron density, $n_e$, from the ACCEPT sample (\citealt{Cavagnolo2009}) and the deprojected profile (Churazov, private comm.),  along with a single-$\beta$ model fit from \citet{Churazov2008}, and a double-$\beta$ model fit from \citet{Matsushita2002}. In order to model the Virgo cluster as accurately as possible, we fit the deprojected data with a double-$\beta$ model. The second panel shows the observed projected temperature profile from the ACCEPT sample, the deprojected data, and a fit from \citet{Churazov2008}. We fit the deprojected data with 
%--------
\begin{eqnarray}
T(r)&=&T_1(r)C(r) + T_2(r)[1-C(r)], \textrm{where} \label{eqn:temp}\\
T_1(r)&=&m_1 \cdot r + T_{01}, \nonumber\\
T_2(r)&=&m_2 \cdot r + T_{02}, \nonumber\\
n_2&=&(m_1-m_2)r\Break + n_1 \Leftrightarrow T_1(r\Break)=T_2(r\Break) ,\nonumber\\
C(r)&=& 1-    \frac{1}{ 1+  \exp \left( -\frac{r-r\Break}{a\Break }\right)  } \nonumber,
  \end{eqnarray}
%=======
which connects two linear functions smoothly.
At large radii, we use a constant temperature, contrary to the fit of  \citet{Churazov2008}. The next panel shows the pressure profiles resulting from the ACCEPT data, the deprojected data,  the \citet{Churazov2008} fits, and our fits to density and temperature. Assuming hydrostatic equilibrium, we can calculate the underlying potential and the associated cumulative DM mass, which is shown in the fourth panel, 
along with the cumulative mass for the ICM atmosphere. 
 Table~\ref{tab:Virgo_parameters} summarises our ICM fit parameters. 
\begin{table}
\caption{ICM parameters for the Virgo cluster model.}
\begin{center}
\begin{tabular}{|l|c|c|}
\hline
density: & \multicolumn{2}{c}{double $\beta$ profile} \\
\hline
core radii $r_{1,2}/\Kpc$: & 1.20 & 23.68\\
core densities $\rho_{01,02}/(\gccm)$:  & $4.42 \cdot 10^{-25}$ & $1.38\cdot 10^{-26}$\\
$\beta_{1,2}$: & 0.42 & 0.52\\ 
\hline
temperature: & \multicolumn{2}{c}{ see Eqn.~\ref{eqn:temp}}\\
\hline
slopes $m_{1,2}/(\K \PC^{-1})$: & -5.6 &  0.\\
$T_{01,02}/(10^7\K)$: & 1.78 & 3.12\\
break radius $r\Break/\Kpc$: & \multicolumn{2}{c}{ 27.7 }\\
break range $a\Break/\Kpc$: & \multicolumn{2}{c}{ 10 }\\
\hline
\hline
\end{tabular}
\end{center}
\label{tab:Virgo_parameters}
\end{table}%

We study two different initial metal distributions, a steep and a flatter one, which are described in Sect.~\ref{sec:metals_ini}.

%**********
\subsubsection{Subcluster}\label{sec:method_subcluster}

The gravitational potential of our subclusters is described by a Hernquist halo (\citealt{Hernquist1990}). We vary the mass and size of the disturber between $0.5$ and $4\cdot 10^{13}M\Sun$ and between 50 and $200\Kpc$, respectively (see Table~\ref{tab:exclude}). At the upper end of our mass range, we are restricted by the condition that we want to model minor mergers only. A mass of $4\cdot 10^{13}M\Sun$ equals the mass of the inner 300 kpc of Virgo. Hence, an interaction with higher mass subclusters would not be a minor merger any more and would leave obvious imprints in the Virgo cluster, which are not observed. The lower end of the mass range is determined such that the subcluster still leaves an imprint that is comparable to the observations.   

We abbreviate the combinations of subcluster mass, size and the pericentre 
and apocentre distance 
of its orbit as M$X$a$Y$dmin$Z$dmax$Q$, where "M$X$" shall indicate a subcluster mass of $X\cdot 10^{13}M\Sun$, a$Y$ a scale radius of $Y\Kpc$, dmin$Z$ the pericentre distance of $Z\Kpc$, and dmax$Q$ the apocentre distance of $Q\Mpc$. E.g. in our fiducial run, M2a100dmin100dmax3, a subcluster with $2\cdot 10^{13}M\Sun$ and scale radius $100\Kpc$ passes the cluster centre with a minimal distance of 100 kpc, 
moving along an orbit where the maximal separation between both clusters was $3\Mpc$.

%**********
\subsubsection{Orbits}  \label{sec:method_orbits}
\begin{figure}
\includegraphics[width=0.45\textwidth]{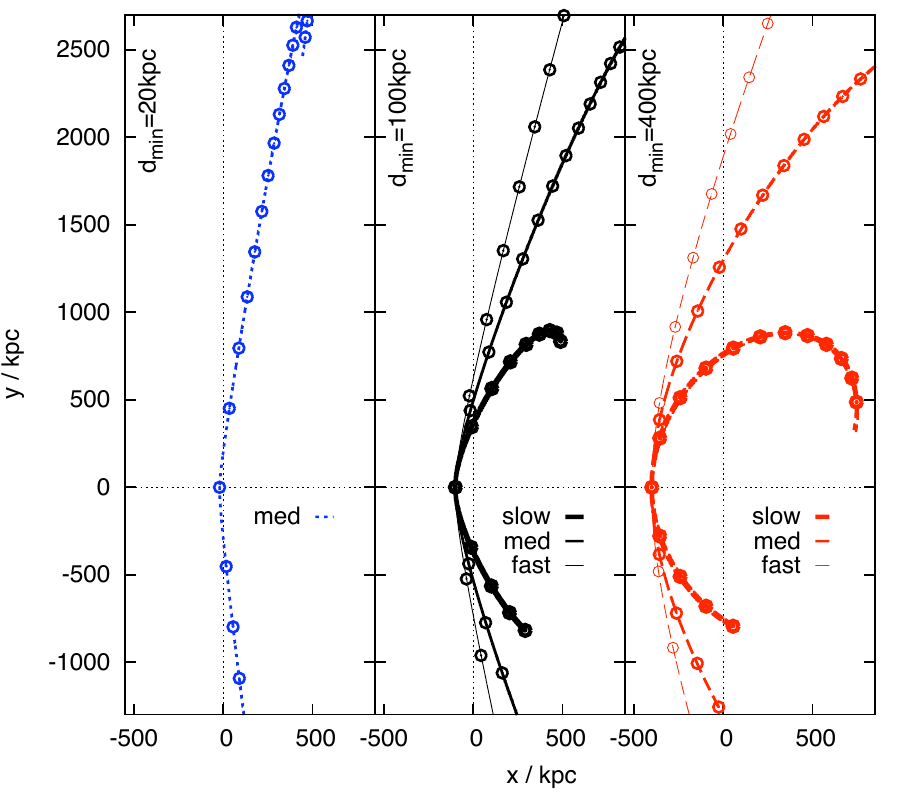}
\centering\includegraphics[width=0.38\textwidth]{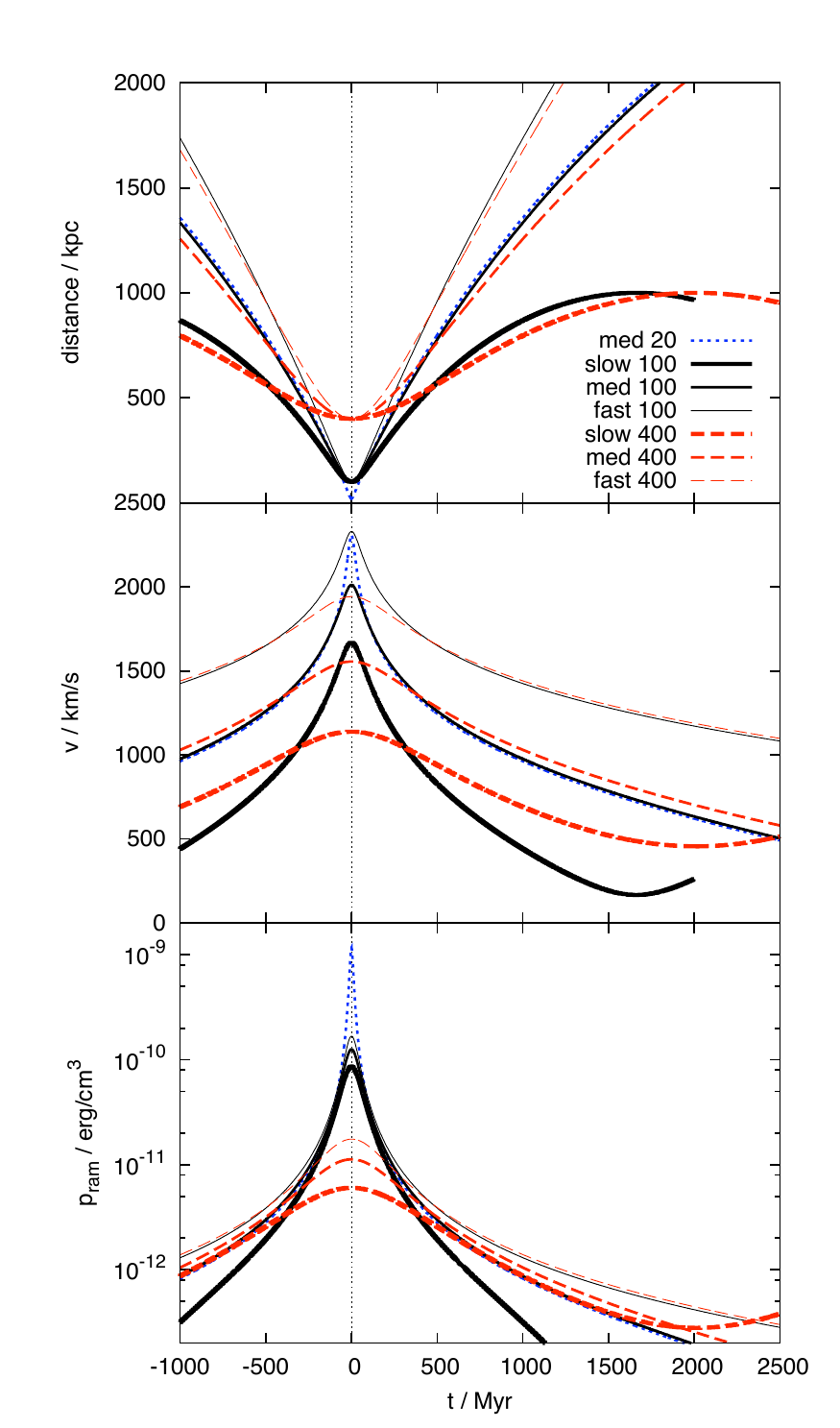}
\caption{Different orbits of the subcluster: The \textbf{top figure} displays the orbits in the $xy$-plane, one pericentre distance per panel. The line thickness codes the orbital velocity: slow, medium and fast orbits arise from an initial separation between subcluster and main cluster of 1, 3, and 10 Mpc, respectively. The circles along the orbit are spaced by $250\Myr$. We initialise our subclusters in the "lower" ($-y$-direction) part of the orbital plane. The pericentre passage happens at $t=0$. The \textbf{bottom figure} displays the evolution of the distance to the cluster centre,  the total velocity, and the ram pressure as a function of time for each orbit (see legend, pericentre distances are given in kpc).}
\label{fig:orbits}
\end{figure}

Slower subclusters on orbits with smaller pericentre distances spend more time close to the cluster core and thus should cause a stronger impact. We test these expectations by modelling the passage of the subcluster along different orbits, whose characteristics are summarised in Fig.~\ref{fig:orbits}.
%FFFFFFFF
%FFFFFFFFF
We study orbits with pericentre distances, $d\Min$, of 20, 100, and 400 kpc. We construct three groups of orbits, namely
"slow", "medium", and "fast, 
which arise from a maximum initial separation between subcluster and main cluster of 1, 3, and 10 Mpc, respectively. 

The subcluster approaches the main cluster core from the $-y$-direction, passes left of the main core at $t=0$, and recedes towards the $+y$-direction.

%% file: compare.tex
%*******************************
\section{Comparison with observations -- inferring merger characteristics} \label{sec:compare}
In our simulations, the gas sloshing proceeds as described in Sect.~\ref{sec:intro_scenario} and is qualitatively independent of subcluster and orbit characteristics. In order to focus the reader's attention to the aims of this work, we give a brief description of the sloshing dynamics in Appendix~\ref{sec:fiducial}.

With the aim of inferring the properties of the responsible subcluster and its orbit, we here present a detailed comparison between true and synthetic observations of the CFs in the Virgo cluster and apply our results also to other CF clusters.

\begin{figure}
\includegraphics[trim=30 -25 275 30,clip,height=4.2cm]{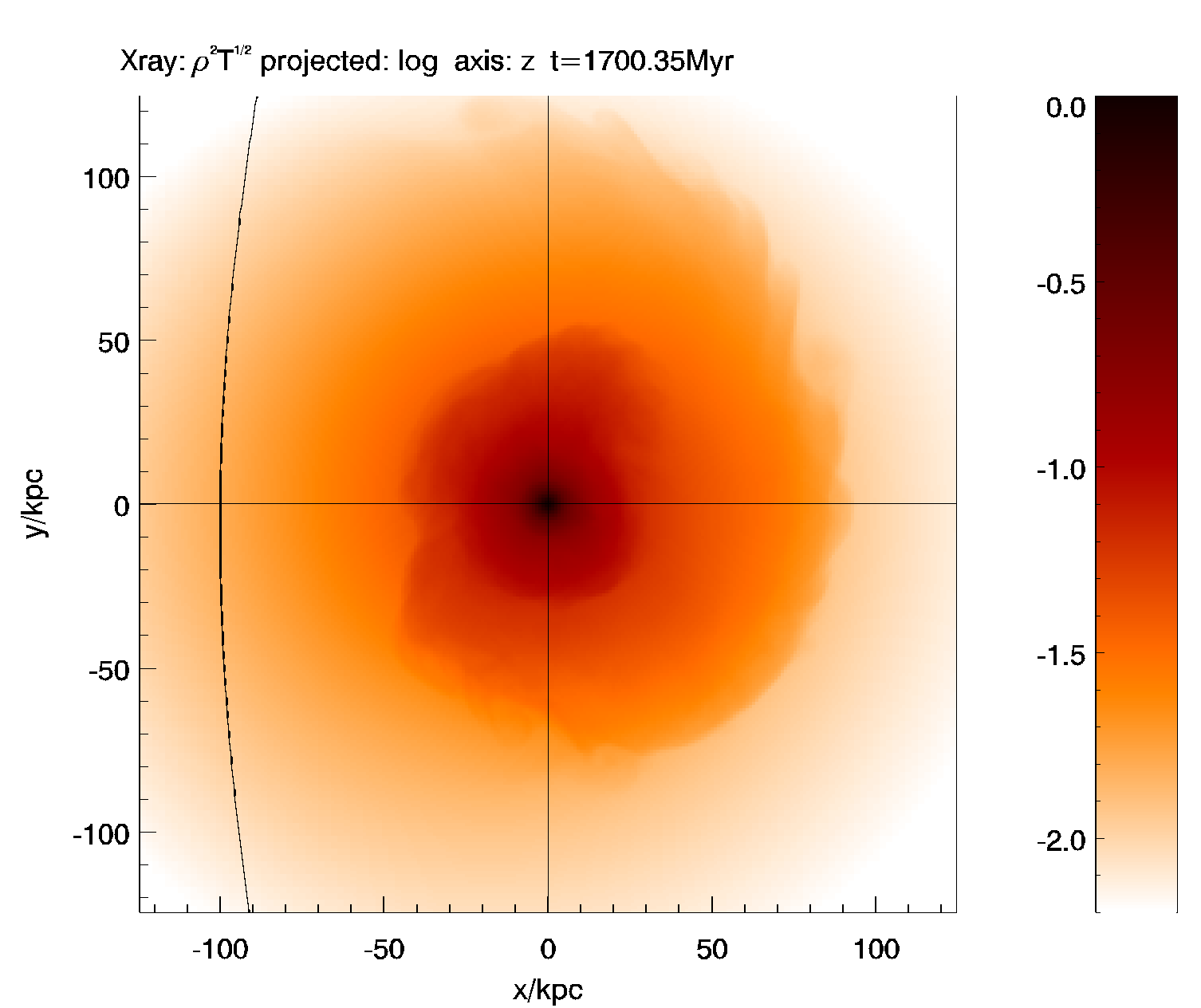}
\includegraphics[trim=1300 -25 100 30,clip,height=4.2cm]{PLOTS/M2.13_I100MAX3_A100_HR_DAMP/proj_Xray_z_size2_0270}
\includegraphics[trim=30 0 215 115,clip,angle=90,height=4.2cm]{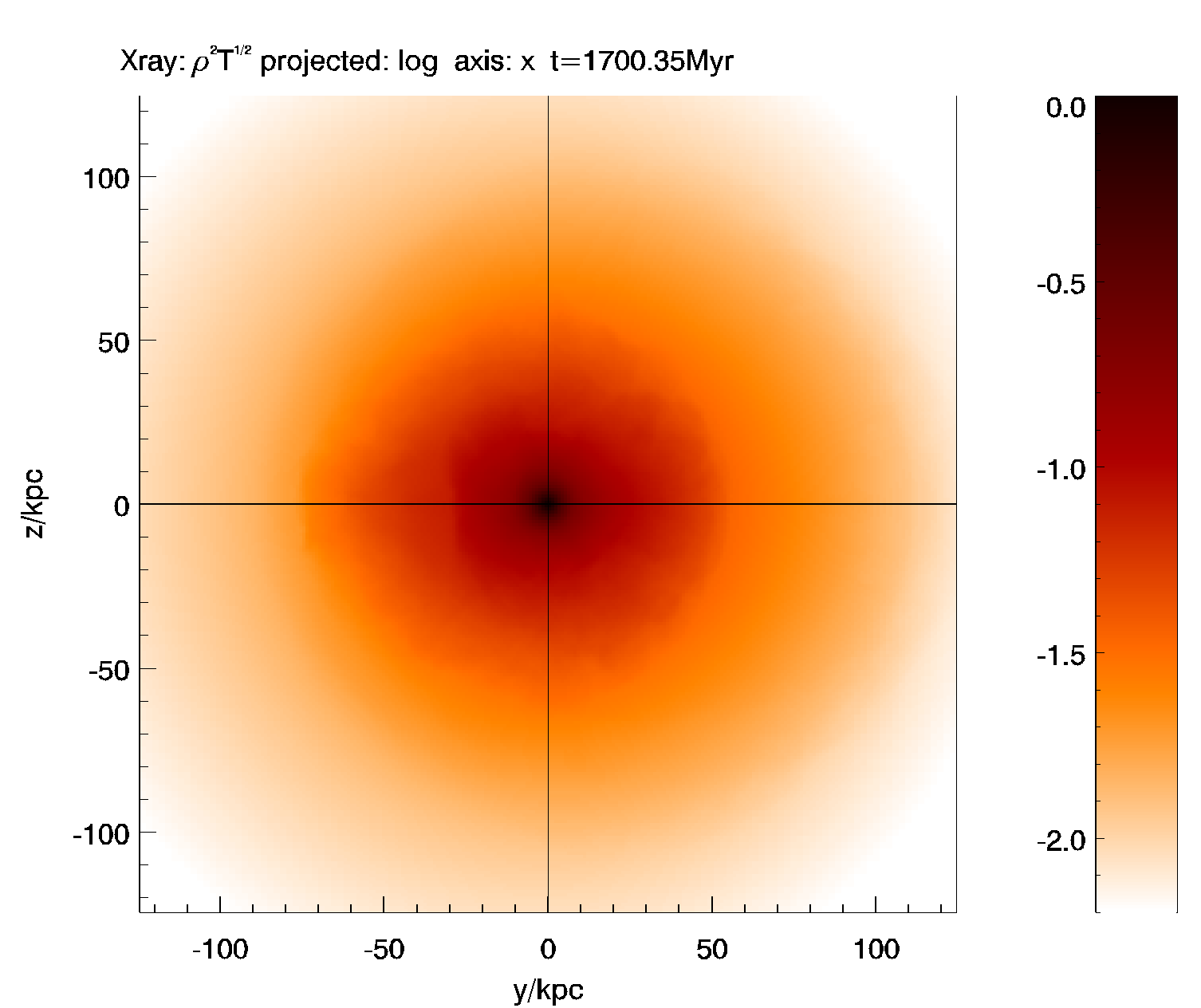}
\newline
\includegraphics[trim=30 -25 275 115,clip,angle=0,height=4.cm]{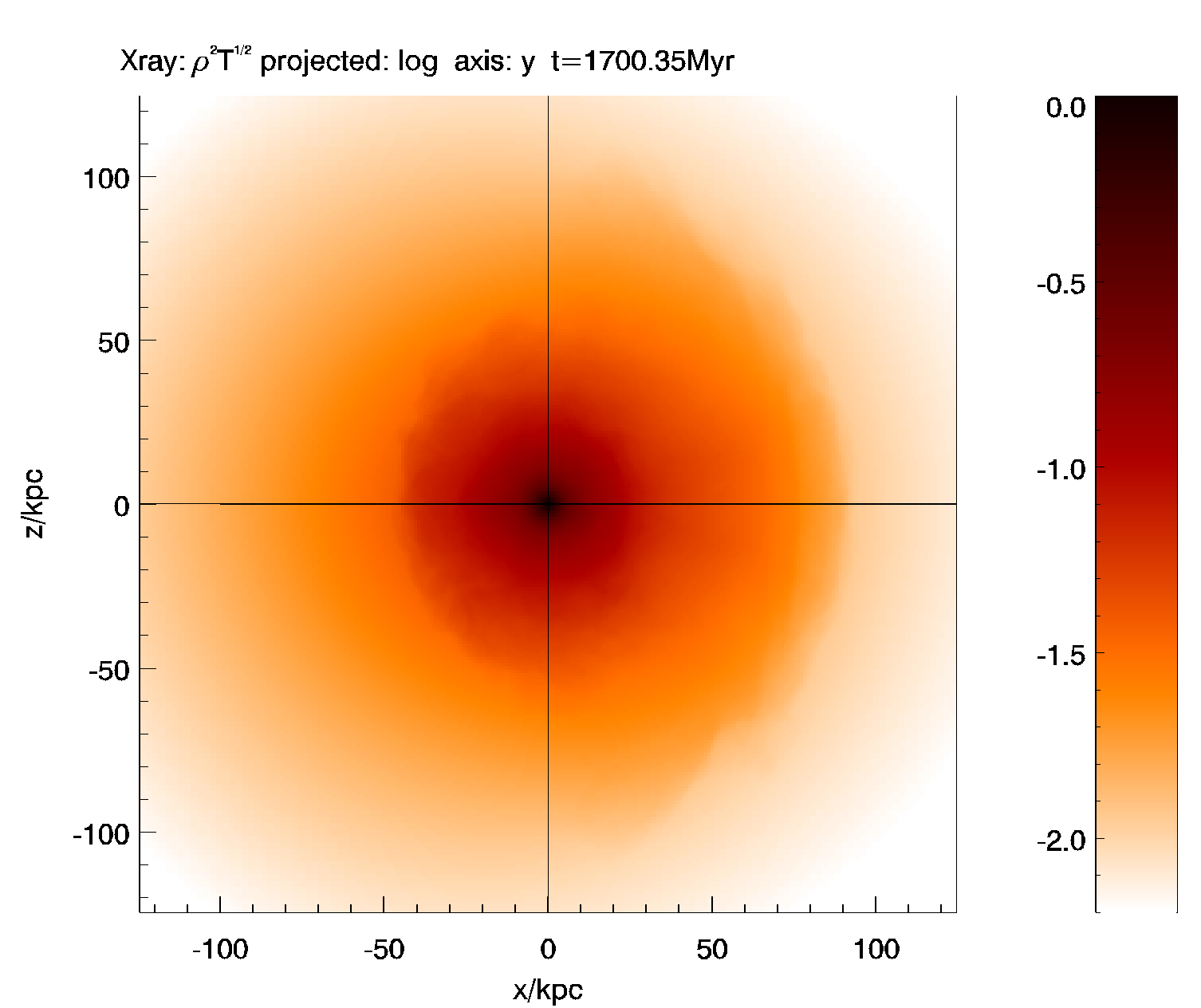}
\hspace{-0.3cm}
\includegraphics[trim=160 275 180 140,clip,height=3.2cm,origin=t,angle=-45]{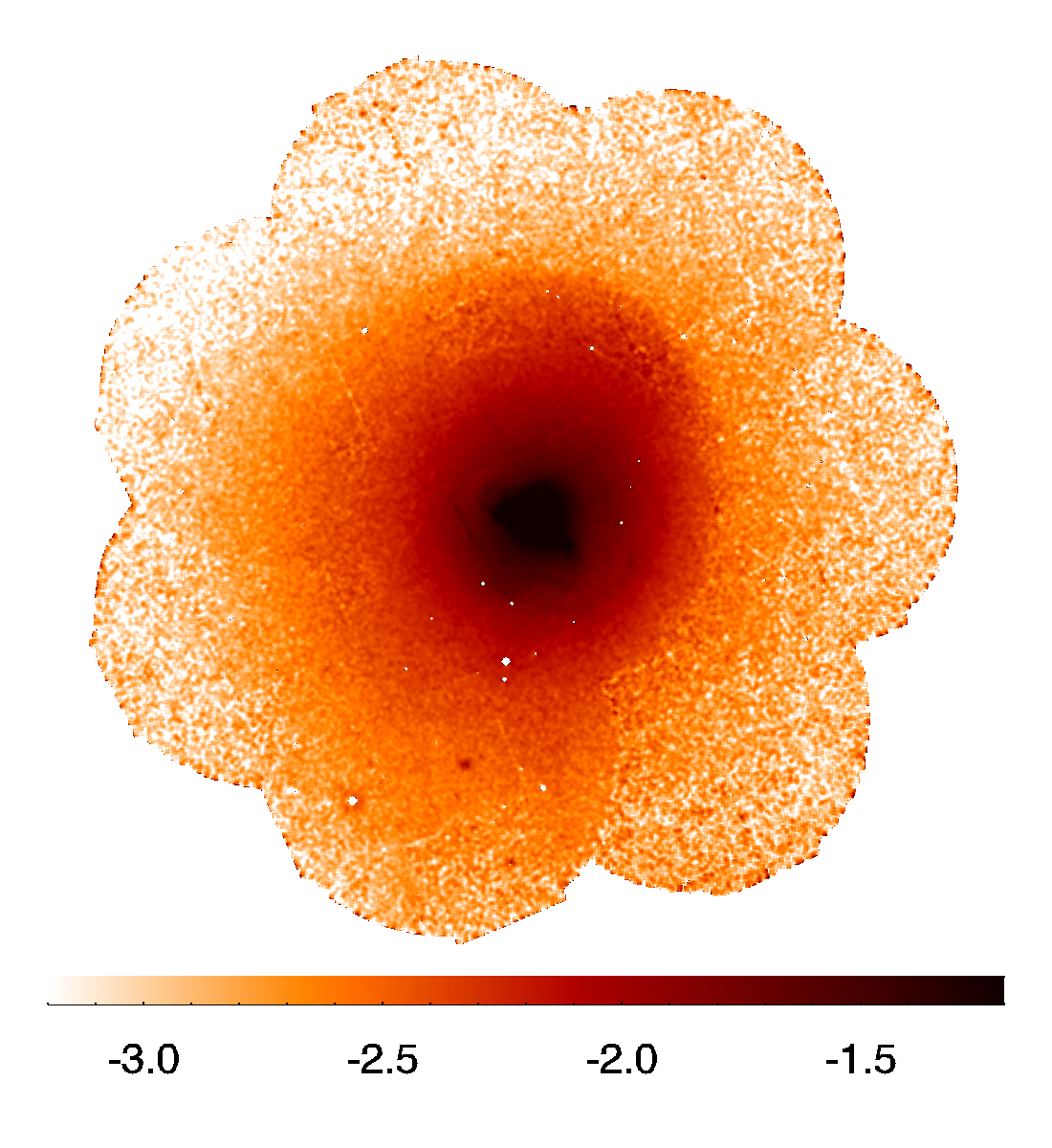}
\caption{
X-ray images: Synthetic images for fiducial run (high-resolution case, compare to Appendix~\ref{sec:resolution}) at $t=1.7\Gyr$ along three LOSs. \textbf{Top left:} Along the $z$-axis, which is perpendicular to the orbital plane. The subcluster orbit is marked by the black line.  \textbf{Bottom left:} Along the $y$-axis, i.e., along the orbit. \textbf{Top right:} Along the $x$-axis. The brightness is given in arbitrary logarithmic units (see colour scale). \textbf{Bottom left:} Observed X-ray image (same as in S10, but consistent colour scale), oriented such that NW is right and NE is up.
}
\label{fig:fiducial-xray}
\end{figure}

\begin{figure}
\begin{center}
\includegraphics[trim=30 -25 275 30,clip,height=4.2cm]{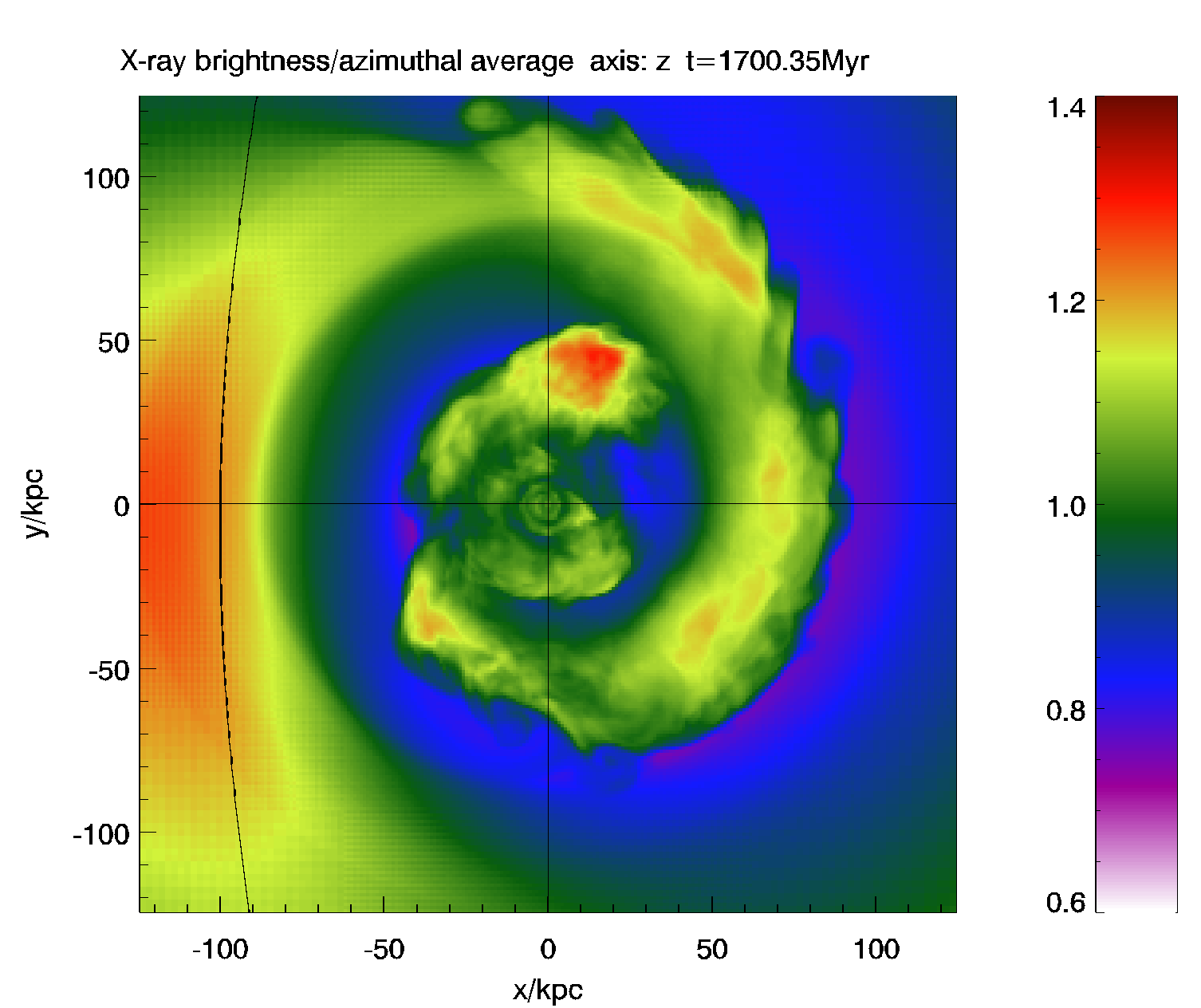}
\includegraphics[trim=1300 -25 100 30,clip,height=4.2cm]{PLOTS/M2.13_I100MAX3_A100_HR_DAMP/proj_Excess_z_size2_0270}
\includegraphics[trim=30 0 215 115,clip,angle=90,height=4.2cm]{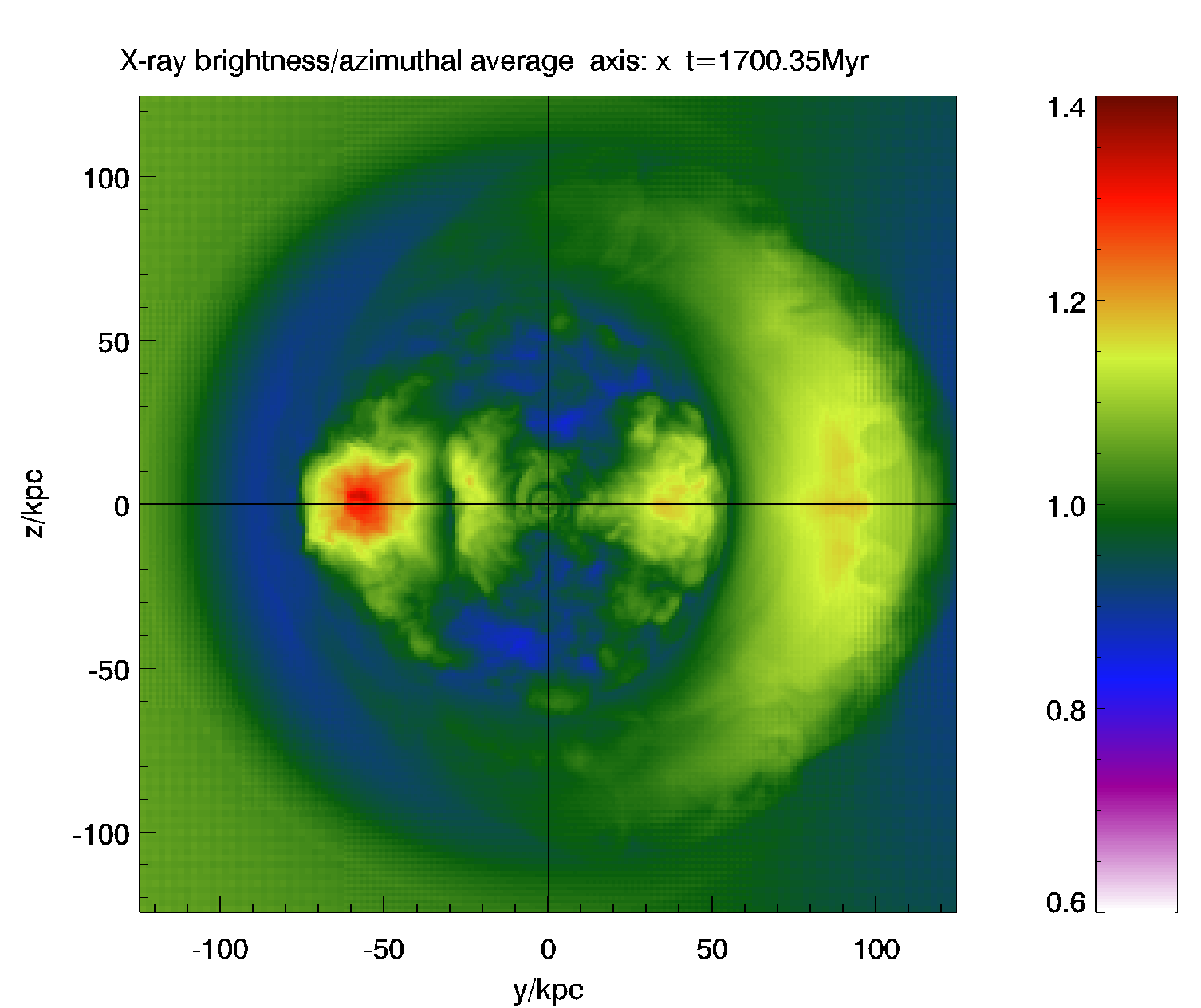}
\newline
\includegraphics[trim=30 -25 275 115,clip,angle=0,height=4.cm]{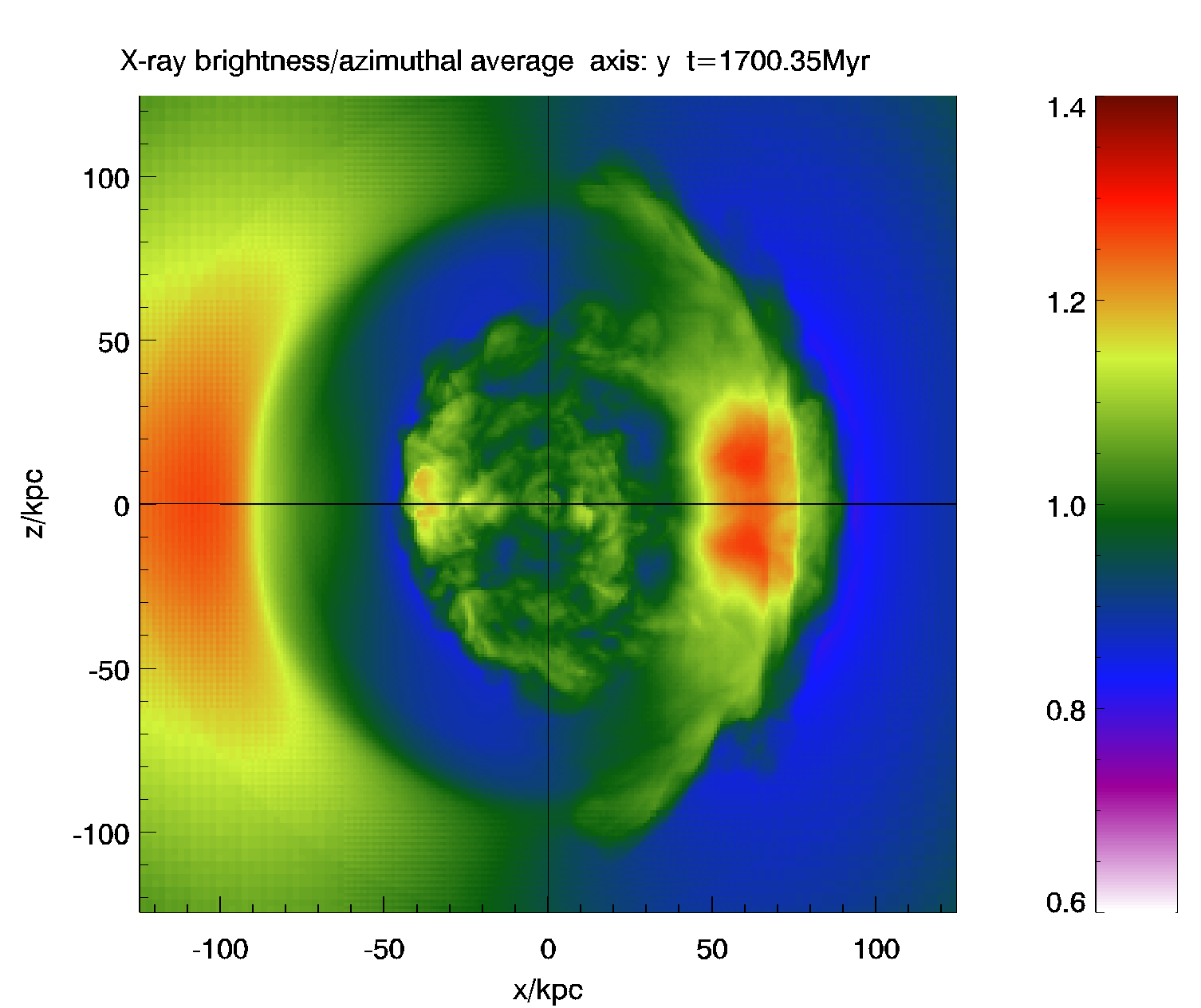}
\hspace{-0.4cm}
\includegraphics[trim=180 300 210 180,clip,height=3.cm,origin=t,angle=-45]{PLOTS/Virgo_residual}
\end{center}
\caption{
Brightness residual maps corresponding to Fig.~\ref{fig:fiducial-xray}. For the observed map, compare to Fig.~\ref{fig:obs_excess}.
}
\label{fig:fiducial-residual}
\end{figure}

Qualitatively, the synthetic observations of all simulation runs are similar.  As the representative case, we show maps of X-ray brightness, brightness residuals\footnote{residuals w.r.t.~the azimuthal average}, and projected temperature for our fiducial run M2a100dmin100dmax3   in Figs.~\ref{fig:fiducial-xray}, \ref{fig:fiducial-residual} and \ref{fig:fiducial_Tproj}, respectively.

The preparation of our synthetic observations is described in Appendix~\ref{sec:synthetic}.

%FFFFFFFFFFFFFFF
%FFFFFFFFFFF
%
%FFFFFFFFFF
%FFFFFFFFFF
%FFFFFFFF
\begin{figure}
\includegraphics[trim=30 -25 275 30,clip,height=4.2cm]{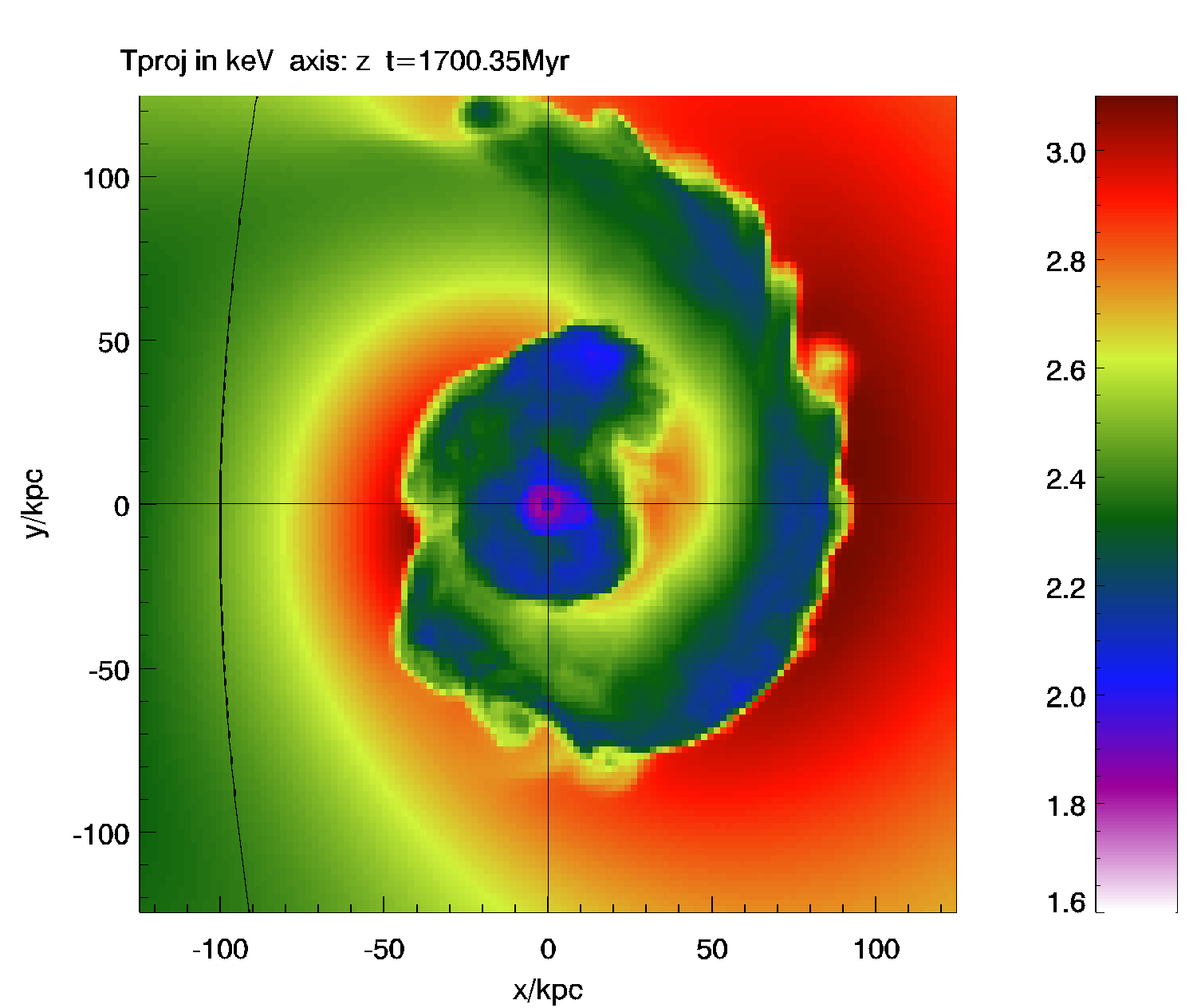}
\includegraphics[trim=1300 -25 100 30,clip,height=4.2cm]{PLOTS/M2.13_I100MAX3_A100_HR_DAMP/proj_Tproj_z_size2_0270}
\includegraphics[trim=30 0 215 115,clip,angle=90,height=4.2cm]{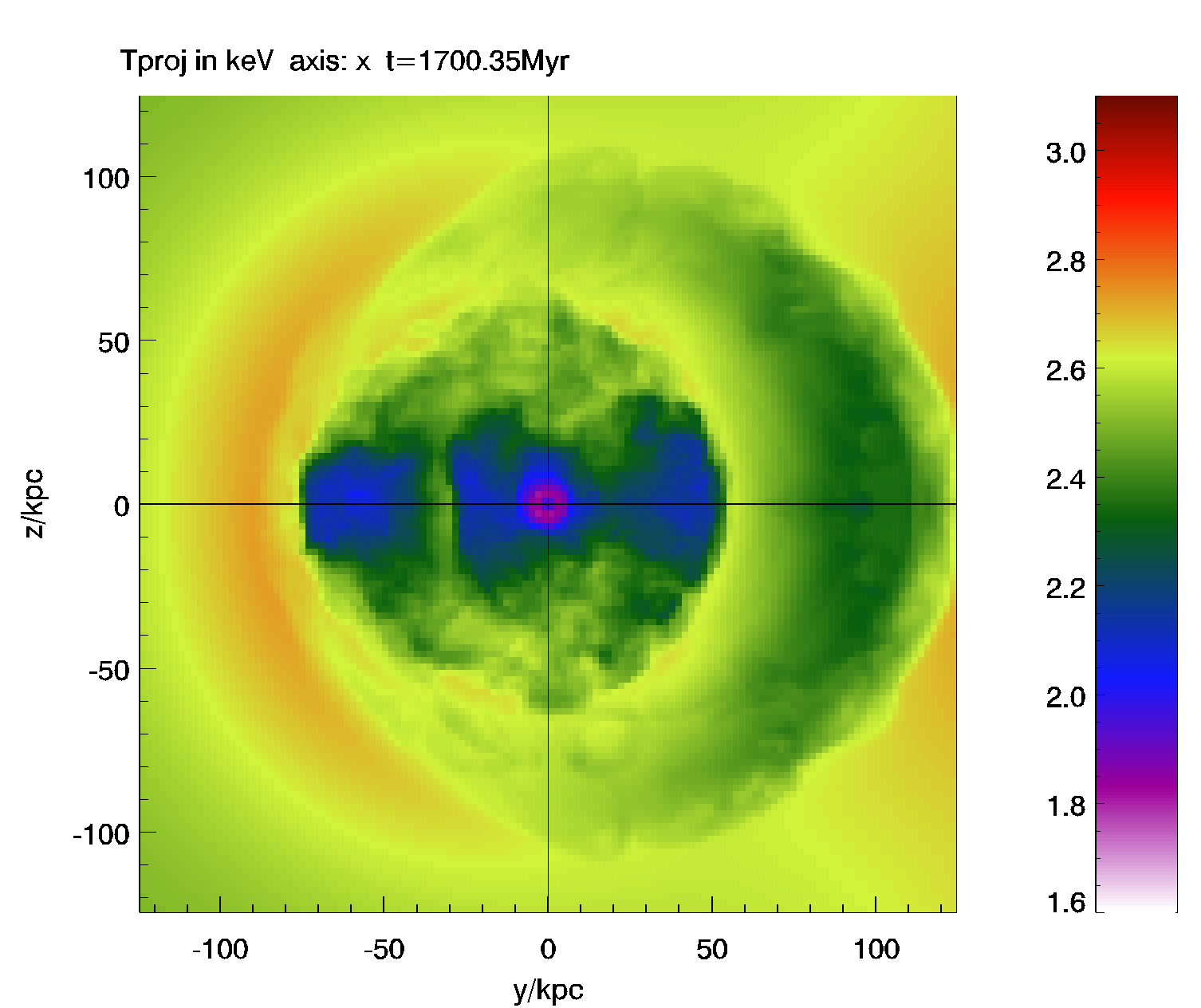}
\newline
\includegraphics[trim=30 -25 275 115,clip,angle=0,height=4.cm]{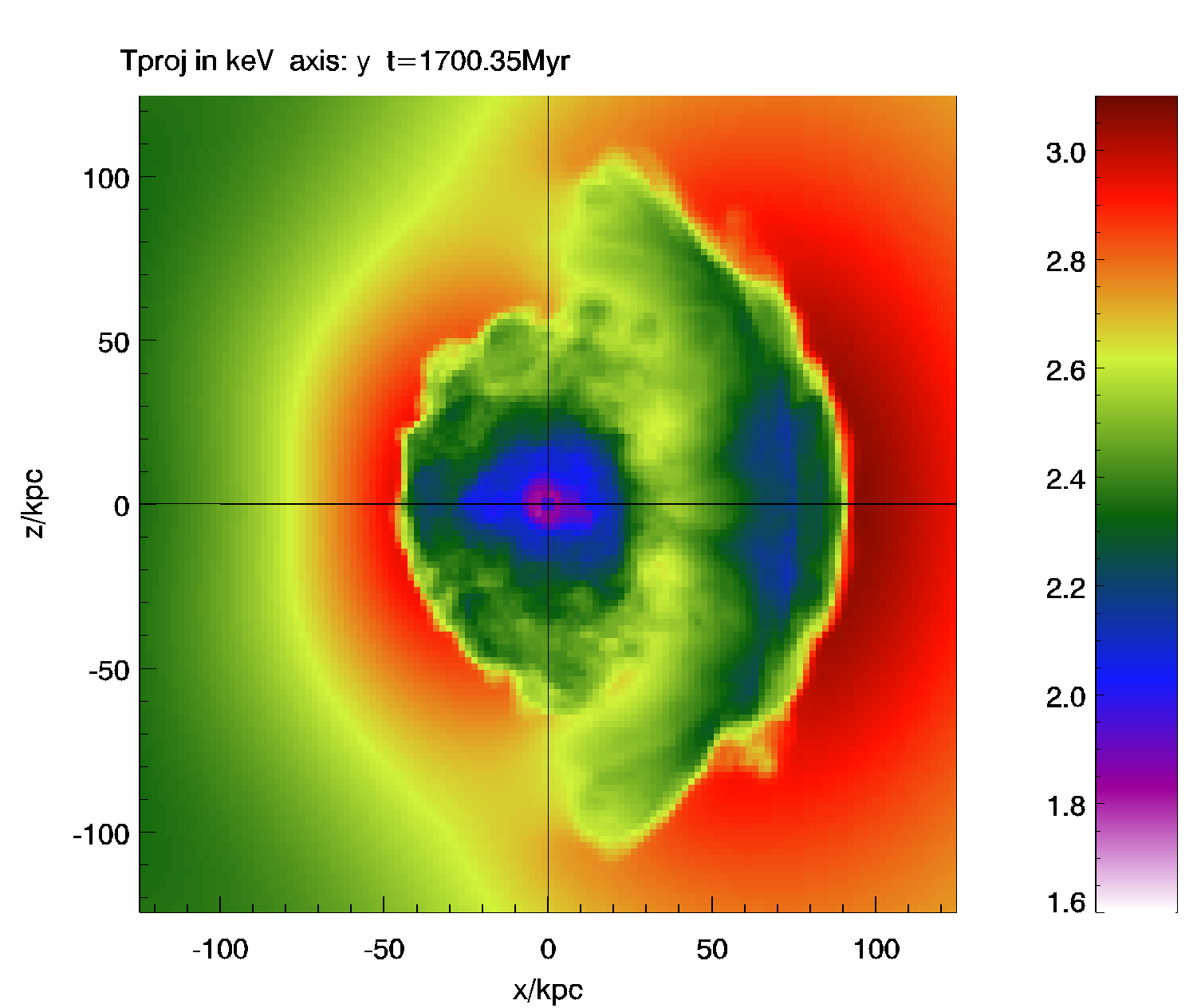}
\hfill\phantom{x}
\caption{
Projected temperature maps corresponding to Fig.~\ref{fig:fiducial-xray}.}
\label{fig:fiducial_Tproj}
\end{figure}
%FFFFFFF

%*******
\subsection{Basic evidence for sloshing scenario}
%
%***
\subsubsection{Brightness and temperature edges} \label{sec:compare_edges}
In the simulations, the gas sloshing produces observable arc-like X-ray brightness edges (Figs.~\ref{fig:fiducial-xray}, \ref{fig:fiducial-residual}) and analogous discontinuities in temperature maps (Fig.~\ref{fig:fiducial_Tproj}). At these edges,  the inner, brighter side is always the cooler one. The first edge appears towards the $+x$ and $+y$-direction  of the cluster core about $0.2\Gyr$ after the pericentre passage (see also Fig.~\ref{fig:slicetemp_massize}). At the opposite side, a front at a smaller radius becomes evident after $0.5\Gyr$. With time, the edges move outwards.  They can be seen along all LOSs. 

Thus, the basic characteristic of the sloshing CFs, namely arc-like brightness edges accompanied by positive temperature jumps on opposite sides of the cluster core, are found in simulations and observations (Fig.~\ref{fig:obs_excess}).

%****
\subsubsection{Spatially congruent cool structures and brightness excess structures} \label{sec:compare_brightcool}
In addition to the actual discontinuities, sloshing produces a characteristic cold pattern in projected temperature maps (Fig.~\ref{fig:fiducial_Tproj}, see also AM06). It takes the form of a single cool spiral winding outwards from the cluster core if the cluster is seen perpendicular to the orbital plane, and the form of cool arcs on opposite sides of the cluster centre if the LOS is parallel to the orbital plane. The arcs are always found at the axis belonging to the orbital plane, never at the perpendicular one.  The outer edges of the cool spiral or cool arcs are the actual CFs. The brightness residual maps show exactly spatially congruent excess patterns (Fig.~\ref{fig:fiducial-residual}). This correspondence of brightness excess and cool regions is found in all known sloshing CF clusters where both data sets exist. In the case of Virgo, the temperature and metallicity maps (\citealt{Simionescu2007}) extend only out to 70 kpc from the cluster centre, but clearly show the SE CF and its inner cold, metal-enhanced arc to be spatially congruent with the brightness excess feature in Fig.~\ref{fig:obs_excess}.

The brightness excess spirals or arcs are the most easily detectable signatures of gas sloshing that are seen even with poor resolution, when the fronts themselves may be smeared out. The morphology of the excess structures may also be more easily detectable than the one of the edges themselves.

%************
\subsection{Morphology and orientation} \label{sec:compare_morph}

%******
\subsubsection{Our line-of-sight (LOS) is perpendicular to the subcluster's orbital plane}  
\label{sec:compare_morph_los}
 The morphology of the observed brightness residual maps presented in Fig.~\ref{fig:obs_excess} depends to some degree on the method of its generation.  A spiral morphology of both, the brightness excess region and the fronts themselves, is clearly seen in the residuals w.r.t.~azimuthal symmetry, and is also supported by the residuals w.r.t. the best-fit $\beta$-model. Hence, from comparison to the simulations (Sect.~\ref{sec:compare_brightcool} and Fig.~\ref{fig:fiducial-residual}), we conclude that our LOS towards the Virgo cluster is approximately perpendicular to the orbital plane of the perturbing subcluster, and focus on this configuration for the further comparison.

We note that we detect the CFs in projections along all grid axes, which was also shown by AM06. Hence, contrary to the suggestion of \citet{Birnboim2010}, the detection of sloshing CFs in a given cluster does not require a favourable viewing angle. However, the signature is slightly weaker if the LOS is parallel to our $x$-axis, i.e.~parallel to the orbital plane but perpendicular to the orbit.

%***
\subsubsection{Orientation on the sky: NNW corresponds to our $+x$-direction} \label{sec:compare_morph_sky}
In the Virgo cluster, the outermost CF is found NNW (labelled "NW" for short) of the cluster core, the second front is on the opposite side (labelled "SE front"). The brightness excess spiral goes anti-clockwise from the centre outwards.  We now rotate the observed brightness residual map such that the orientation of the observed brightness excess spiral matches the one in our synthetic maps (Fig.~\ref{fig:fiducial-residual}). The best match, regarding the orientation of the excess spiral as well as the brightness excess distribution at the outer boundary of the observed field of view, is achieved by identifying the NW front  with our $+x$-direction or up to $45\degree$ clockwise from this. The same orientation is also favoured when comparing the radii of the NW and SE front between observation and simulation (see Sect.~\ref{sec:compare_radius}). For further comparison, we identify the direction of the NW front with our $x$-axis.

The current position of the perturbing subcluster is discussed in Sect.~\ref{sec:compare_identify}.

%*******
\subsection{The size of CF spiral constrains the age to $\mathbf{\sim 1.5\Gyr}$} 
\label{sec:compare_radius}
\begin{figure*}
\centering\includegraphics[width=0.9\textwidth]{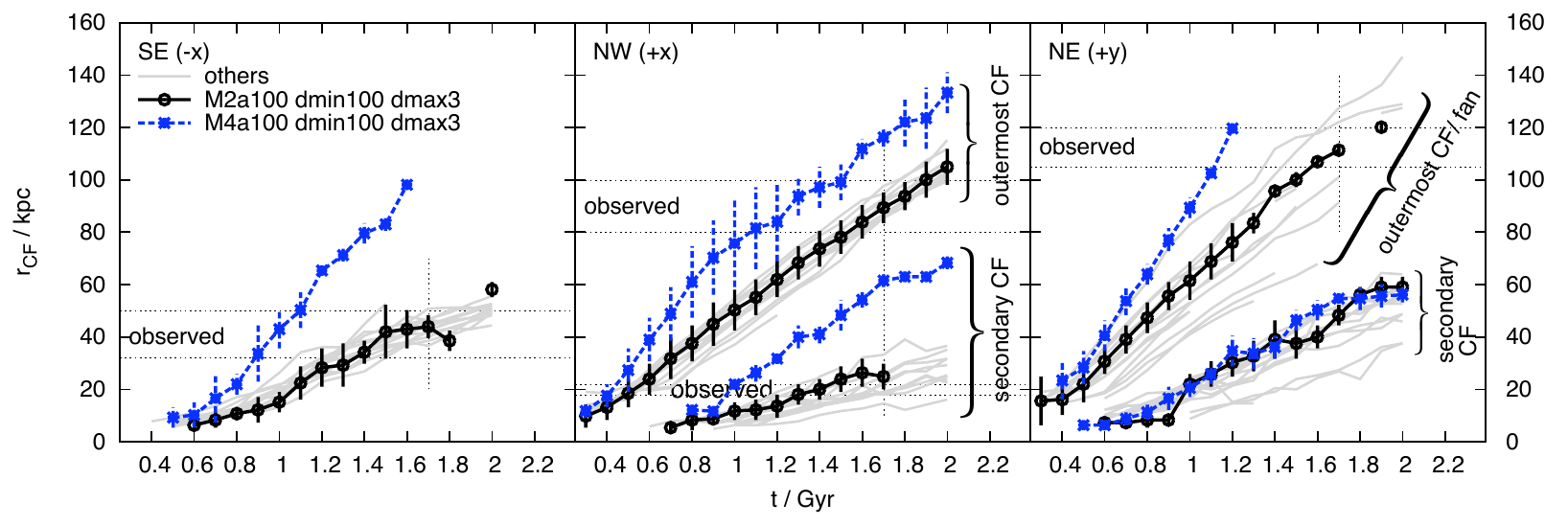}
\caption{
CF radii as a function of time. From temperature profiles towards the directions $+x$ (NW), $-x$ (SE), and $+y$ (NE), we derive the radius (see Sect.~\ref{sec:deriveCFradius}) of the outermost CF.  For the NW and NE directions, we also show the radius of the second CF or cold fan, see labels in the middle and rhs panel.  The thin black horizontal lines mark the radius of the observed CF structure. The thin vertical black line marks the time of $1.7\Gyr$. We show the results of all simulations runs in light grey lines and highlight the fiducial run, M2a100dmin100dmax3 (see legend). The CF radii depend mostly on time and very little on subcluster and orbit characteristics, except for the cold fan in the NE.  Only in the case where the most massive subcluster passes close to the Virgo core, M4a100dmin100dmax3 (see legend), are the CF radii significantly larger than in all other cases. The error bars indicate the width of the fronts. 
}
\label{fig:rcf_time}
\end{figure*}

In Fig.~\ref{fig:rcf_time}, we measure the radii and widths of the CFs (see Sect.~\ref{sec:deriveCFradius}) in all simulations into the SE, NW, and NE direction and plot their evolution as a function of time. 
%
%FFFFFFFFFF
%FFFFFFFFFF
%
For all subclusters and orbits (except M4a100dmin100dmax3), the CF radii depend mostly on the age (i.e.~the time since pericentre passage) and very little on the characteristics of the subcluster and the orbit. There is a weak trend that configurations with less impact (small mass, large scale radius, large pericentre distance, fast passage) result in marginally smaller CF structures.  Hence, we plot the results for all simulations in Fig.~\ref{fig:rcf_time} by grey lines and highlight only the fiducial run (M2a100dmin100dmax3) and the exception M4a100dmin100dmax3 (discussed below). The radii differ between different subclusters and orbits only for the outermost CF in the NE ($+y$) direction, the direction the subcluster recedes to. However, going anti-clockwise from the $+y$-axis,  the sharp outer edge of the cool spiral turns into a flattening gradient in temperature (Figs.~\ref{fig:slicetemp_massize} and \ref{fig:fiducial_Tproj}), i.e.~the true CF turns into a cool "fan". The radius of the outer boundary of the cool fan depends on the subcluster and the orbit characteristics, but also here we find a high degree of degeneracy. 

The only exceptional subcluster and orbit configuration is M4a100dmin100dmax3, where the rather massive subcluster is close to disrupting the cool core. Here, at a given time,  the CF radii are significantly larger.

Apparently, the sloshing itself is governed mainly by the underlying potential of the Virgo cluster. Thus, we can derive the age of the CFs in Virgo from the CF radii. In Fig.~\ref{fig:rcf_time}, we mark the radii of the observed Virgo CFs by horizontal black lines (inner and outer edge of each CF, see also Table~\ref{tab:cfquantities}). For an age of $1.7\Gyr$, our simulations reproduce the size of the observed brightness excess spiral in NW and SE direction simultaneously. In the NE direction, most of the subclusters which do not reproduce the contrasts of the X-ray brightness and the temperature across the CFs correctly (see Sect.~\ref{sec:compare_contrasts} and Table~\ref{tab:exclude}) also produce the cool fan at an either too small or too large radius. The strongest impact subcluster M4a100dmin100dmax3 does not match the CF radii in the NW and SE simultaneously at any given time and any reasonable orientation. For the remaining configurations, alternative orientations of our synthetic maps on the sky fail to simultaneously match both observed CF radii, supporting our earlier constraint on the orientation (Sect.~\ref{sec:compare_morph_sky}).

Given the simplicity of our model, the size of the observed CF structure is reproduced remarkably well at $t=1.7\Gyr$. The direct comparison (Roediger \& ZuHone, in prep.) between rigid potential simulations such as the ones used here and full hydro+Nbody simulations shows that the rigid potential simulations overestimate the age of CF by $\sim 200 \Myr$.  Hence, we conclude that the CFs in the Virgo cluster are about $1.5\Gyr$ old. This more than doubles the estimate derived in S10.

%*******
\subsection{Contrasts across the edges constrain the subcluster mass/size/orbit} \label{sec:compare_contrasts}
With time, the CFs move outwards, and the contrasts of all quantities across them decrease. The age of the CFs is already fixed by their radius.  Thus, we  compare our simulation results at simulation time $t=1.7\Gyr$ (see Sect.~\ref{sec:compare_radius}) to the observations and use the contrasts to constrain the subcluster properties. In brief, subclusters with more mass, smaller scale radius, small pericentre and apocentre distance cause higher contrasts. 

%FFFFFFFFFF
\begin{figure}
\begin{center}
\includegraphics[width=0.49\textwidth]{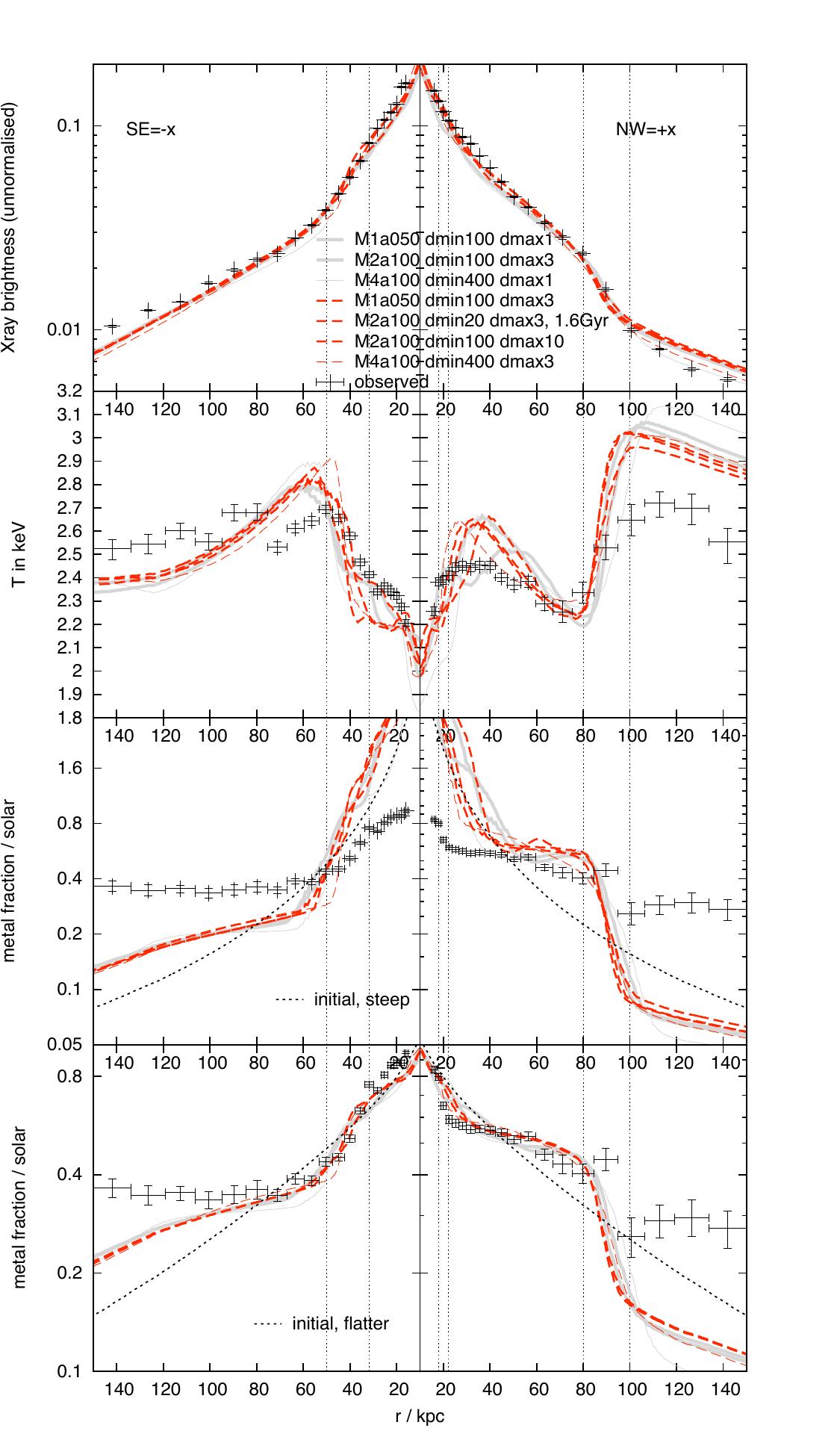}
\caption{
Comparison of azimuthally averaged profiles of X-ray brightness, projected temperature, and projected metallicity for best runs.  The thin vertical lines mark the observed CFs. The dashed (red) lines are for the cases that match also the overall X-ray brightness profiles best. The light grey lines collect the cases that match the profiles at the CFs but show small deviations in the inner X-ray brightness profiles.\newline
For the metallicity, we show the results for two initially different metal distributions: a steep one that follows the light profile of M87 (third row),  and an initially flatter one  (fourth row, see Sect.~\ref{sec:metals_ini}).\newline
}
\label{fig:bestprof}
\end{center}
\end{figure}
%FFFFFFFFFF
%
%TTTTTTTT
\begin{table*}
\caption{
Quantitative characteristics of observed and simulated CFs (fiducial case).  The observed values are derived from the profiles of S10. For the simulation, we give the values as derived by the automatic CF detection described in Appendix~\ref{sec:deriveCFradius}. In brackets, we give the resulting X-ray brightness contrast and temperatures assuming the same radii as in the observation.}
\begin{center}
\begin{tabular}{l||cc||cc}
                         &    \multicolumn{2}{c}{NW}  &     \multicolumn{2}{c}{SE}\\
quantity                       & observation & simulation   & observation & simulation \\
\hline
radius in kpc                               &  80 - 100     &    83 - 95   (80 - 100)   &  32 - 50         &       40 - 48    (32 - 50)       \\
X-ray brightness jump                &  2.28 - 2.46 &    1.72    ( 2.1 )            &  2.11 - 2.18  &    1.51    ( 2 )         \\
Temperature inside CF in keV   & 2.29 - 2.34  &      2.24    (2.2)            & 2.4 - 2.43           &        2.43  (2.4)        \\
Temperature outside CF in keV & 2.58 - 2.72  &         ($< 3$)                & 2.68  - 2.71         &           $ <2.8 $            \\
Metallicity jump                         & 1.15 - 2.      &                                     &1.5 - 1.8         &                    \\
\hline\hline
\end{tabular}
\end{center}
\label{tab:cfquantities}
\end{table*}%
%TTTTTTTTTTT
%

From radial profiles like in Fig.~\ref{fig:bestprof}  (see Appendix~\ref{sec:derive_profiles} for their calculation), we measure the projected temperature and X-ray brightness at the  inner and outer edge of the NW and SE fronts, and derive the contrast of the X-ray brightness as the ratio of the values at the inner and outer edge (see Appendix~\ref{sec:deriveCFradius}). The results derived in this manner from the observation as well as the fiducial simulation are listed in Table~\ref{tab:cfquantities}. 

Our automatic CF detection (see Sect.~\ref{sec:deriveCFradius}) leads to smaller CF widths and, consequently, to lower X-ray contrasts over the front, because part of the general decrease of  brightness over the CF width is not  included. If we assume for a CF width the same values as in the observation, the simulation nearly reproduces the observed jump in X-ray brightness. Also the temperatures just inside the CFs are reproduced well, whereas the over-prediction outside of the CFs is an artefact of the simulation method (Sect.~\ref{sec:method_PRA}).

%TTTTTTTT
\begin{table*}
\caption{List of simulation runs and their (dis)agreement with observations: Subcluster mass (column 1), subcluster scale radius (col.~2), pericentre distance of orbit (col.~3), apocentre distance of orbit (col.~4),  result of two comparison criteria  (col.~5, 6; exclamation marks indicate strong mismatch, brackets weak mismatch, "-" indicates no mismatch). The following comments (col.~7) apply to some simulations: (a) No simultaneous match of CF radii, close too disruption of core. (b) Perturbed X-ray profiles. (c) Too strong secondary CF in NW, brightness deficit inside 40 kpc in SE.\newline
The first and second group are our best-match simulations, configurations in the second group differ from the observations in details. The configurations in the third group do not fit the observation, remarkably the mismatch occurs consistently in more than one criterion.}
\begin{center}
\begin{tabular}{lllllll}
$M\Sub/ $             & $a\Sub$ &  $d\Min$ &  $d\Max$    & contrasts at NW  & CF radius          & \\
$10^{13}M\Sun$ & $/\Kpc$   &   $/\Kpc$   &  $/\Mpc$ & and SE CF too \ldots & in NE too \ldots & comment \\
\hline
1 &   50   & 100    & 3   & - &  ((small)) &  \\
2 & 100   &   20    & 3   & - &      -        &  \\
2 & 100   & 100    & 10 & - & ((small)) &  \\
4 & 100   & 400    & 3   & -  &  (small)  &  \\
\hline
1 & 50   & 100    & 1   &        -        &     -         &  (c) \\
2 & 100 & 100    & 3   &       -         &      -       &  (c)\\
4 & 100 & 400    & 1   & ((strong))  & (large)  &  (c)\\
\hline
4 & 100   & 100    & 3   & ((strong))& large & (a), (b) \\
2 & 100   & 100    & 1   & (strong) & (large) &  (b) \\
4 & 200   & 400    & 3   & weak  & small!  &  \\
4 & 100   & 400    & 10 & weak  & small!  &  \\
2 & 100   & 400    & 3   & weak! & small!  &  \\
1 & 100   & 20     & 3    & weak! & (small) &  \\
1 & 100   & 100    & 3   & weak! & small   &  \\
0.5 & 25  & 100    & 3   & weak  & small   &  \\
\hline\hline
\end{tabular}
\end{center}
\label{tab:exclude}
\end{table*}%
%TTTTTTTTTTT

We derive the contrasts for all our simulations, compare them to the observed ones, and list the outcome of the comparison in Table~\ref{tab:exclude}. The runs in the bottom group in this table produce either too strong or too weak contrasts. Interestingly, these runs also predict a different  radius of the CF/cool fan towards the NE (Fig.~\ref{fig:rcf_time}), where too weak contrasts correspond to a too small radius and vice versa. Hence, we can reject these runs consistently from, both, the size and shape of the CF structure in maps and from the contrasts across the CFs. For completeness, we compare their synthetic X-ray brightness and temperature profiles to the observed ones in Fig.~\ref{fig:nomatchprof}.

%*******
\subsubsection{Degeneracy of subcluster+orbit characteristics} \label{sec:otheroptions}
Not only our fiducial run, but seven of our runs (first two groups in Table~\ref{tab:exclude}) reproduce the observed profiles nearly equally well. We plot projected radial profiles for  X-ray brightness, temperature, and metallicity towards the NW and SE direction for all these runs in Fig.~\ref{fig:bestprof}. 

%***
\paragraph{Similarities}
Like the profiles, also the synthetic maps of all quantities differ remarkably little between these runs. We show brightness residual maps of all best-match cases in the appendix in Figs.~\ref{fig:best1} and \ref{fig:best2}.  All of them reproduce the observed brightness residual map equally well. 

%***
\paragraph{Differences}
Distinguishing observationally between the best-match configurations is very difficult. 

A close inspection of the profiles reveals that in some cases (second group in Table~\ref{tab:exclude}, grey lines in Fig.~\ref{fig:bestprof}) there is a brightness deficit compared to the observation around radii of 20 or 30 kpc. This is due to a well-formed secondary CF towards the NW inside this radius, which is not detectable in the observed X-ray profiles. Also the remaining runs predict this secondary CF, but exhibit only a small deviation from the observed X-ray profiles. Although the Virgo X-ray profile does not show this CF, both, the temperature and the metallicity exhibit a steep gradient in the NW profile at $20\Kpc$. It is also evident in the metallicity map of \citet{Simionescu2007}. This is exactly where the simulations predict a secondary CF in the NW (see also Fig.~\ref{fig:rcf_time}). The observational signature of this CF may have been disturbed to some degree by the nearby active galactic nucleus (AGN) in the Virgo centre. In the simulations, the differences regarding the secondary NW CF are so subtle that we cannot disregard any of our seven best-match cases on this basis.

The large-scale asymmetry described in more detail in  Sect.~\ref{sec:compare_largescale} provides the most promising means to break the degeneracy: it is strongest for massive subclusters with large pericentre distances and weakest for the smallest pericentre distances (Figs~\ref{fig:best1} and \ref{fig:best2}). However, also here the differences are subtle and may be hard to distinguish from intrinsic asymmetries in Virgo.

%*******
\subsubsection{Comment on metallicity}
Given that the metallicity contrast across the CF depends strongly on the initial (unknown) metal distribution, it is not straightforward to use the observed contrast as a constraint on sloshing dynamics. However,  for our initially flatter metallicity profile (see Sect.~\ref{sec:metals_ini}), our simulated metallicity profiles match the observed ones inside $100 \Kpc$ (Fig.~\ref{fig:bestprof}) well. Outside $\sim 100\Kpc$, our simulations produce a significantly lower metallicity compared to the observed values and, consequently, a too strong contrast across the NW front. This is due to an inaccurate initial metallicity at these large radii. Adding initially a metallicity floor outside $\sim 100\Kpc$ would reduce the contrast across the NW front without changing the contrast of the SE front, which is at a smaller radius. As the sloshing increases the metallicity outside $\sim 100\Kpc$ in the SE direction, this constant floor would be turned into a slightly rising metallicity profile, just as observed in this region.

%***********
\section{Observable features of evolved CFs} \label{sec:newfeatures}
The basic observable properties of sloshing CFs as discussed in previous work and in the previous section are
\begin{itemize}
\item  spatially congruent cool, brightness excess features in form of a one-armed spiral or arcs on opposite sides of the core, depending on the LOS  (Sect.~\ref{sec:compare_brightcool})
\item brightness and temperature discontinuities, i.e.~the actual CFs, at the outer edges of the cool spirals or arcs (Sect.~\ref{sec:compare_edges})
\item steep positive gradients in temperature profiles and steep negative ones in X-ray brightness and metallicity at the radii of the CFs (Sect.~\ref{sec:compare_contrasts}).
\end{itemize}
We here refine and extend this list and our understanding of the features.

%****
\subsection{Alternating structure of cool+brightness excess and warm+brightness deficit} \label{sec:compare_alternate}
Sloshing produces a characteristic staggered pattern: Regions of lower temperature and  surface brightness excess alternate with brightness deficit regions of higher temperature, either in a spiral or staggered arcs pattern depending on the LOS. This pattern can be seen in maps (Figs.~\ref{fig:fiducial-residual}, \ref{fig:fiducial_Tproj}) as well as in profiles (Fig.~\ref{fig:bestprof}). It even extends beyond the outermost CF. It is seen in all simulations and also in observations: in Virgo  (Fig.~\ref{fig:obs_excess}), A496 (\citealt{Dupke2007}), A2142 (\citealt{Markevitch2000}), MS 1455.0+2232 and  RX J1720.1+2638 (\citealt{Mazzotta2008}). 

The cool regions are congruent with the brightness excesses because the cool gas is associated with a higher gas density, which leads to a higher X-ray brightness and vice versa. The same pattern can be deduced from the mock observations of AM06.

%*************
\subsection{Asymmetry on large scales} \label{sec:compare_largescale}
From our simulations, we also derive synthetic maps out to 500 kpc from the cluster centre.
At first glance, the large-scale X-ray images appear approximately symmetrical and undisturbed as soon as the subcluster has left the field of view. Except for the central spiral excess, there is no obvious trace of the interaction. 
%
%FFFFFFFFFFFFFFF
\begin{figure}
\includegraphics[trim=30 -25 275 30,clip,height=4.2cm]{PLOTS/M2.13_I100MAX3_A100_HR_DAMP/proj_Excess_z_size0_0270}
\includegraphics[trim=1300 -25 100 30,clip,height=4.2cm]{PLOTS/M2.13_I100MAX3_A100_HR_DAMP/proj_Excess_z_size0_0270}
\includegraphics[trim=30 0 215 115,clip,angle=90,height=4.2cm]{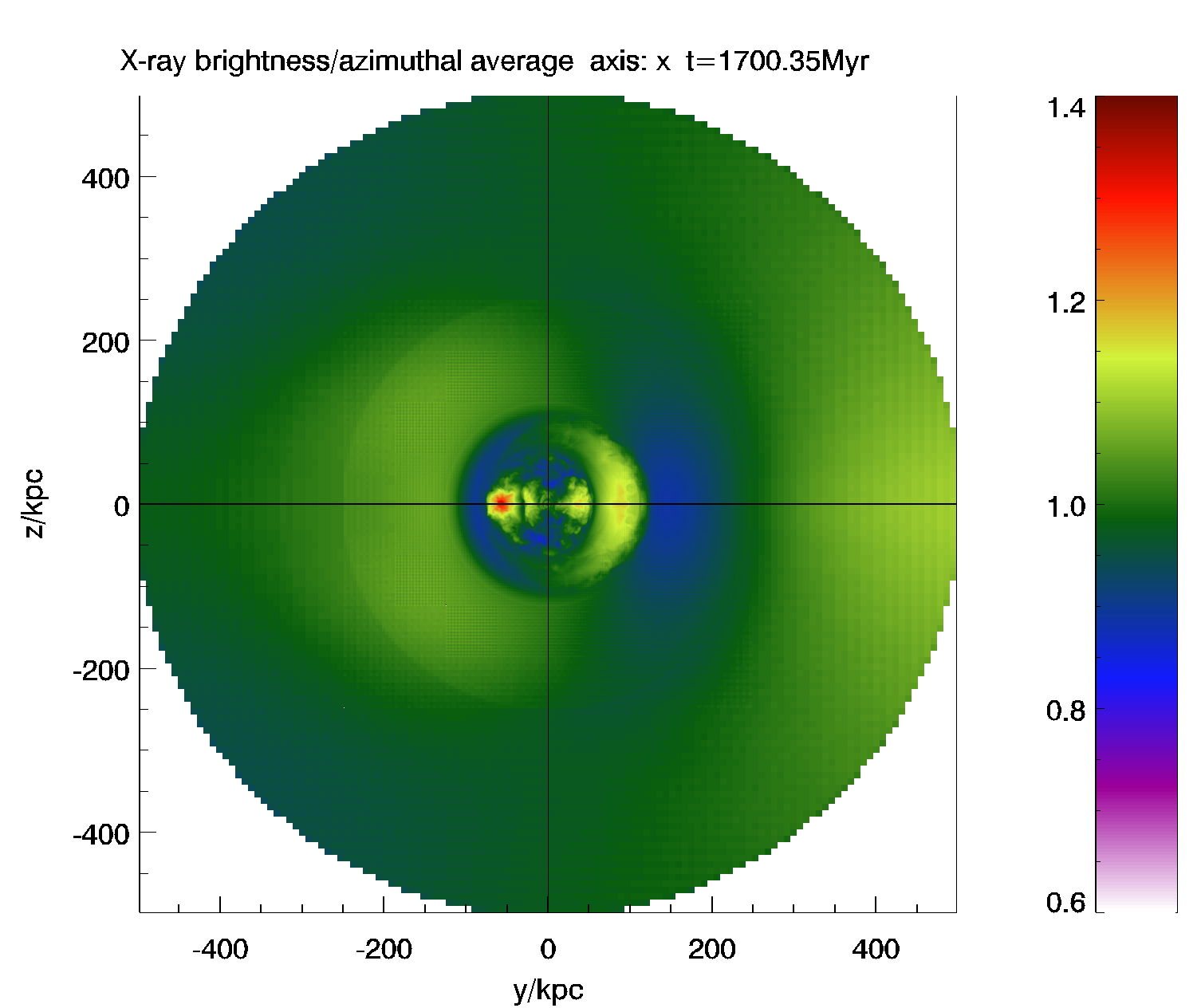}
\newline
\includegraphics[trim=30 -25 275 115,clip,angle=0,height=4.cm]{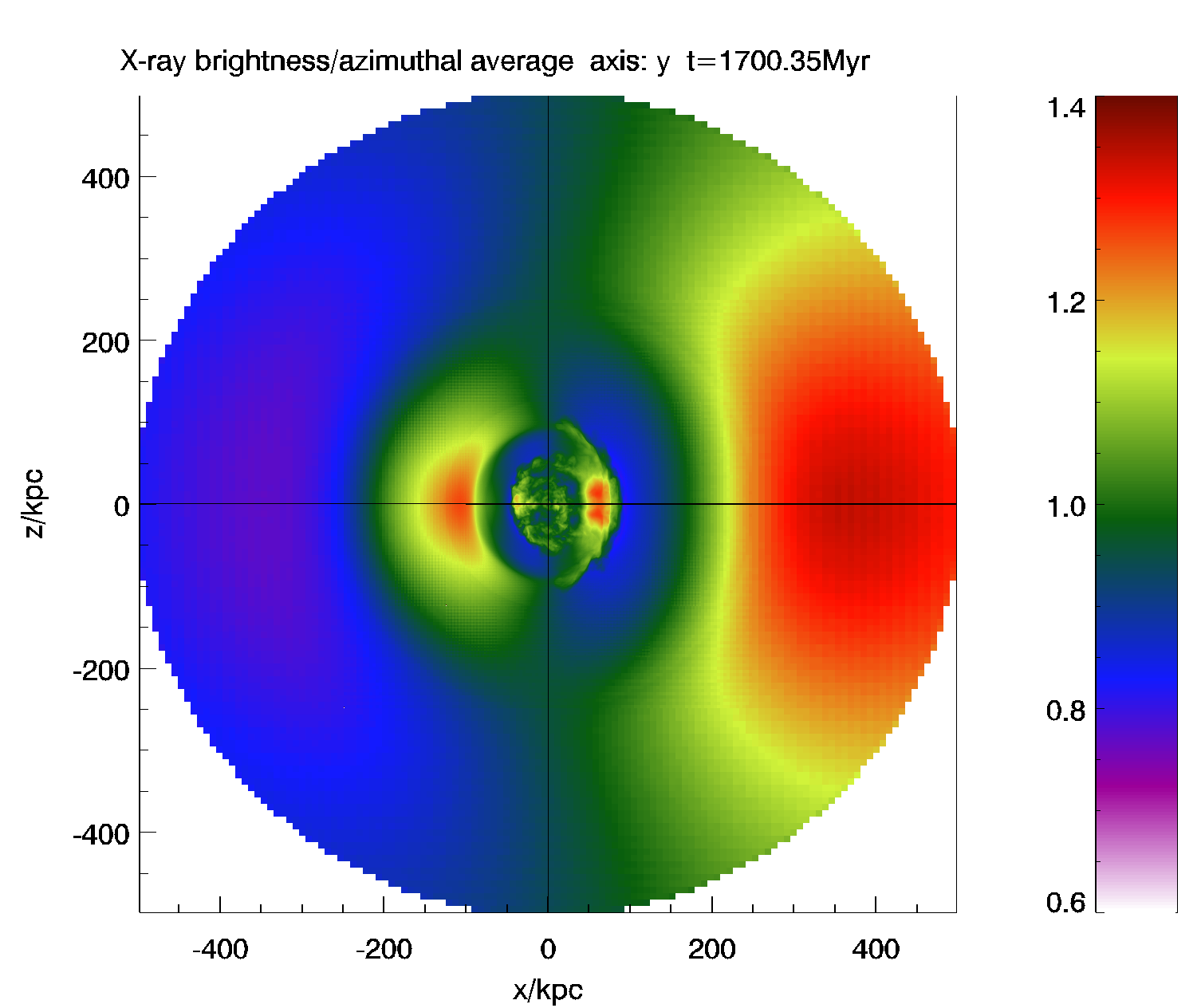}
\hspace{-0.3cm}
\includegraphics[trim=120 275 120 100,clip,height=3.cm,origin=t,angle=-45]{PLOTS/Virgo_residual}
\caption{
Brightness residual maps for a large field of view, corresponding to Fig.~\ref{fig:fiducial-xray}. For the observed map, compare to Fig.~\ref{fig:obs_excess}. The red circle in the top left panel indicates the size of the observed field of view.
}
\label{fig:fiducial_large}
\end{figure}
%FFFFFFFFFFF
%
The residual maps, however, (Fig.~\ref{fig:fiducial_large}) reveal that there is a clear asymmetry in the X-ray brightness. The central brightness excess spiral continues from the $+y$-direction anti-clockwise, forming a brightness excess fan. The outer boundary of this fan is not a sharp edge. Relative to the orbit of the subcluster, the fan is located  at the same side of the main cluster core as the subcluster's pericentre. Outside of the fan, there is a large brightness deficit region, whereas on the opposite side, there is a large brightness excess region outside $\sim 200\Kpc$.

The large-scale asymmetry is strongest in the orbital plane, perpendicular to the cluster orbit. Also the metallicity is asymmetrical, where higher metallicity is associated with brightness excess. The asymmetry originates from the offset of the main cluster core w.r.t.~the overall cluster towards the pericentre of the orbit. The same effect takes place in the full hydro+Nbody simulations (AM06, their Fig.~4).

The detection of such a  large-scale asymmetry could be used to distinguish the standard sloshing scenario from other scenarios, which could cause different large-scale signatures.

%****
\subsubsection{Large-scale asymmetry in Virgo}
The observed residual map of Virgo extends to about 170 kpc. Inside this range, it agrees with the predicted asymmetry. It will be interesting to compare our prediction to larger-scale observations. The detection of such a large-scale feature would greatly help to constrain the orientation of our simulated maps on the sky. However, Virgo is a dynamical cluster, still in the state of assembly, so that also other processes will influence the symmetry or asymmetry at this length scale. 

%****
\subsubsection{Large-scale asymmetry in A496, Perseus, 2A 0335+096}
The cluster A496 (\citealt{Tanaka2006}) shows a clear brightness excess (and also cooler) region $\sim 400 \Kpc$ towards N-NW, from which the authors derive that the subcluster passed the main cluster core from NW to SE. Our simulations indicate that the subcluster moved in W-E direction, passing the cluster core in the S. The orientation of the spiral formed by the CFs observed by \citet{Dupke2007} indicates for  a subcluster motion from W to E.  This excess region also matches the orientation of the central CFs described in \citet{Dupke2007}, where the outermost CF is towards N. Additionally, \citet{Tanaka2006} report another CF about 200 kpc S-SE of the cluster centre, which is evident as a brightness excess region in their residual map. Our simulations suggest that this is the cool brightness excess fan, which should be and is found opposite of the large-scale excess region. Thus, our simulations seem to reproduce also this cluster qualitatively. A496 will be an interesting target for future simulations aiming at a quantitative reproduction of the observed features.

The residual and temperature maps for the Perseus cluster (\citealt{Churazov2003}) qualitatively show exactly the same features: a cool brightness excess at large radius towards E, a large brightness deficit region at large radii towards W. The large brightness excess region at smaller radius W of the core could be the cool fan. North of the cluster core, there is a faint connection between the large-scale excess in the East and the fan in the W, which is also seen in our simulations. This argues for a subcluster orbit from N to S, passing W of the cluster core. We note that there is an inner brightness excess spiral in Perseus that does not seem to continue the outer structures, which could indicate a second sloshing event. 

The features in the large-scale images of the cluster 2A 0335+096 (\citealt{Tanaka2006}) show considerably more substructure, but are still consistent with a brightness excess fan towards the NW and a large-scale excess towards the SE. The fan could be the continuation of the central cool spiral seen by \citet{Sanders2009_2a}. If this interpretation is correct, the subcluster should have moved from SW towards NE, passing the cluster core in the NW.  \citet{Tanaka2006} discuss cooling or ram pressure stripping or turbulent viscous stripping as the origin of this cool brightness excess region. Our simulations suggest that they could be a by-product of the passage of a subcluster that was already stripped before pericentre passage.

%****
\subsection{Substructure in profiles}\label{sec:compare_profilestructure}
Radial profiles taken towards the CFs  (Fig.~\ref{fig:bestprof}) show a consistent structure in both, simulations and observations: 
\begin{itemize}
\item When approaching the CF from the inside, the temperature decreases or forms a plateau. Also the metallicity shows a plateau.
\item The front itself is manifested as a steep increase in temperature and a steep decrease in metallicity and X-ray brightness. The apparent width of the CFs is due to azimuthal averaging over concentric ring segments whose boundaries do not coincide with the CF. The contrast of any quantity across a CF, defined as the ratio of the quantity at the inner and outer edge of the CF, always includes the intrinsic variation over this radial range.
\item Outside the CF, the temperature again decreases, i.e.~there is a warmer rim enclosing each CF. 
\end{itemize}

The same structure is found in other CF clusters, as far as data quality allows its detection. The decrease of temperature with radius inside the CFs may be a speciality of the compactness of the Virgo cluster, but plateaus in temperature inside the CF are found for the N front in A496 (\citealt{Dupke2007}), the SW front 2A 0335+096 (\citealt{Sanders2009_2a}), in A1795 (\citealt{Markevitch2001}), A2142 (\citealt{Markevitch2000}), A2204 (\citealt{Sanders2005a2204}), MS 1455.0+2232 (\citealt{Mazzotta2008}), NE front in NGC5098 (\citealt{Randall2009ngc5098}), probably in A1201 (\citealt{Owers2009a1201}). A metallicity plateau is seen in A2204 (\citealt{Sanders2009a2204}). Indications of a warm rim outside the CF are seen in A1795, A2204 and A2142. However, most published data does not resolve the temperature profiles well enough to clearly detect these features. Occasionally, only the temperature jump across the CF is measured. 

For future observations we suggest  to derive the full profiles from the cluster centre beyond the CFs. This will enable a better comparison to simulations, and tighter constraints on the identification of the responsible subcluster. Also an alignment of the spectral extraction rings with the CFs could be helpful in measuring accurate contrasts.

%% file: metals.tex
%************
\section{Metal transport} \label{sec:metals}
While sloshing redistributes the central gas, it must also redistribute the heavy elements throughout the central region. In addition to producing metallicity discontinuities at the CFs, which are observed in Virgo (Fig.~\ref{fig:bestprof}),  sloshing may also be able to broaden the metal distribution throughout the cluster centre. We study this process by tracing the evolution of an initially centrally peaked metallicity distribution in our fiducial run. We find the evolution of the metal distribution to be tightly coupled to the overall dynamics. Simulations that produce similar temperature structures, also produce similar metallicity structures.

%*********
\subsection{Initial metal distribution} \label{sec:metals_ini}
We study an initially steep and a flatter radial metal density profile. In the steep case, we follow \citet{Rebusco2006}, who assume that originally the central metal distribution in a cluster follows the light profile of the central galaxy (here M87, \citealt{Kormendy2009}), but is broadened by diffusion caused by turbulent gas motions. It is also debatable whether the metal distribution about 2 Gyr ago should follow the current light profile of M87. If the metal peak in the Virgo centre was produced by stellar mass loss and SNIa, then the O/Fe profile should not be flat as observed (S10). More likely, the metal distribution originates from the assembly of M87 and has already undergone some redistribution. Hence, we also consider  a flatter metal profile which approximates the observed metal density inside the NW CF. In both cases, we fit the initial metal density by a deprojected Sersic profile (\citealt{Sersic1963}, \citealt{Prugniel1997}) with the parameters listed in Table~\ref{tab:met_distr}.

%
%TTTTTT
\begin{table}
\caption{Parameters for deprojected Sersic profiles describing the steep (fitted to M87 light) and the flatter metal distribution.}
\begin{center}
\begin{tabular}{|c|c|c|}
\hline
 & steep ($\propto$ M87 light) & flatter \\
 \hline
 scale radius & 54.56 kpc & 120 kpc \\
 Sersic index & 11.84 &  2 \\
 \hline
 \hline
\end{tabular}
\end{center}
\label{tab:met_distr}
\end{table}%
%TTTTTTT

The evolution of 3D metallicity profiles across the CFs is shown in Fig.~\ref{fig:evolprofs}. Metallicity discontinuities appear exactly at the CFs, but the contrast across them is clearly higher for the initially steep profile.

%*********
\subsection{Comparison to observations} \label{sec:metals_obs}
%
%FFFFFFF
\begin{figure}
\begin{center}
\includegraphics[trim=30 -25 305 0,clip,height=4.8cm]{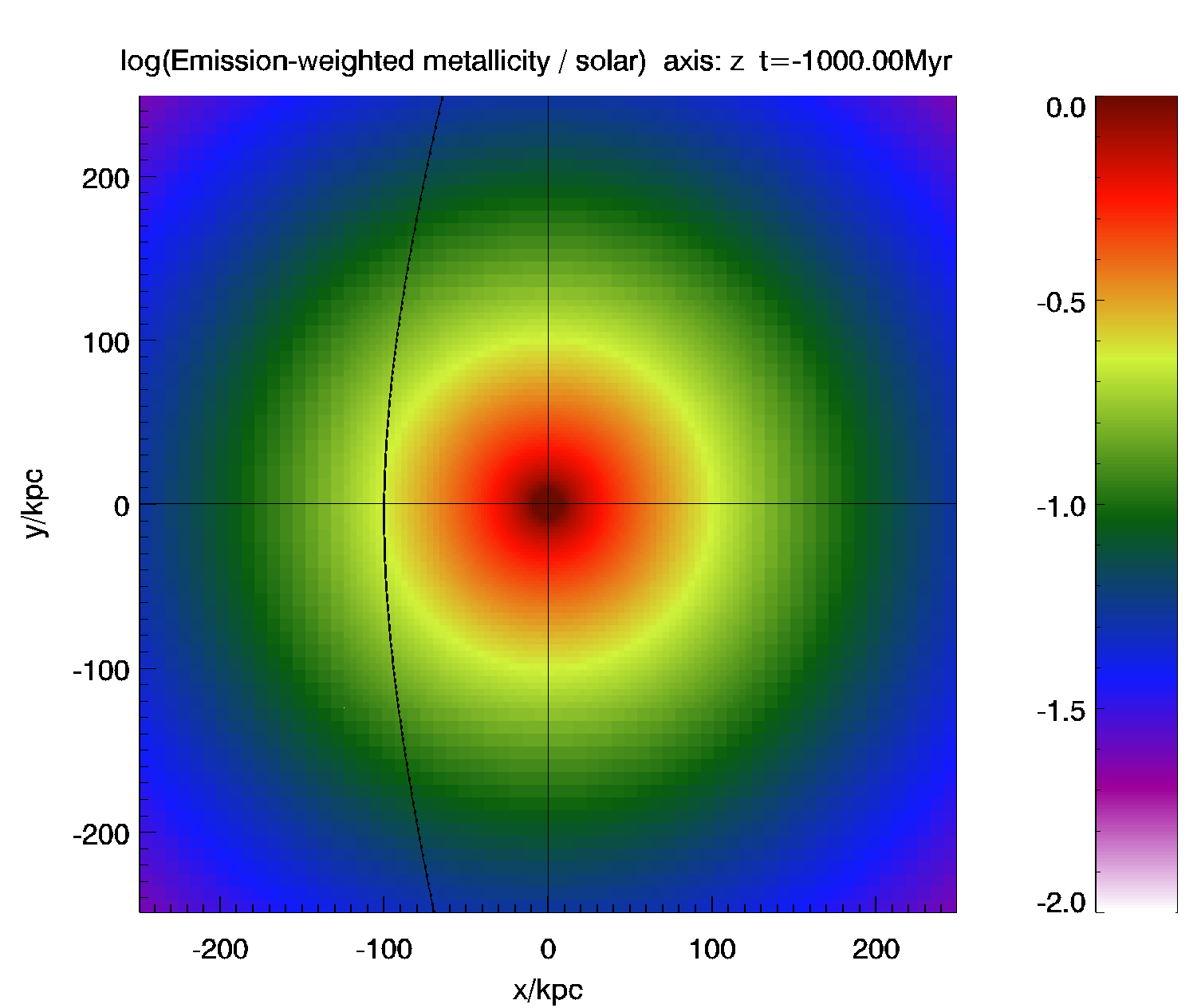}
\hspace{-1.15cm}
\includegraphics[trim=1280 -25 0 0,clip,height=4.8cm]{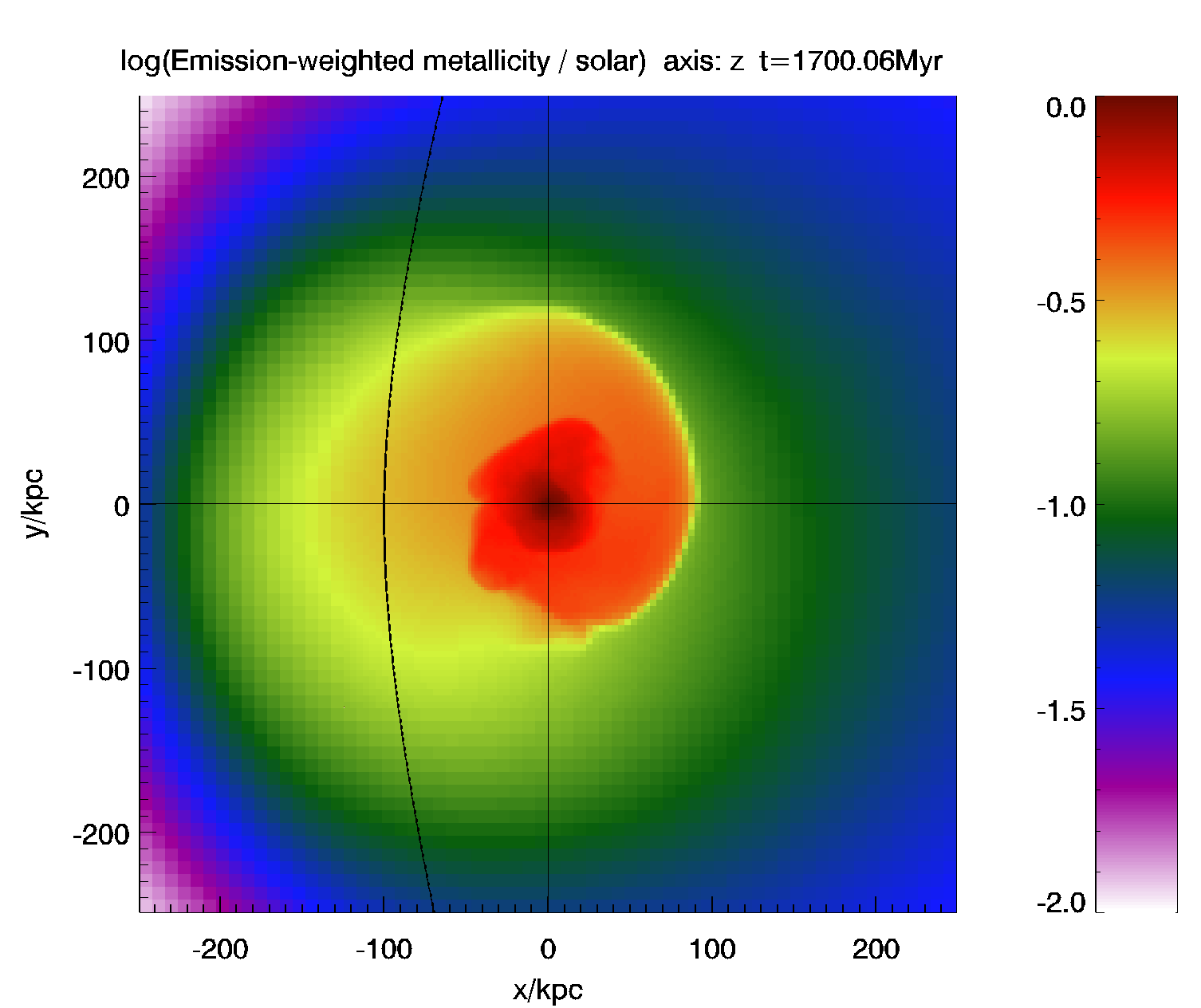}
\hspace{-0.25cm}
\includegraphics[trim=90 -25 275 0,clip,angle=0,height=4.8cm]{PLOTS/M2.13_I100MAX3_A100_LR_DAMP_Metflat/proj_Met_z_size1_0270.png}
\caption{Synthetic projected metallicity map (in solar units) for initial state (left) and $1.7\Gyr$  after subcluster pericentre passage (right), for the case of the flatter initial metal distribution.}
\label{fig:metmap} 
\end{center}
\end{figure}
%FFFFFF
%
Projected metallicity maps (Fig.~\ref{fig:metmap}) display a spiral structure coincident with the one seen in the temperature and brightness residual maps (Figs.~\ref{fig:fiducial-residual}, \ref{fig:fiducial_Tproj}).

The two bottom panels in both subfigures of  Fig.~\ref{fig:bestprof} compare the simulated to the  observed metallicity profiles. The steep gradients at all three CFs are clearly reproduced, so are the plateaus inside the CFs and the NW-SE asymmetry (see Sect.~\ref{sec:newfeatures}). In case of the initially steep profile, the metallicity contrasts across the fronts are higher than observed, and the metallicity in the central $\sim 50\Kpc$ remains  too steep compared to the observations. This shows that the redistribution by sloshing alone is not sufficient to transform a steep metal peak originating from stellar mass loss of the central galaxy into the observed flattened one. Additional processes are required (see e.g.~\citealt{Rebusco2005,Rebusco2006}, and discussion in Sect.~\ref{sec:metals_diffusion}).

For the initially flatter case, we achieve a good agreement between observations and simulations (see Sect.~\ref{sec:compare_contrasts}). Therefore, we will concentrate on this case for the further analysis.

%*********
\subsection{Broadening of the global metal distribution} \label{sec:metals_global}
%
%****
\subsubsection{General} 
% 
%FFFFFFF
\begin{figure}
\includegraphics[width=0.5\textwidth]{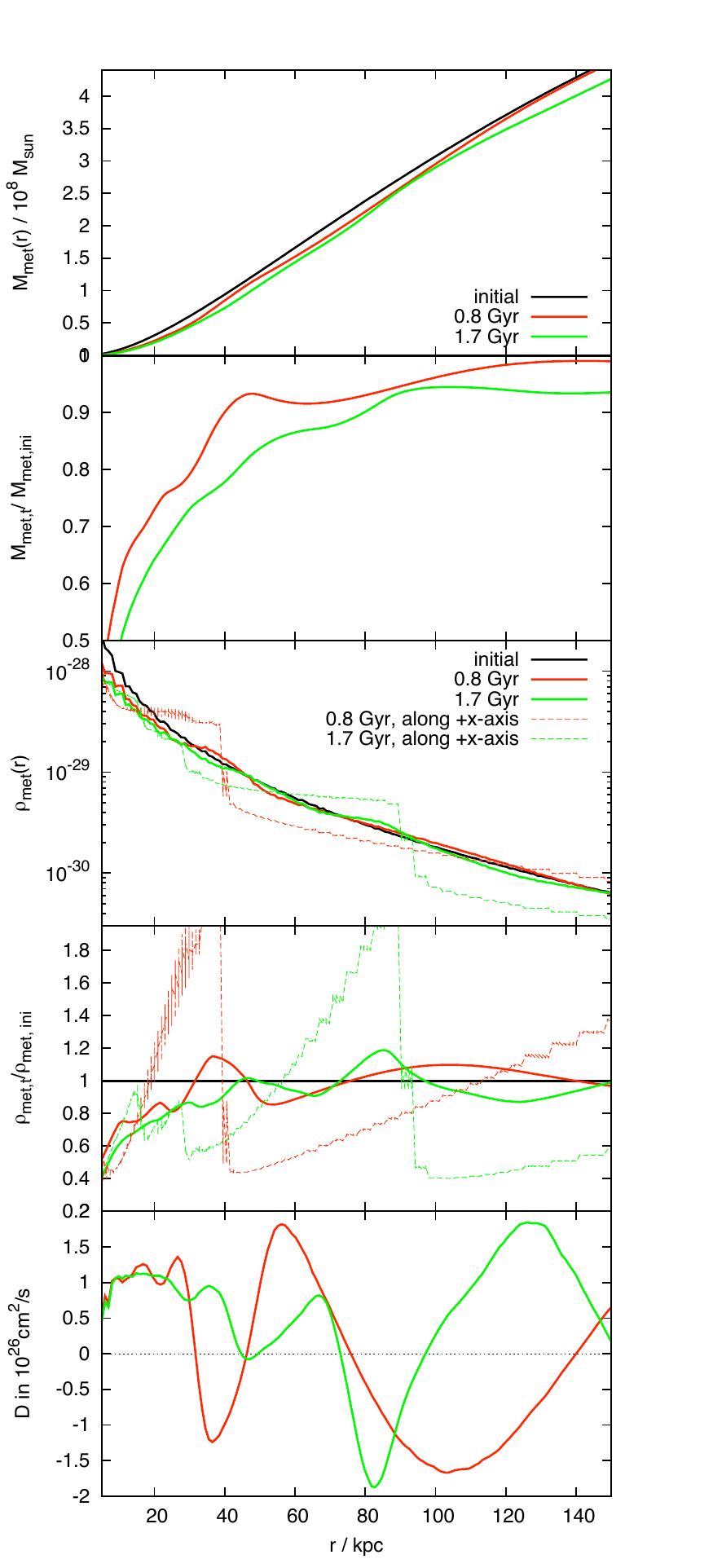}
\caption{Effect of sloshing on the metal distribution. \textbf{Top:} cumulative metal mass for initial time and $t=0.8, 1.7\Gyr$. \textbf{Second panel:} Relative change of cumulative metal mass: profiles from previous panel divided by initial distribution. \textbf{Third:} spherically averaged metal density (thick lines) for the initial state and $t=0.8, 1.7\Gyr$. The thin lines show the metal density along the $+x$-axis of the simulation grid. \textbf{Fourth:} Relative change of metal density profiles: profiles from the previous panel divided by the initial profile. \textbf{Fifth:} diffusion constant derived from the profiles in the third panel.}
\label{fig:metevol} 
\end{figure}
%FFFFFF
We derive the cumulative metal mass and the spherically averaged metal density for different timesteps and compare them to the initial state (Fig.~\ref{fig:metevol}). Sloshing does not only influence the metal distribution towards the CFs, but broadens the overall metal distribution.
Within 1.7 Gyr after pericentre passage,  the central metal density is depleted by about 25\%, while the metal density between 75 and 95 kpc increases by about 10 to 20\%. This is somewhat less than the 20 to 30\% derived in S10, but there only the difference between the NW and the SE sector was compared, while we considered the global broadening.  Sloshing is, however, not a symmetric process, and the metal redistribution is much stronger if considered along a line directly towards a CF. Along the x-axis of our simulation box, the metal density is enhanced by a factor of 1.5 to 2 within the 20 kpc inside of the CF (third and 4th panel of Fig.~\ref{fig:metevol}).

%****
 \subsubsection{Diffusion description} \label{sec:metals_diffusion}
Assuming the heavy elements in a cluster core originate mainly from stellar mass loss of the central galaxy and are redistributed by turbulence, \citet{Rebusco2005,Rebusco2006} derive an effective diffusion constant of the order of $10^{29}$ cm$^2 /$s for several clusters, and $0.8\cdot 10^{29}$ cm$^2 /$s for Virgo. \citet{Graham2006} find  a value of $ 10^{28}$ cm$^2 /$s for the central region of the Centaurus cluster, but considerably smaller values outside $25\Kpc$. \citet{Roediger2007bubbles} have studied the metal mixing induced by AGN-inflated buoyantly rising bubbles, which drag metal-rich gas from the centre to larger radii. They derived effective diffusion coefficients that depended strongly on cluster radius, ranging from $10^{29}$ cm$^2 /$s at about $10\Kpc$ from the cluster centre to about $10^{26}$ cm$^2 /$s at $50\Kpc$ radius and $10^{22}$ cm$^2 /$s at $100\Kpc$ radius. 

In order to quantify the effectiveness of sloshing in metal redistribution, we calculate the effective diffusion coefficients from our simulations by comparing the spherically averaged density profiles at $t=0.8, 1.7\Gyr$ to the initial state  (see \citet{Roediger2007bubbles} for details). The result is shown by the thick lines in the bottom panel of Fig.~\ref{fig:metevol}. It makes clear that diffusion is not a good approximation here, because we derive negative, hence unphysical, diffusion constants in some parts. This occurs e.g.~inside the major CF, i.e.~at those radii where sloshing deposits the metals it has removed from elsewhere. True diffusion would act differently: metals moved from the centre to some larger radius would not remain there but continue to diffuse outwards until a flat distribution is achieved. In contrast, sloshing removes metals  from different radii and "piles them up" inside the CFs. 

In the inner part we derive a positive diffusion constant of the order of  $10^{26}$ cm$^2 /$s between 10 and 40 kpc, which is less than what was derived for AGN activity. This is not surprising, because the AGN sits in the very centre of the cluster and creates the strongest impact there. In contrast, the very central gas will be affected least by sloshing because it is held more tightly by the deep central gravitational potential. Thus, AGN activity and gas sloshing are complementary processes, the former being more effective in the very cluster centre, the latter at larger radii. In a realistic case, a combination of both could be very effective. 
Furthermore, the metal mixing by AGN induced buoyant bubbles is, on average, a continuous process, because a series of bubbles is inflated by intermittent AGN activity. The sloshing studied here happens in the aftermath of a single event. Hence, the diffusion efficiency averaged over the whole event is rather low. If the effective diffusion constant  is averaged over a much shorter time, e.g.~50 to 200 Myr, it is at least a factor of 10 higher. Also repeated minor merger events may enhance the efficiency of metal redistribution.

%% file: gas.tex
%**************************************
\section{Subclusters with gaseous atmospheres} \label{sec:gas}

Previous work (AM06) stresses that the subcluster needs to be gas-free to produce sloshing CFs but leave the overall cluster appear relaxed. Having constrained the mass and size of the responsible subcluster, we want to test if a sufficient subcluster can be completely ram pressure stripped significantly prior to pericentre passage.

%**********
\subsection{Initial setup} \label{sec:gaseous_ini}

We fill the fiducial subcluster M2a100 either with an extended or with a compact atmosphere. For the subcluster atmosphere we assume a  constant  temperature with a central drop described by 
%-------
\begin{equation}
T(r)= T_0\;  \frac{1+(r/r_{T0})^3}{D_T+(r/r_{T0})^3},
\end{equation}
%======
where $T_0$ is the temperature of the subcluster gas at large radii, $T_0/D_T$ the central temperature, and $r_{T0}$ the scale radius of the cooler core. Demanding hydrostatic equilibrium, we can calculate the density and pressure profile for the subcluster gas. For the simplest case of a constant temperature, $T(r)=T_0$, these are
%-------
\begin{equation}
\frac{p(r)}{p_0}=\frac{\rho(r)}{\rho_0}=\exp\left(  \frac{m_p G M\Sub}{k T_0} \left[\frac{1}{r+a\Sub} - \frac{1}{a\Sub} \right]  \right),
\end{equation}
%======
where $m_p$ is the mean particle mass in the subcluster atmosphere, and $M\Sub$ and $a\Sub$ are the mass and scale radius of the Hernquist potential describing the subcluster DM distribution. The slope in density and pressure is very sensitive to the temperature of the subcluster gas. The subcluster atmosphere truncates where its pressure drops below the ambient pressure of the main cluster's ICM. This truncation condition determines the size and total gas mass of the subcluster. We adopt an ambient ICM pressure of $5\cdot 10^{-13}\Presunit$, typical for the Virgo cluster at $1\Mpc$ (Fig.~\ref{fig:iniprofs}). Then we chose the temperature and density of the subcluster gas such that the subcluster gas pressure drops below the ambient ICM pressure at a radius such that, in both cases, the subcluster contains about the same amount of gas in this pressure limit. Table~\ref{tab:subclustergas} summarises the parameters for both atmospheres.
%TTTTTT
\begin{table}
\caption{Parameters for the gaseous subcluster atmospheres (see Sect.~\ref{sec:gaseous_ini} for details).}
\begin{center}
\begin{tabular}{|c|c|c}
\hline
 & extended & compact \\
 \hline
 $T_0/10^7\K$ & 2  & 1\\
$D_T$ & 2& 1.5 \\
$r_{T0}/\Kpc$ & 30 & 30\\
$\rho_0/\gccm$ & $8.8\cdot 10^{-27}$ & $1.1\cdot 10^{-25}$\\
\hline\hline
\end{tabular}
\end{center}
\label{tab:subclustergas}
\end{table}%
%TTTTTTT

%**********
\subsection{Result}

On our fiducial orbit, the compact atmosphere is not completely ram pressure stripped. On the contrary, its dense gas interacts strongly with the Virgo core gas and destroys the cool core completely. 

%FFFFFFFF
\begin{figure}
\begin{center}
\includegraphics[width=0.45\textwidth]{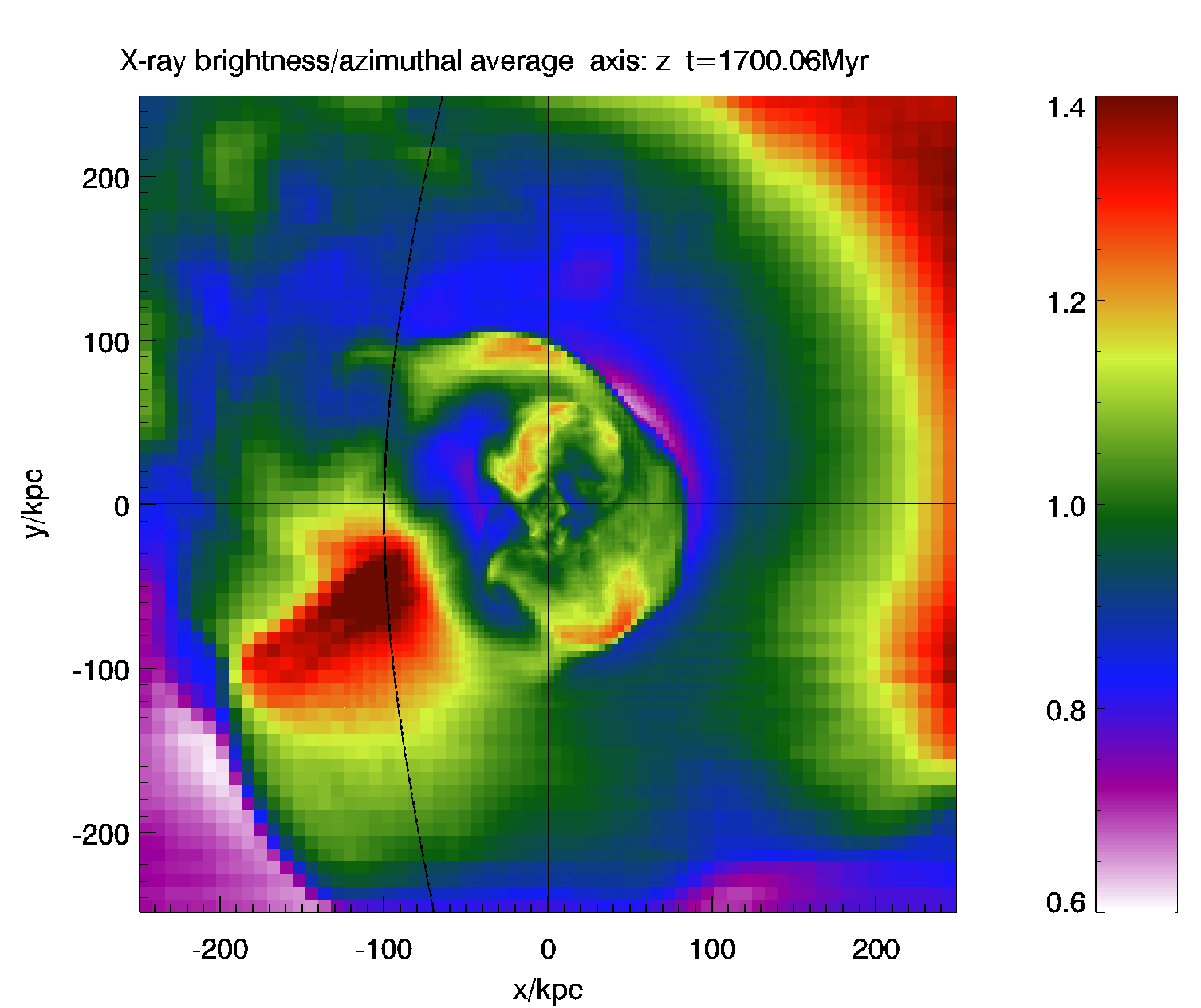}
\newline
\end{center}
\caption{
Fiducial orbit, fiducial subcluster M2a100, filled with extended gaseous atmosphere: The residual map and projected temperature map at $t=1.7\Gyr$ are similar to the case without gas (Fig.~\ref{fig:fiducial-residual}), although more irregular.}
\label{fig:gas_ext}
\end{figure}
%FFFFFFFFF
The extended atmosphere is  ram-pressure-stripped just at the pericentre passage. However, its less dense gas cannot penetrate the central 100 kpc of the Virgo core, where sloshing can go on in a similar manner as in the gas-free cases. 
The resulting CF structures are also similar (Fig.~\ref{fig:gas_ext}), although more irregular. The CFs on the NW and SE sides are found at the same positions as in the gas-free case. The X-ray brightness profiles are, however, more distorted than in the gas-free cases and in the observation, reflecting the more irregular CF structure. The ability of the gaseous atmosphere to disturb the sloshing is sensitive to the balance between the gas density and the ram pressure the subcluster experiences. For example, a subcluster with a half as dense atmosphere is already ram pressure stripped 400 kpc before the pericentre passage, and the central sloshing proceeds as in the initially gas-free case.

The regularity of the observed brightness excess structure suggests that the subcluster responsible for the CFs in the Virgo cluster has lost its gas well before its pericentre passage.  Alternatively, the subcluster may have passed the cluster core at a large pericentre distance, such that the subcluster atmosphere does not penetrate the main cluster core, where it would erase the sloshing signature. AM06 present such a case with a mass ratio of 5:1 and a pericentre distance of about 400 kpc. The CF and cool spiral observed in Abell 1644 (\citealt{Johnson2010}) seems to be such a case. However, the fact that Abell 1644 is a major merger (mass ratio 2:1) requires a rather large pericentre distance to prevent the disruption of the cool cores. In both cases, either the early loss of the gaseous atmosphere or the passage at a large distance,  traces of stripped subcluster gas might be found at larger distances from the cluster centre.

%% file: shock.tex
%****************
\section{Sloshing triggered by a galaxy's bow shock} \label{sec:bowshock}

Any process that offsets the ICM  in a "bulk-like" manner will cause subsequent sloshing and formation of CFs. An alternative to displacing the ICM by gravitational interaction with a subcluster is the passage of a shock, proposed by \citet{Churazov2003}. These authors presented a simulation of an idealised plane shock passing a hydrostatic cluster. 

\begin{figure}
\begin{center}
\includegraphics[trim=0 0 0 -0,clip,height=4.9cm]{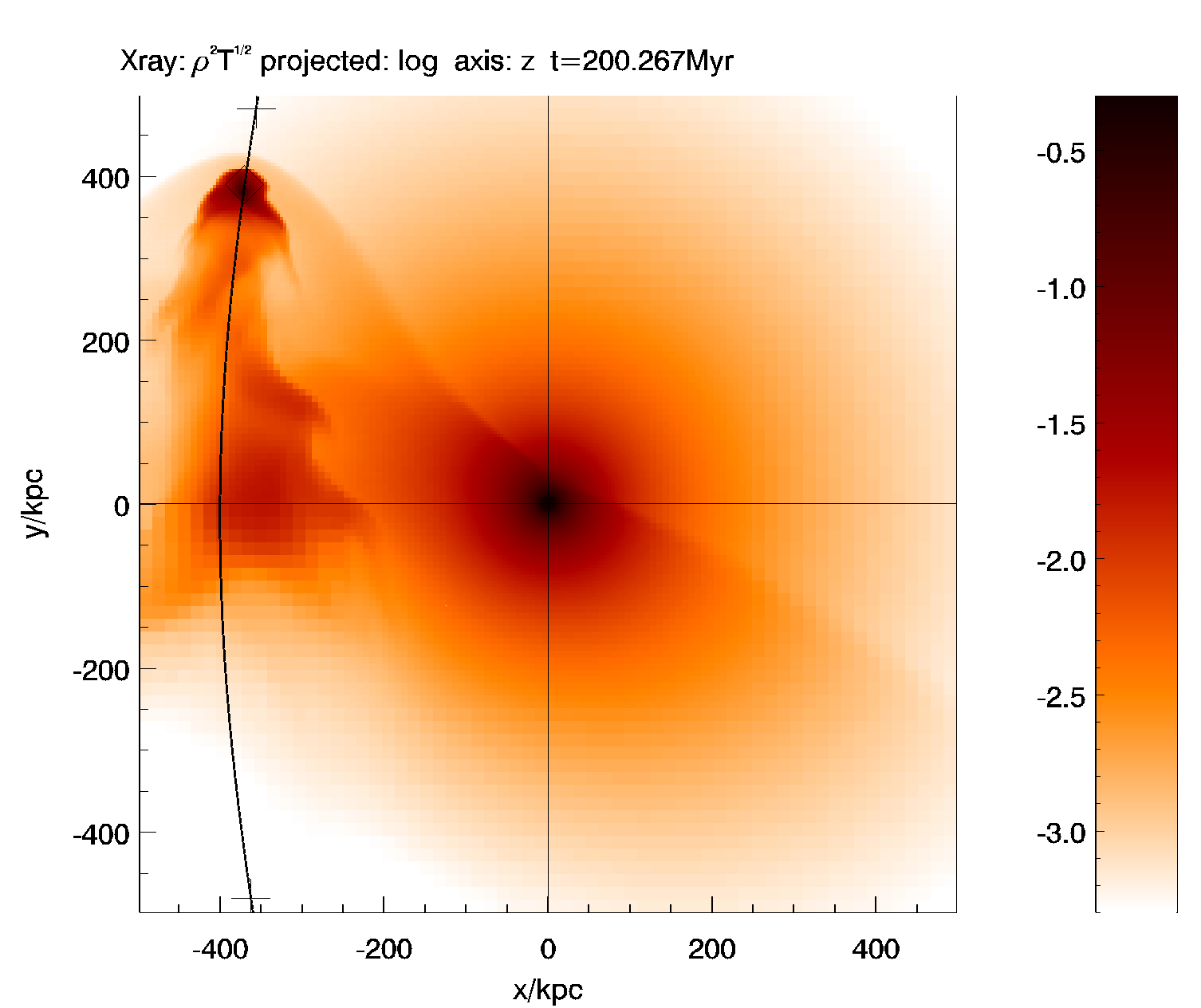}
\includegraphics[trim=0 0 0 -0,clip,height=4.9cm]{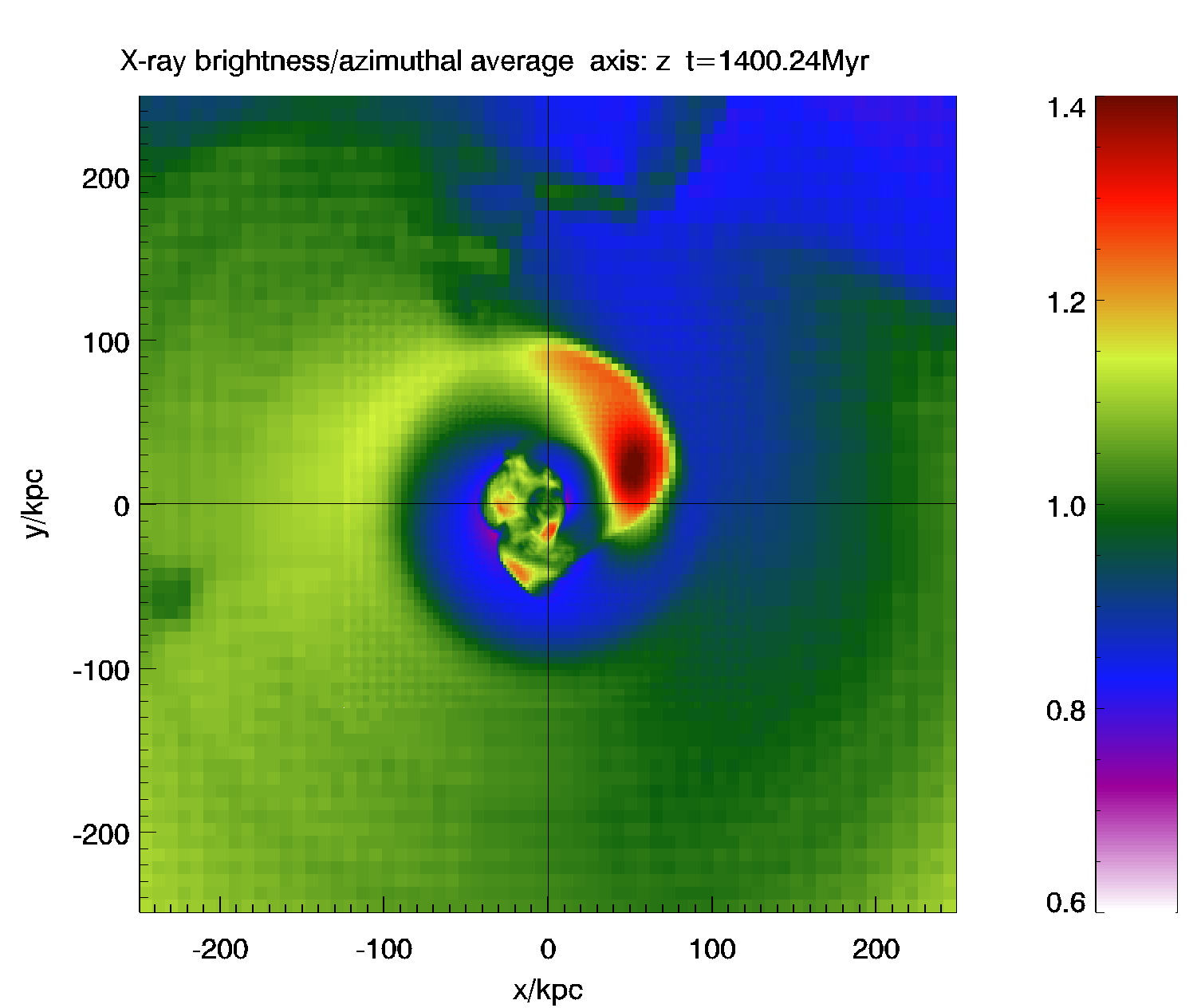}
\end{center}
\caption{
Sloshing triggered by a galaxy's bow shock. \textbf{Top:} Large-scale X-ray image at the moment when the bow shock crosses the cluster centre. The ram pressure stripped galaxy can be seen at the lhs. \textbf{Bottom:} Brightness residual map out to 250 kpc at 1.4 Gyr after the galaxy's pericentre passage. 
}
\label{fig:bowshock}
\end{figure}

We want to advance the idea of a shock as the cause of sloshing CFs and study the impact of a bow shock of a massive galaxy.
Such a scenario is also attractive when regarding merger rates: passages of a $2\cdot 10^{12}M\Sun$ galaxy are more frequent than passages of $> 10^{13}M\Sun$ subclusters.
Thus, we let a galaxy pass the Virgo core at a distance of 400 kpc (orbit "dmin400 fast" in Fig~\ref{fig:orbits}). The large pericentre distance ensures that its ram-pressure stripped tail does not appear in images of the central region.  
A high velocity is needed to produce a sufficiently strong shock. 
The galaxy potential is described by a Hernquist halo with $2\cdot 10^{12}M\Sun$ and scale radius 30 kpc. Initially, the galaxy contains a gaseous halo of $3\cdot 10^6\K$ and central density $10^{-24}$g cm$^{-3}$.  The exact distribution of the gas in the galaxy does not matter as long as the galaxy can retain some gas and hence a bow shock up to pericentre passage (see Fig.~\ref{fig:bowshock}). The galaxy may also be a spiral galaxy. The bow shock passes the Virgo core about 200 Myr after the closest approach of the galaxy. 
The shock normal points towards the diagonal between the $+x$- and $+y$-directions, which is also the main direction of the triggered sloshing. Fig.~\ref{fig:bowshock} also displays a brightness residual map at 1.2 Gyr after shock passage, i.e.~1.4 Gyr after pericentre passage of the galaxy. In the inner 120 kpc, the synthetic maps resemble the ones of the classic scenario (Fig.~\ref{fig:fiducial-residual}) except that they appear to be rotated by about $45\degree$ anti-clockwise. In order to match the observed map, we identify the direction of the NW CF with the diagonal between the $+x$- and $+y$-directions. 
%
%FFFFFFFF
%FFFFFFFFF
%
%
%FFFFFFFF
\begin{figure}
\centering
\includegraphics[width=0.46\textwidth]{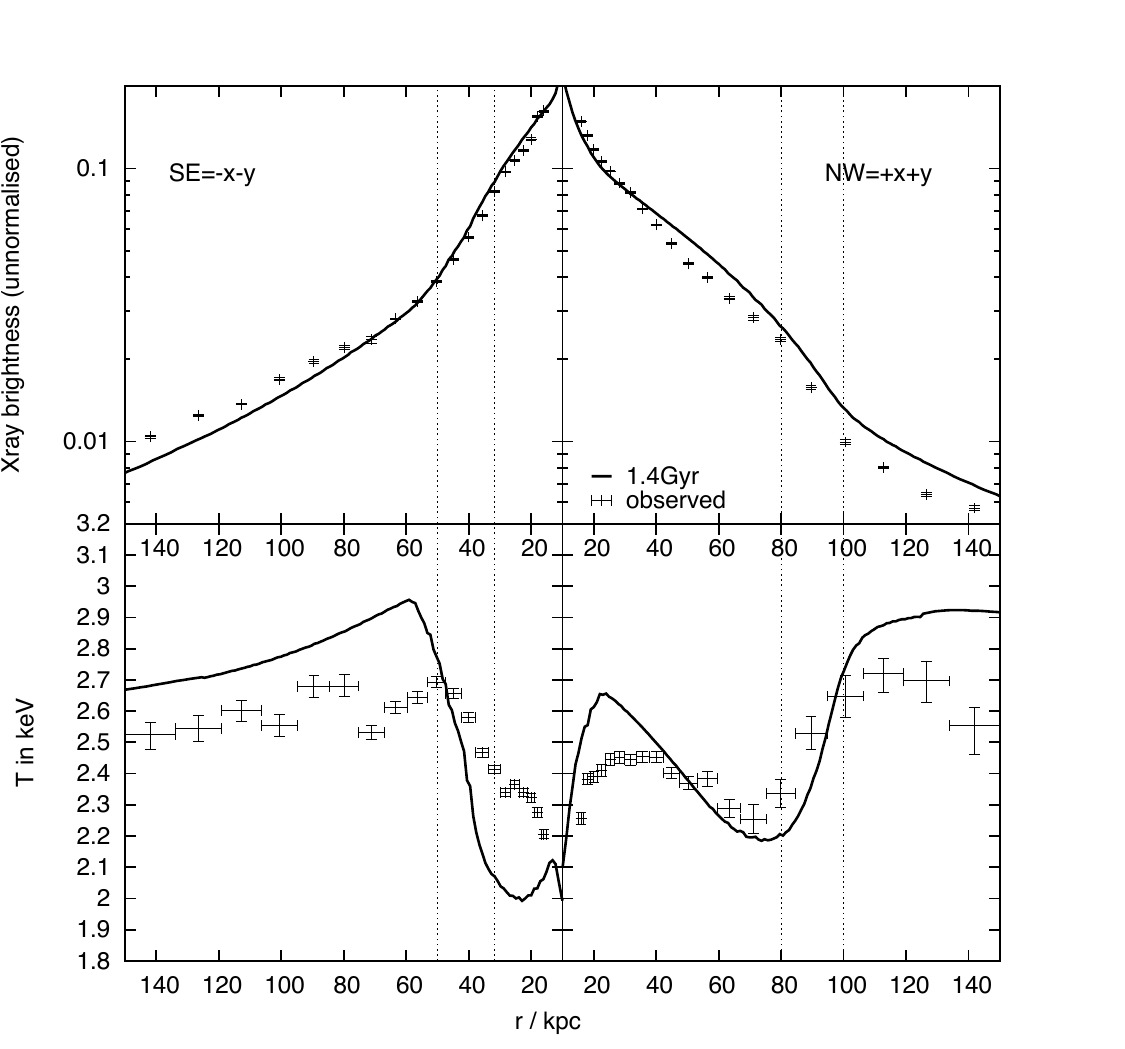}
\caption{
Projected and azimuthally averaged profiles across the CFs after bow shock passage: X-ray brightness, projected temperature (compare to Fig.~\ref{fig:bestprof}). We match the orientation of our simulated maps on the sky such that the direction of the observed NW CF is towards the diagonal between the $+x$ and $+y$-axes.
}
\label{fig:bowshock_prof}
\end{figure}
%FFFFFFFFF
%
Figure~\ref{fig:bowshock_prof} compares the observed and synthetic profiles of X-ray brightness and projected temperature at this moment. The radii of all CFs are matched simultaneously, and the contrasts across the NW front agree well. The simulated ones across the SE front are too strong. Also the X-ray profile on the NW side slightly deviates from the observed one. However, these details will also depend on the details of the interaction with the passing galaxy. Thus, we conclude that also the bow shock of a galaxy passing the Virgo core as explained above could have caused the observed CF structures in the Virgo cluster. 

The best way to distinguish between this and the classic scenario is provided by the larger-scale structure: Both, the cool brightness excess fan between about 100 and 170 kpc SE of the cluster core, and the pronounced brightness excess outside 200 kpc NW of the Virgo core (see Sect.~\ref{sec:compare_largescale}) would be absent in the galaxy bow-shock scenario.
 In this case, a large-scale view may reveal traces of the gas stripped from the galaxy. Where exactly this may be found would also depend on the original gas distribution inside the galaxy and thus the galaxy's ram pressure stripping history.

A possible candidate galaxy is discussed in Sect.~\ref{sec:compare_identify}.

%% file: discussion.tex
%****************
\section{Discussion} \label{sec:discussion}
%

%*******
\subsection{Reliability of our model} \label{sec:discuss_rigidpots}
The strongest concern about our model is the restriction to rigid potentials. However, in a direct comparison to full hydro+Nbody simulations, Roediger\&ZuHone (in prep.) show that the rigid potential approximation reproduces the orientation of the sloshing signature correctly. From the size of the CF structure, the  rigid potential approximation overestimates the age of the minor merger slightly, by about 200 Myr. The temperature inside the CFs as well as the density inside and outside of the CFs is reproduced well. In our comparison to the Virgo data, we have used only these well-reproduced characteristics. 

 %*******
\subsection{Identifying the responsible subcluster} \label{sec:compare_identify}
According to the merger geometry derived in  Sect.~\ref{sec:compare_morph}, the subcluster that perturbed the Virgo core has moved from $\sim$W-SW to E-NE, passing the Virgo core in the South about $1.5\Gyr$ ago. Consequently, the current position of the subcluster (see Fig.~\ref{fig:orbits}) corresponds to about 1.5-2.5 Mpc E-NE of the Virgo centre. The uncertainty in distance is due to the ill-constrained velocity of the subcluster. Our test particle orbits do not take into account dynamical friction, which will cause the subcluster to be at a somewhat closer distance than derived from Fig.~\ref{fig:orbits}. Thus, our simulations suggest the perturbing subcluster about 1-2 Mpc eastward of the Virgo centre. The subcluster should still be moving away from the Virgo centre.

The region in question is just outside the Virgo cluster catalogue (\citealt{Binggeli1985,Binggeli1987,Binggeli1993suppl}). An inspection of  sky maps reveals the following candidates: M60 (radial velocity\footnote{Heliocentric radial velocities are taken from  the NASA/IPAC Extragalactic Database (NED).} $1117\Kms$, Virgo mean is $957\Kms$), which is situated about 1 Mpc E-SE of the Virgo centre; further north from here are NGC 4654 (radial velocity $1046\Kms$), NGC 4639 (radial velocity $1018\Kms$), NGC 4659 (radial velocity $480\Kms$), and NGC 4689 (radial velocity $1616\Kms$) still further North. All of these except  NGC 4659 have a radial velocity close to the Virgo cluster mean, suggesting that their orbits are close to the plane of the sky. The massive elliptical galaxy M60 contains about $10^{13}M\Sun$ within 76 kpc (\citealt{Shen2010}), which is close to the mass we require. However, it still contains a gaseous atmosphere of its own. It might have passed the Virgo core at a large enough distance to prevent its ram pressure stripped gas to show up in the central part. In this case, its bow shock could have assisted the gravitational triggering of the sloshing.  Alternatively, it may have acquired a new gaseous halo within the last 1.5 Gyr.

If the CF signatures in the Virgo cluster were caused by the bow shock of a galaxy (see Sect.~\ref{sec:bowshock}), the disturbing galaxy should now be found about 1.5-2 Mpc N-NE of the Virgo core. 
A possible candidate is the peculiar S0 galaxy M85, about 2 Mpc N of Virgo. Its heliocentric radial velocity is $730\Kms$, which is again close to the Virgo mean. This galaxy even shows some recent star formation in its centre that may have been triggered by the interaction with the ICM, but may also be due to interaction with its neighbour  NGC 4394 (see \citealt{Sansom2006} and references therein).

%*******
\subsection{Stability of CFs}
CFs are expected to be associated with shear motions, which should lead to the formation of Kelvin-Helmholtz instabilities (KHIs) and possibly to the destruction of CFs. We indeed observe this instability in the later stages (after 1 Gyr) in our simulations, although it is not omnipresent. 
Moreover, we find that the CFs are never destroyed completely, but are reformed continuously. 
Observing the KHI might be intrinsically difficult: in Appendix~\ref{sec:resolution} we show that in projection the CFs appear less disturbed in the case of the highest resolution, although here the KHI is clearly present. With such high resolution (CFs resolved with 0.5 kpc), the KHI takes place at small scales, which are averaged out by projection. Thus, CFs may appear sharp in projection although the KHI is taking place.

%%*******
\subsection{Impact of sloshing on the radio lobes}
\begin{figure*}
\includegraphics[trim=0 0 400 0,clip,height=5.5cm]{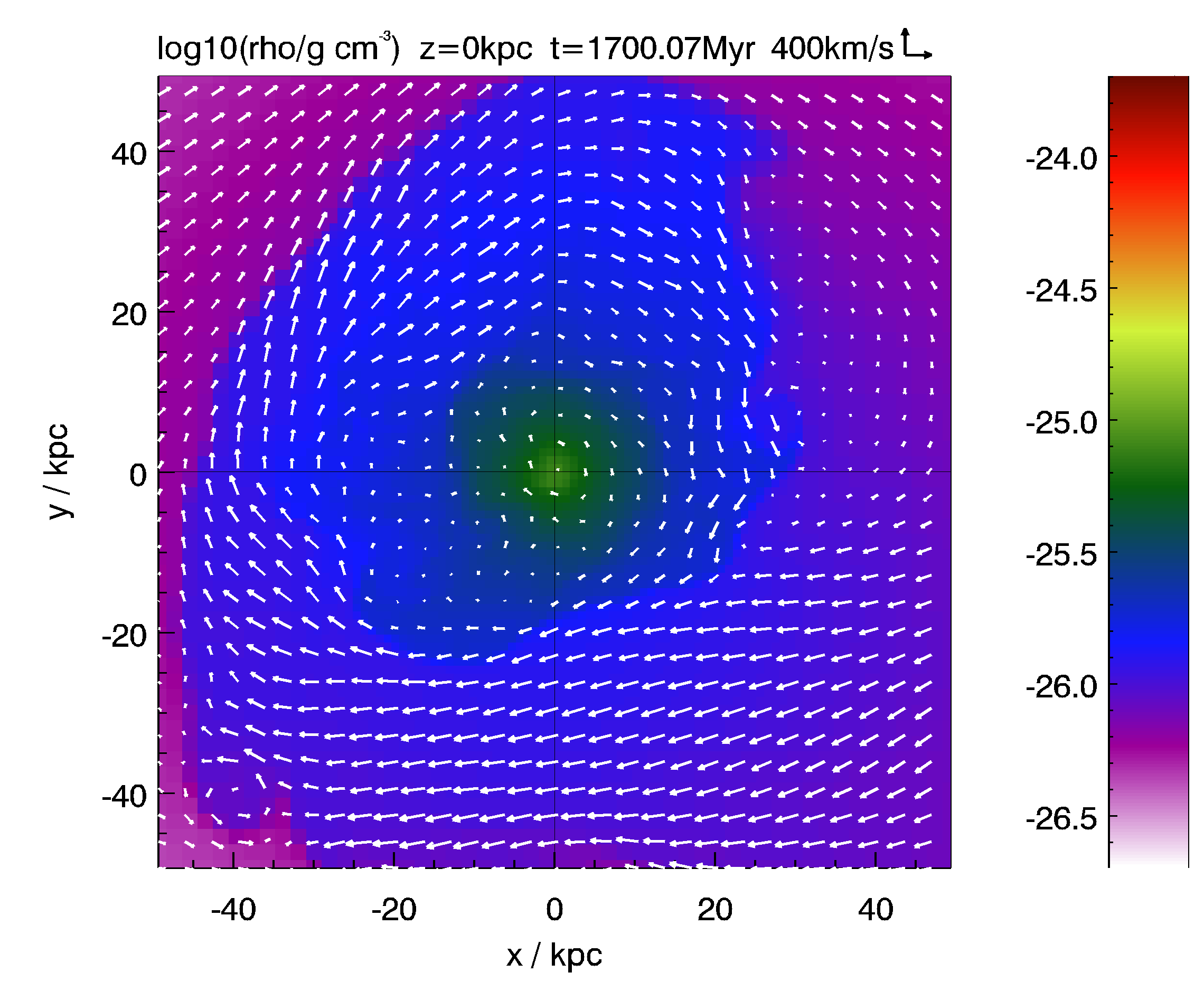}
\includegraphics[trim=100 140 0 0,clip,origin=c,angle=-90,height=5.9cm]{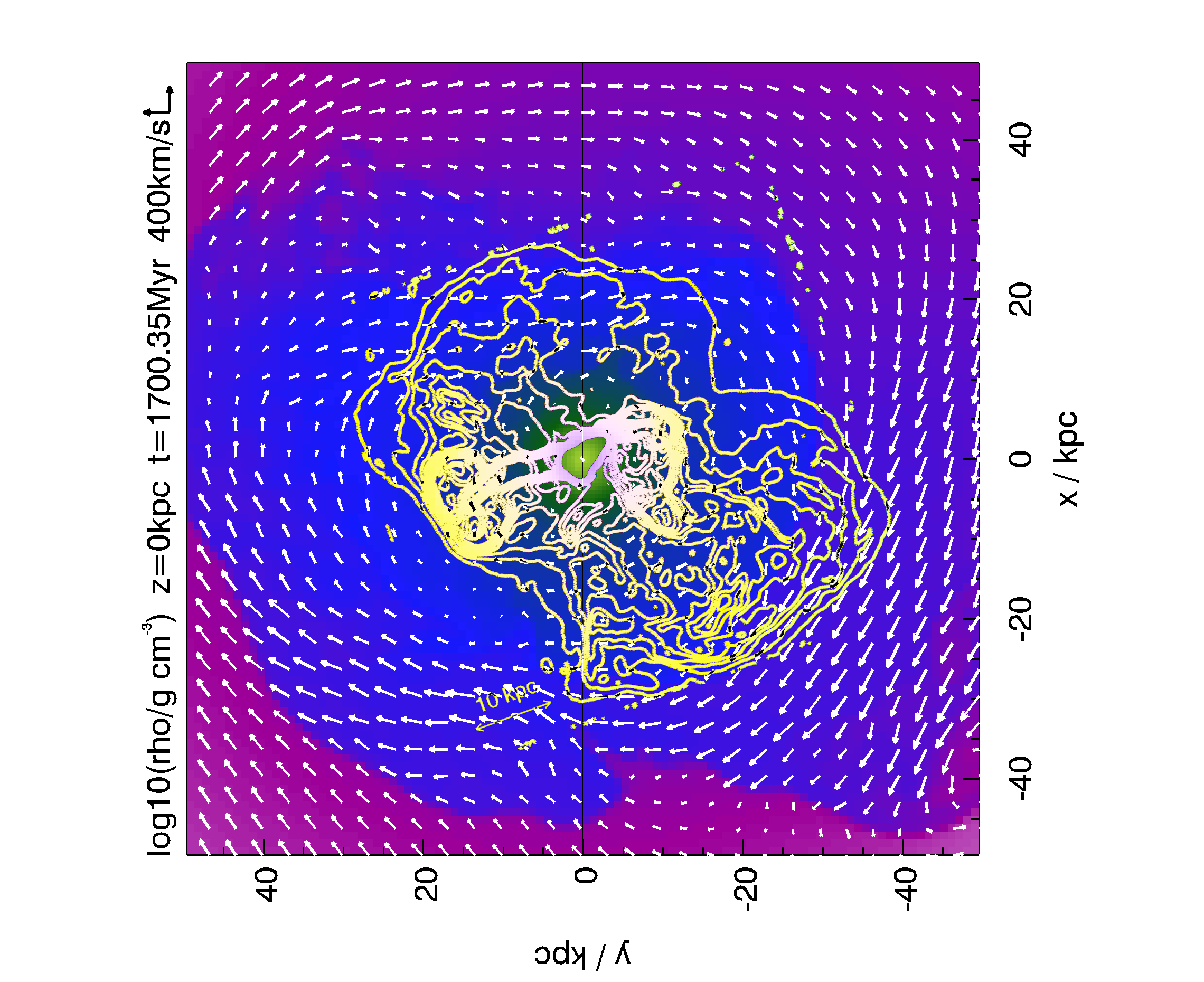}
\includegraphics[trim=200 0 0 0,clip,height=5.5cm]{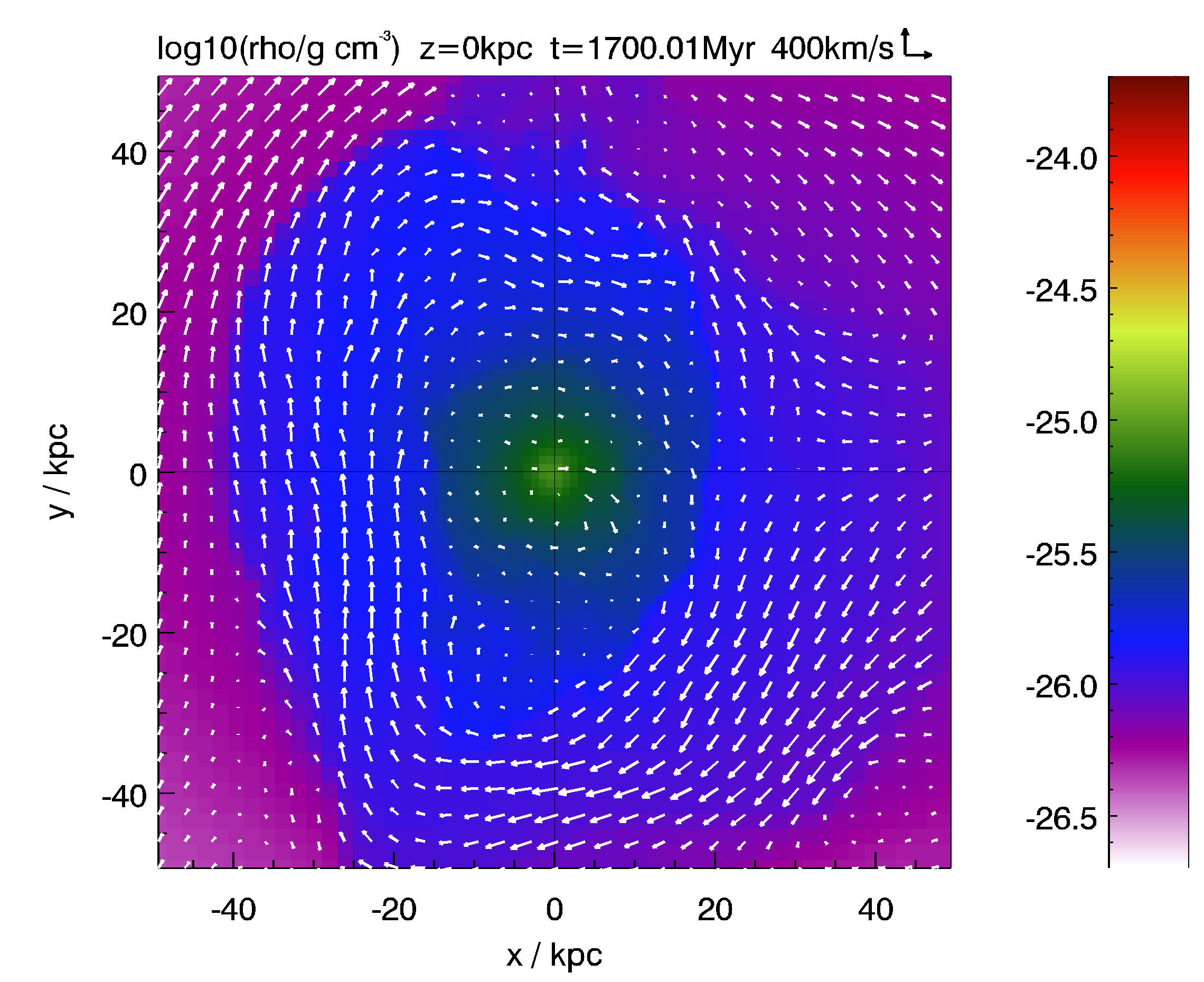}
\vspace{-0.8cm}
\caption{Zoom-in of the velocity field (white arrows) and density field (colour-coded in $\gccm$) at $t=1.7\Gyr$ in the orbital plane of the subcluster. We show of our best-match cases: M1a50dmin100dmax3 (left), M2a100dmin100dmax3 (middle), M4a100dmin400dmax3 (right). In the middle panel, we overlay the observed radio lobes (\citealt{Owen2000}).}
\label{fig:bend_arms}
\end{figure*}

The central AGN of the Virgo cluster has induced a rich structure inside the inner $\sim 30 \Kpc$ of the ICM, including radio lobes and cavities. The inner radio lobes show a peculiar morphology: while the eastern arm extends straight out to about 20 kpc, the western arm is bent sharply southwards at about $15 \Kpc$ from the cluster centre. S10 discussed the possibility that the western arm is bent due to the same sloshing motions that cause the CFs. 
Furthermore, the outer radio lobes are rotated about $50\degree$ clockwise w.r.t.~the inner lobes.
In Fig.~\ref{fig:bend_arms} we compare the central velocity fields from three of our best-match simulations to the observed radio lobes. In our images, N-NW is along the $+x$-direction, and we have rotated the observed radio contours accordingly. In the central 15 kpc, the velocity field is weak and unlikely to modify the AGN outflow. Outside $15\Kpc$, 
the velocity field exhibits a clear clockwise rotational pattern, but the details depend on the subcluster and the orbit characteristics. At the western side, the flow moves into the same direction as the bending of the jet arm, but the reason of the sharp bend is not obvious. However, the rotational flow pattern could have rotated the outer, and older, radio lobes to their current orientation within $\sim 100\Myr$, which is comparable to their ages derived by \citet{Owen2000} and \citet{Werner2010}.  The latter authors also argue that both, the inner and outer radio lobes extend considerably along our LOS. Thus, the radio lobes might be affected be the sloshing only mildly. The details of the interplay of the AGN activity and gas sloshing have to be studied with a set of dedicated simulations.
%FFFFFFFFFFFFFFF
%FFFFFFFFFFF
%

%% file: summary.tex
%****************
\section{Summary} \label{sec:summary}
We perform a set of hydrodynamical simulations of minor-merger induced gas sloshing and the subsequent formation of cold fronts. For the first time, these simulations aim at reproducing the characteristics of the cold fronts and associated structures not only qualitatively, but also quantitatively. In this paper, our target is the Virgo cluster. 

By comparing synthetic and real observations in great detail, we show that the sloshing scenario can reproduce the known characteristics: a pair of cold fronts, their radius, the contrast in X-ray brightness, projected temperature, and metallicity across the fronts, and  a spiral-shaped brightness excess. 
The comparison suggests the presence of a third cold front 20 kpc N-NW of the cluster centre, whose detection was unsure so far.

Furthermore, we identify additional,  so far unreported,  features typical for sloshing cold fronts. (i) X-ray brightness excess regions alternate with brightness deficit regions. Exactly spatially congruent structures are found in temperature and metallicity maps, where brightness excesses correspond to cool, metallicity enhanced regions and vice versa. This alternating behaviour is also evident when over-plotting profiles across opposite cold fronts. (ii)  Inside each cold front, the temperature profile is constant or radially decreasing, and it is  accompanied by a plateau in metallicity. (iii) On the outside, the cold fronts are bordered by a warm rim. (iv) On a larger scale of a few 100 kpc, there is a 
typical large-scale brightness asymmetry, which will be helpful in constraining the orbit disturbing subcluster. 
We can trace these new features not only in Virgo, but also in other clusters exhibiting sloshing cold fronts. 

Our constraints on the disturbing subcluster are the following: From the size of the observed cold front structure, we constrain the age of the cold front structure in Virgo to about 1.5 Gyr. The presence of the spiral-shaped brightness excess suggests our line-of-sight to be nearly perpendicular to the orbital plane. The contrasts of X-ray brightness and temperature across the fronts,  exclude subcluster masses below $10^{13}M\Sun$ and above $\sim 4\cdot 10^{13}M\Sun$. 
The subcluster mass, its size and the pericentre distance of the orbit are degenerate. The limits of the possible range are ($2\cdot 10^{13}M\Sun$, 100 kpc pericentre distance), and ($4\cdot 10^{13}M\Sun$, 400 kpc pericentre distance). The disturbing subcluster must currently be located east of the Virgo centre. Its orbital velocity is no well-determined parameter, thus its current distance can be constrained between 1 and 2 Mpc only. We suggest M60,  currently about $1\Mpc$ east of the Virgo centre as the most likely culprit.

Additionally, we quantify the metal redistribution caused by sloshing in Virgo. We demonstrate that sloshing alone cannot transform a peaked metal distribution proportional to the stellar mass density in M87 to the observed flattened one. We also demonstrate that diffusion is inept to describe this particular redistribution process.

To complete our analysis, firstly we show that the disturbing subcluster could indeed be completely ram pressure stripped before pericentre passage as required by the standard scenario, and thus leaves no traces but the observed ones in Virgo. However, a compact gaseous atmosphere might survive beyond the pericentre passage and leave additional features in the Virgo ICM. Secondly, we demonstrate that the bow shock of a fast galaxy passing the Virgo cluster at $\sim 400\Kpc$ distance also causes sloshing and leads to very similar cold front structures. The disturbing galaxy would be located about 2 Mpc north of the Virgo centre, M85 being a possible candidate.

%% file: appendix.tex
%************
\section{Resolution test} \label{sec:resolution}
%************

The formation mechanism of CFs is accompanied by shear motions which are prone to KH instability. In this instability, the small-scale modes grow fastest, hence, in simulations, their evolution can be resolution-dependent. 
Furthermore, the Virgo cluster potential is very steep in the inner few kpc, which is not  resolved sufficiently in our standard-resolution simulations. This results in a warm ICM core in the inner 5 kpc after the passage of the subcluster, which is absent in the high-resolution cases. However, this very central region does not influence our results. 

\begin{figure*}
\begin{center}
\includegraphics[trim=0 0 600 0,clip,height=5.8cm]{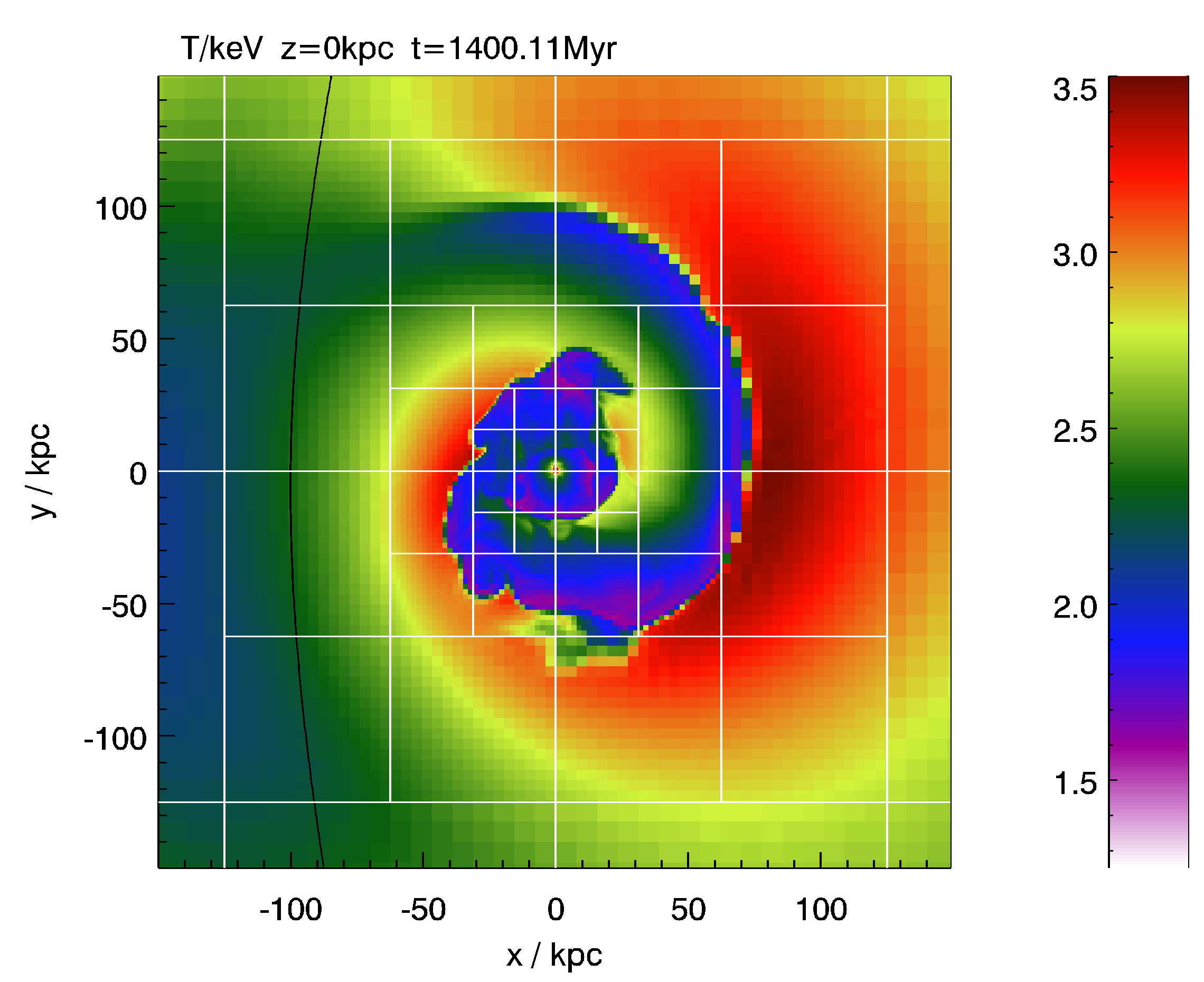}
\includegraphics[trim=340 0 600 0,clip,height=5.8cm]{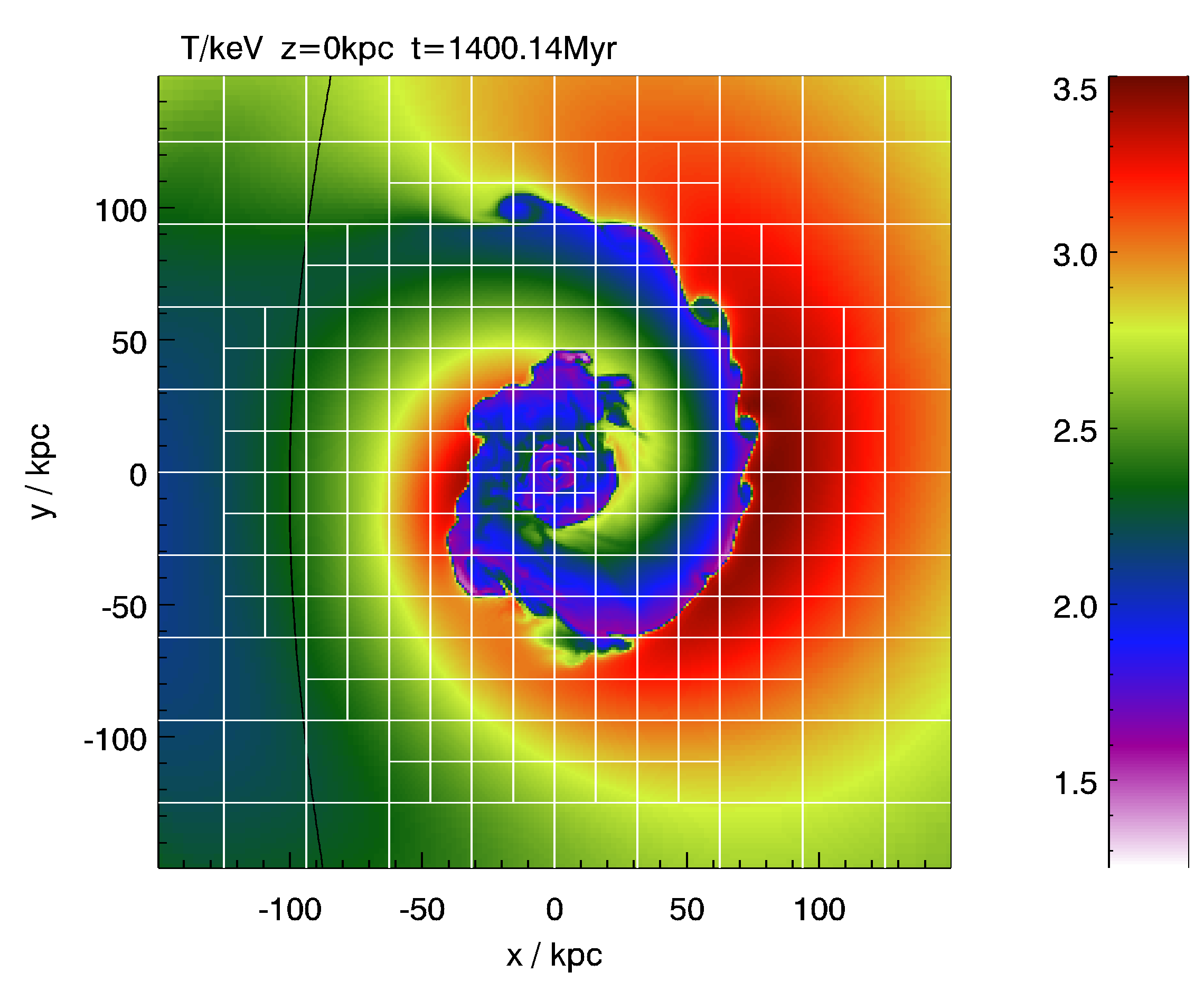}
\includegraphics[trim=340 0 0 0,clip,height=5.8cm]{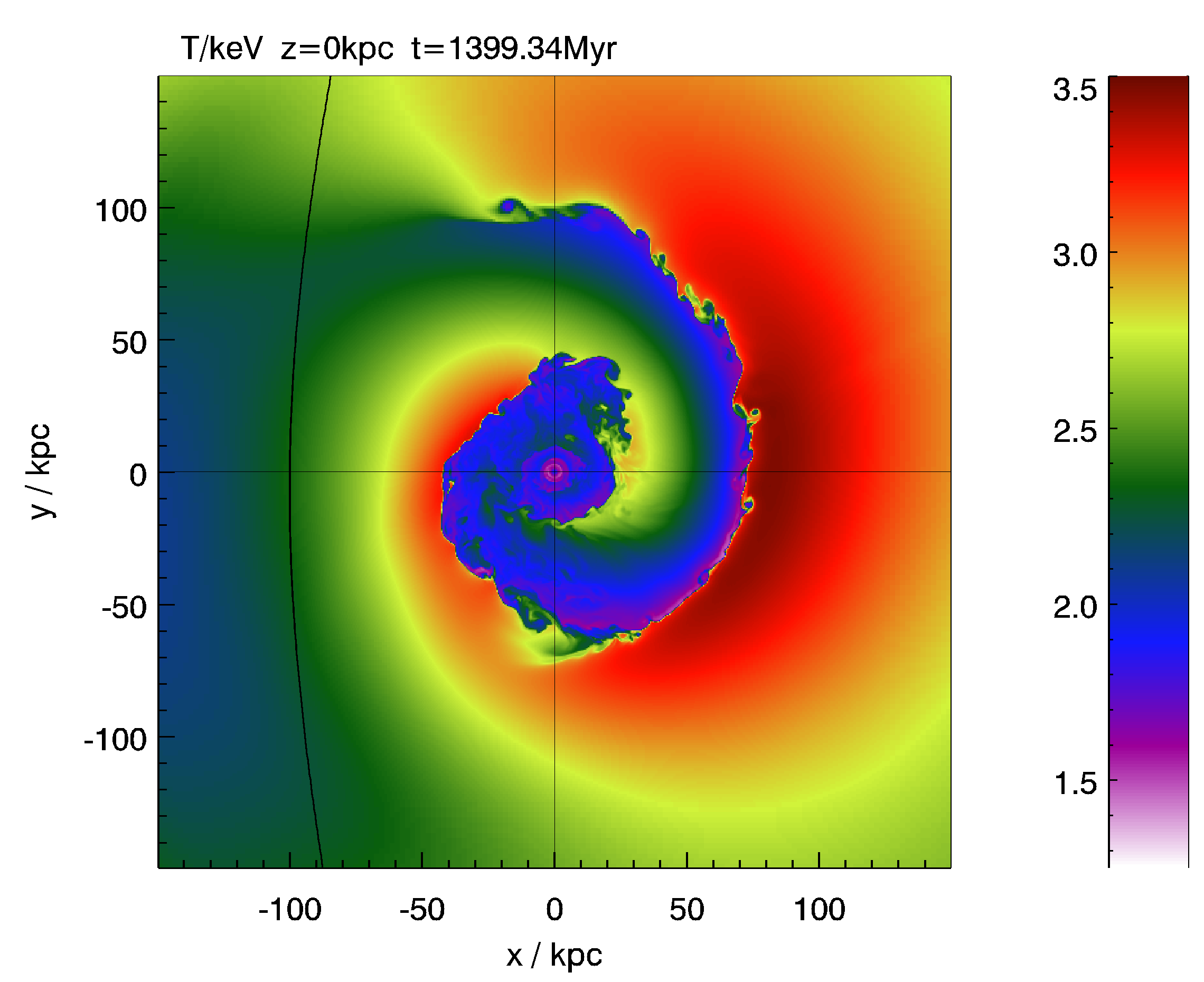}
\hspace{0.2cm}
\newline
\caption{Impact of resolution in fiducial run: Colour-coded temperature slices in the orbital plane. White boxes demonstrate the resolution, one box corresponds to one "block" of  $16^3$ cells. Three different resolutions are shown: \textbf{Left:} standard resolution, innermost blocks have a resolution of 1 kpc. \textbf{Middle:}  High resolution, innermost blocks have a resolution of 0.5 kpc. \textbf{Right:} Very high resolution, same block distribution as middle panel, but each block contains  $32^3$ grid cells, and the innermost blocks have a resolution of 0.25 kpc.}
\label{fig:res_Tslices}
\end{center}
\end{figure*}

We test our fiducial run in 3 resolutions. Figure~\ref{fig:res_Tslices} displays temperature slices in the orbital plane for our standard resolution, a higher resolution case where the resolution is at least a factor 2 better at each position, and a very high resolution which is again a factor of 2 better everywhere.
%
%FFFF
%FFFFFFF
%
The overall structure, regarding size and contrast across CFs, is independent of resolution. The fronts are discontinuities within the resolution for all resolutions.
Even with higher resolution, clear KH rolls appear only after 1 Gyr after core passage.  At no time the CFs are completely disrupted.  With better resolution, KHI takes place at smaller scales. The sloshing motion continuously reforms the CFs.

%
%FFFF
\begin{figure*}
\includegraphics[trim=0 0 350 100,clip,height=5.cm]{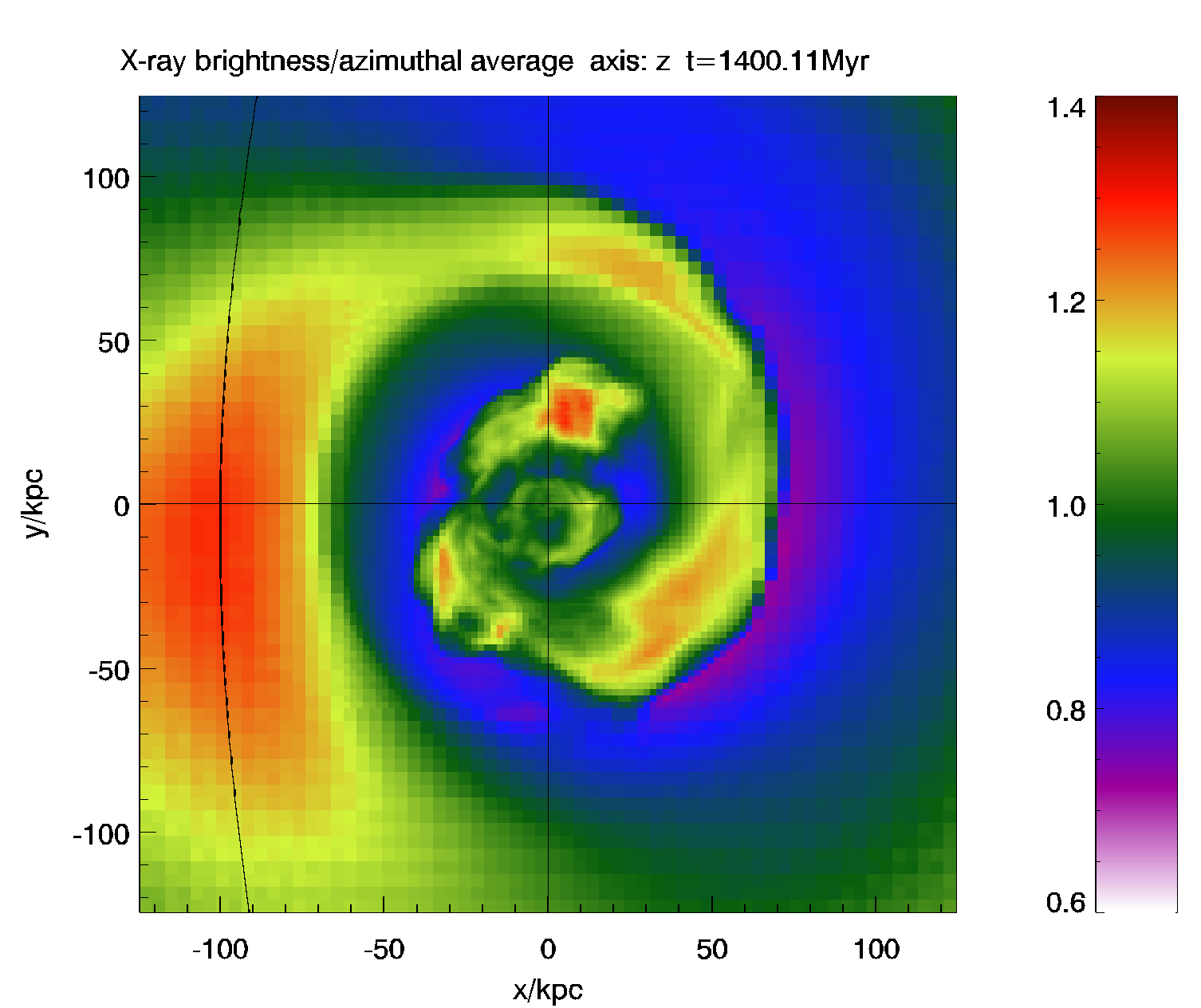}
\includegraphics[trim=180 0 350 100,clip,height=5.cm]{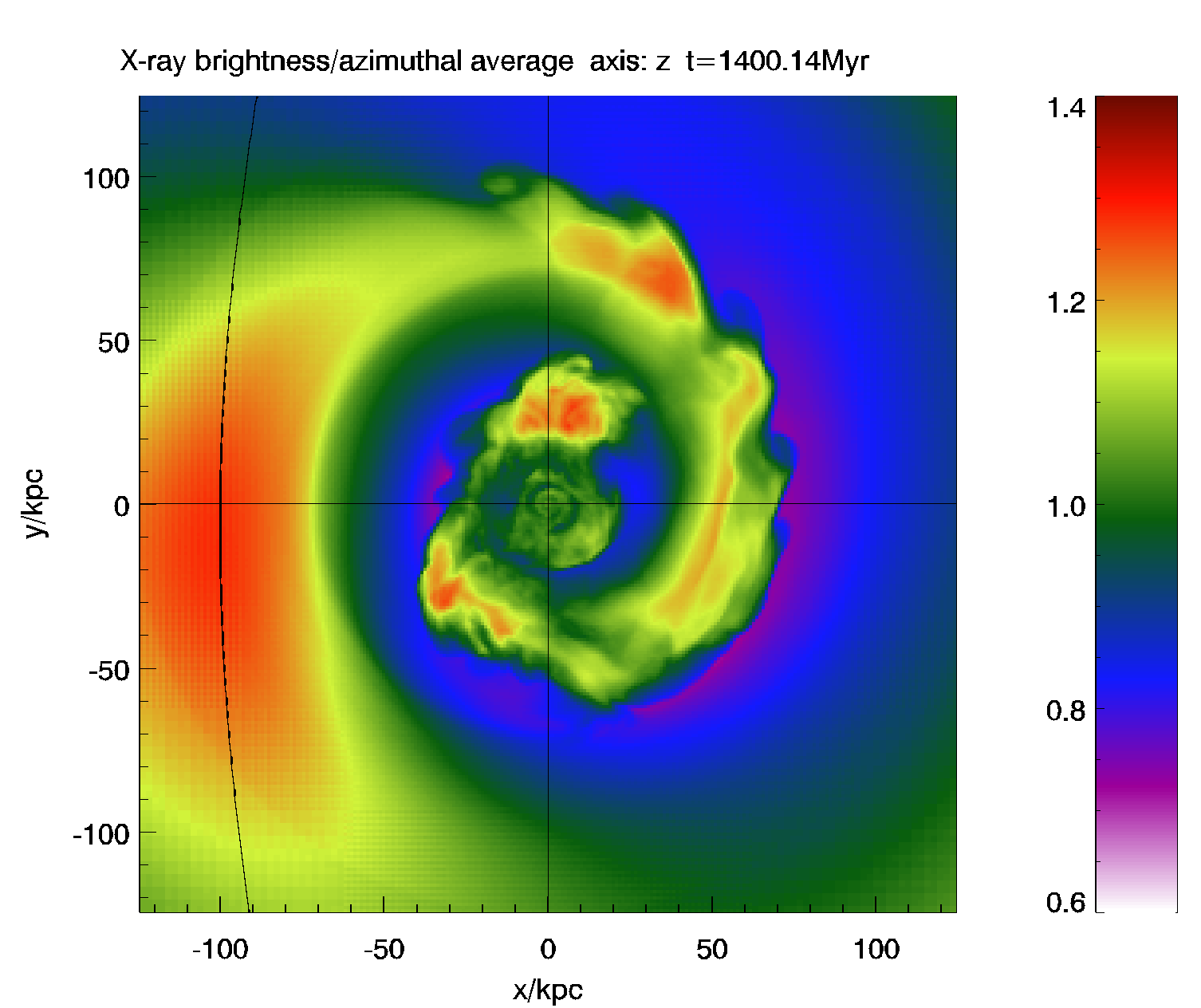}
\includegraphics[trim=180 0 0 100,clip,height=5.cm]{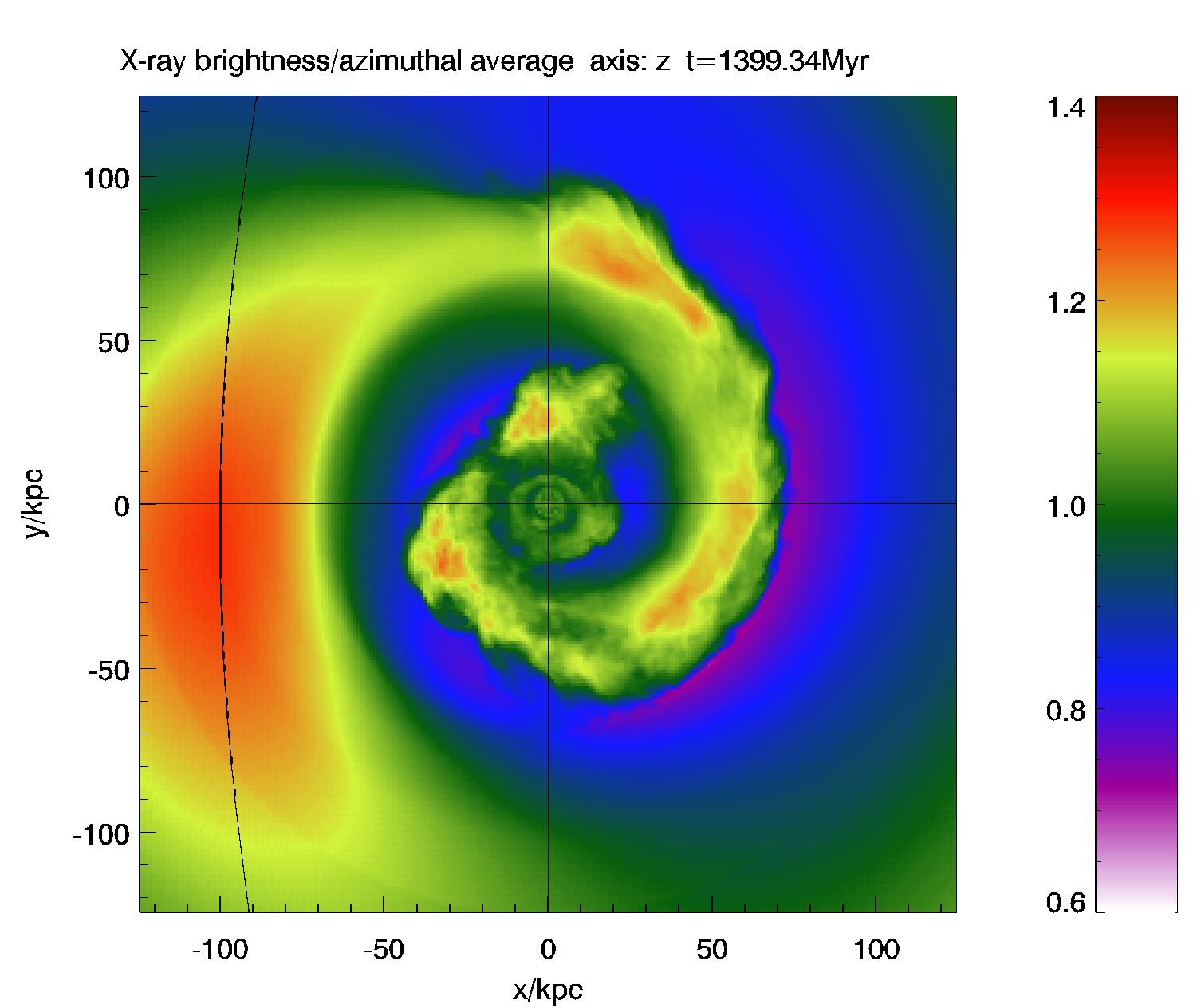}
\newline
\caption{Synthetic brightness residual maps corresponding to Fig~\ref{fig:res_Tslices}.}
\label{fig:res_proj}
\end{figure*}
\begin{figure}
\begin{center}
\includegraphics[width=0.5\textwidth]{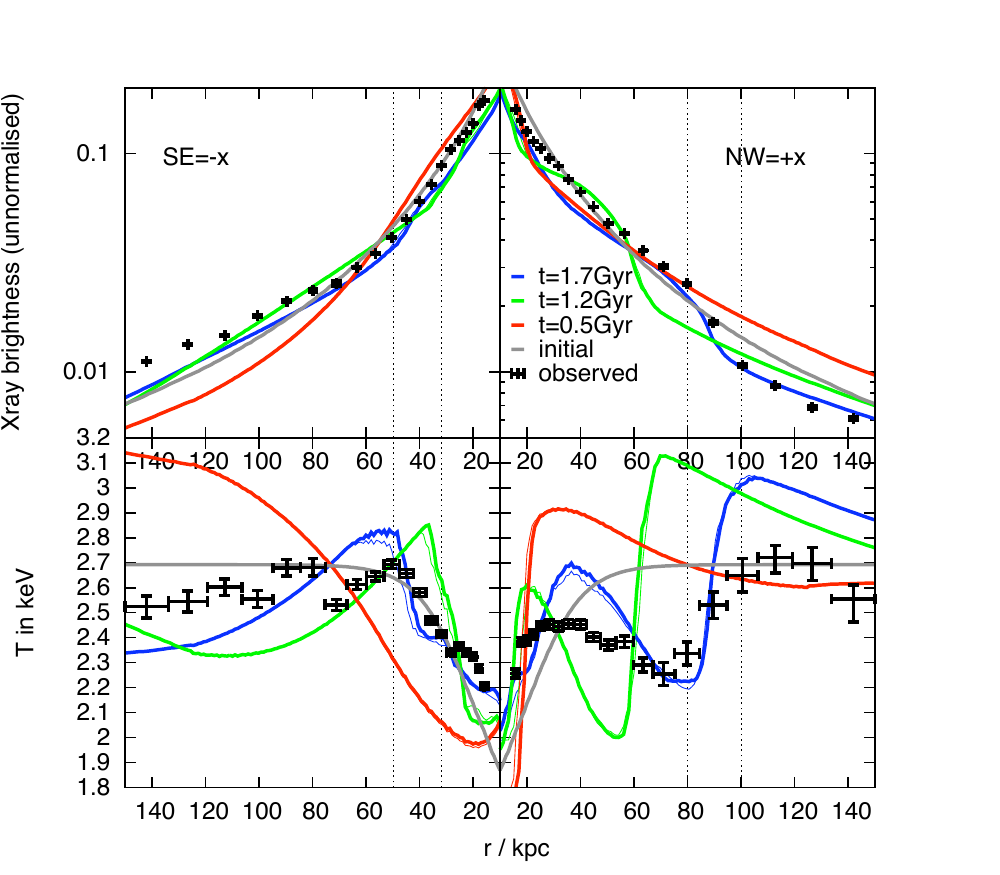}
\caption{Projected, azimuthally averaged profiles of X-ray brightness and projected temperature (like in Fig.~\ref{fig:bestprof}), at $t=0.5, 1.2, 1.7\Gyr$ (see legend), for two resolutions (thick lines for high resolution, thin lines for standard resolution, compare Fig~\ref{fig:res_Tslices}).}
\label{fig:res_profs}
\end{center}
\end{figure}

%FFFFFF
In projection, the KHIs are less pronounced (see X-ray images, residual images, projected temperature maps in Fig.~\ref{fig:res_proj}), for the very high resolution, they are hardly detectable.
The large-scale signature reported in Sect.~\ref{sec:compare_largescale} does not depend on resolution. Neither do the azimuthally averaged profiles of X-ray brightness, projected temperature and metallicity (Fig.~\ref{fig:res_profs}).

%FFFFFFFFFFFF
%FFFFFFFFFFFFFFF

%% file: fiducial.tex
%**********************************
\section{Evolution and dynamics of sloshing in Virgo}
 \label{sec:fiducial}
  \label{sec:fiducialevolution}
%**************************
%
%FFFFFFFF
\begin{figure}
\includegraphics[trim=    0 80 100 0,clip,height=0.16\textheight]{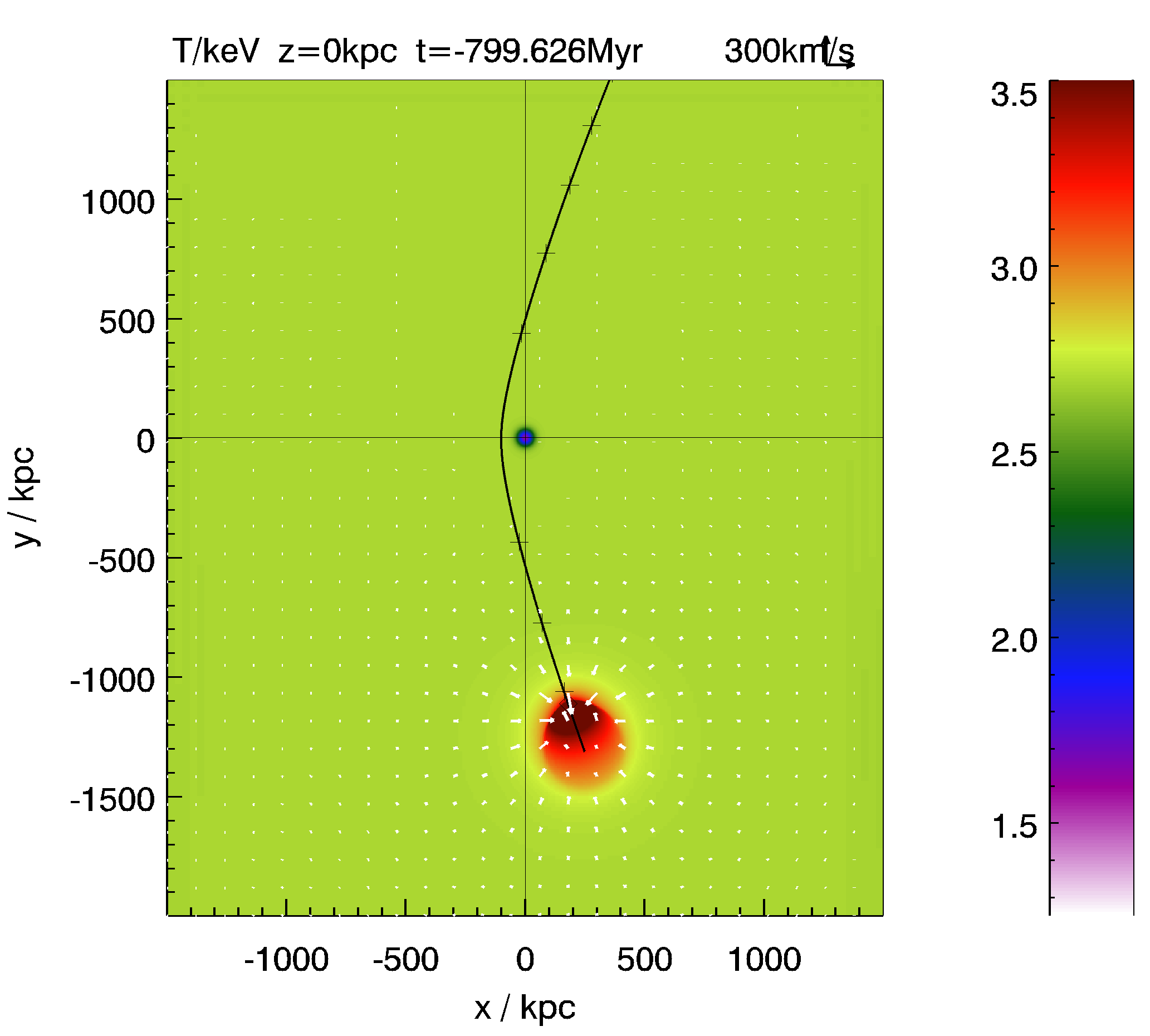}
\includegraphics[trim=    0 80 630 0,clip,height=0.16\textheight]{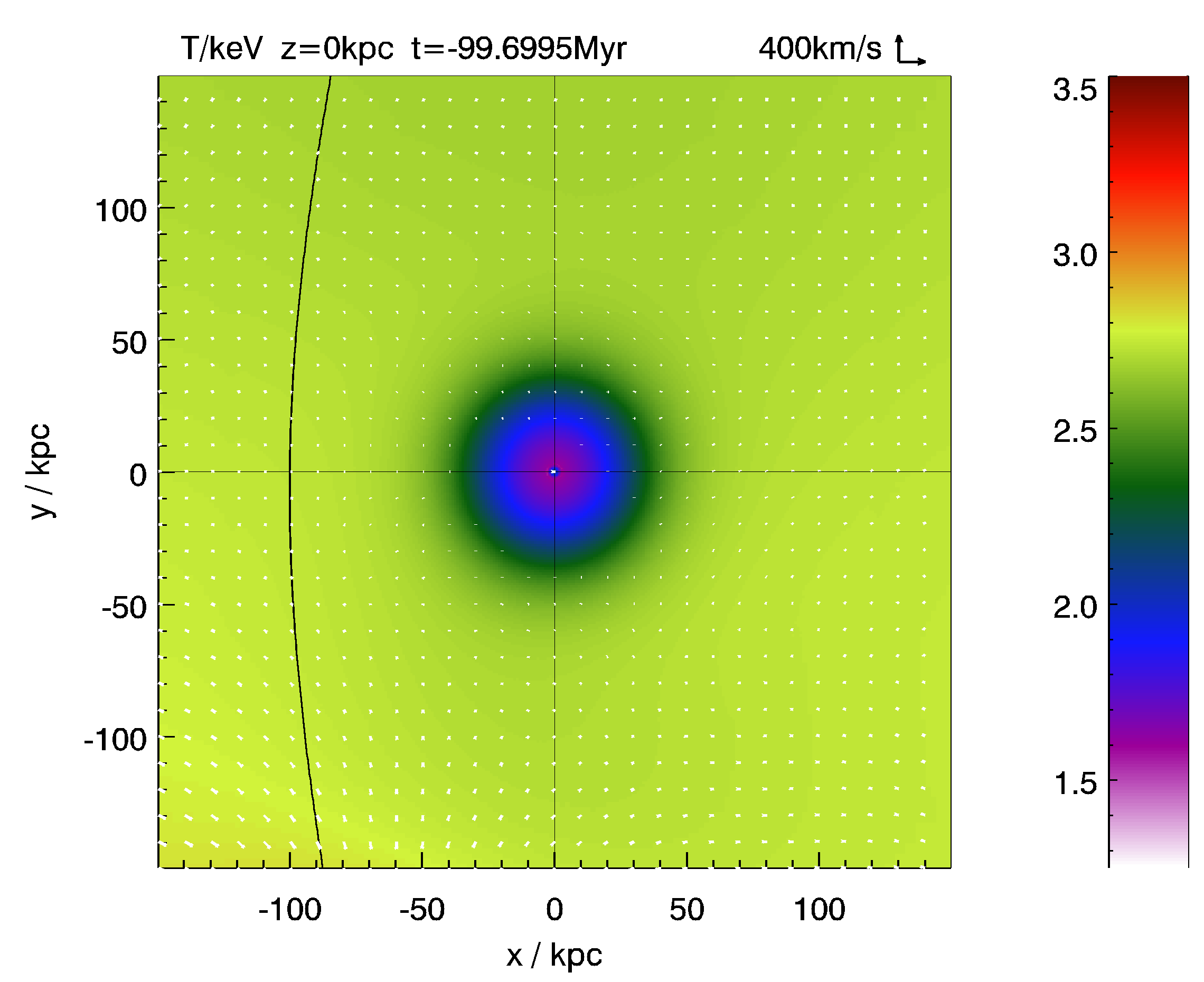}\newline
\includegraphics[trim=    0 80 330 0,clip,height=0.16\textheight]{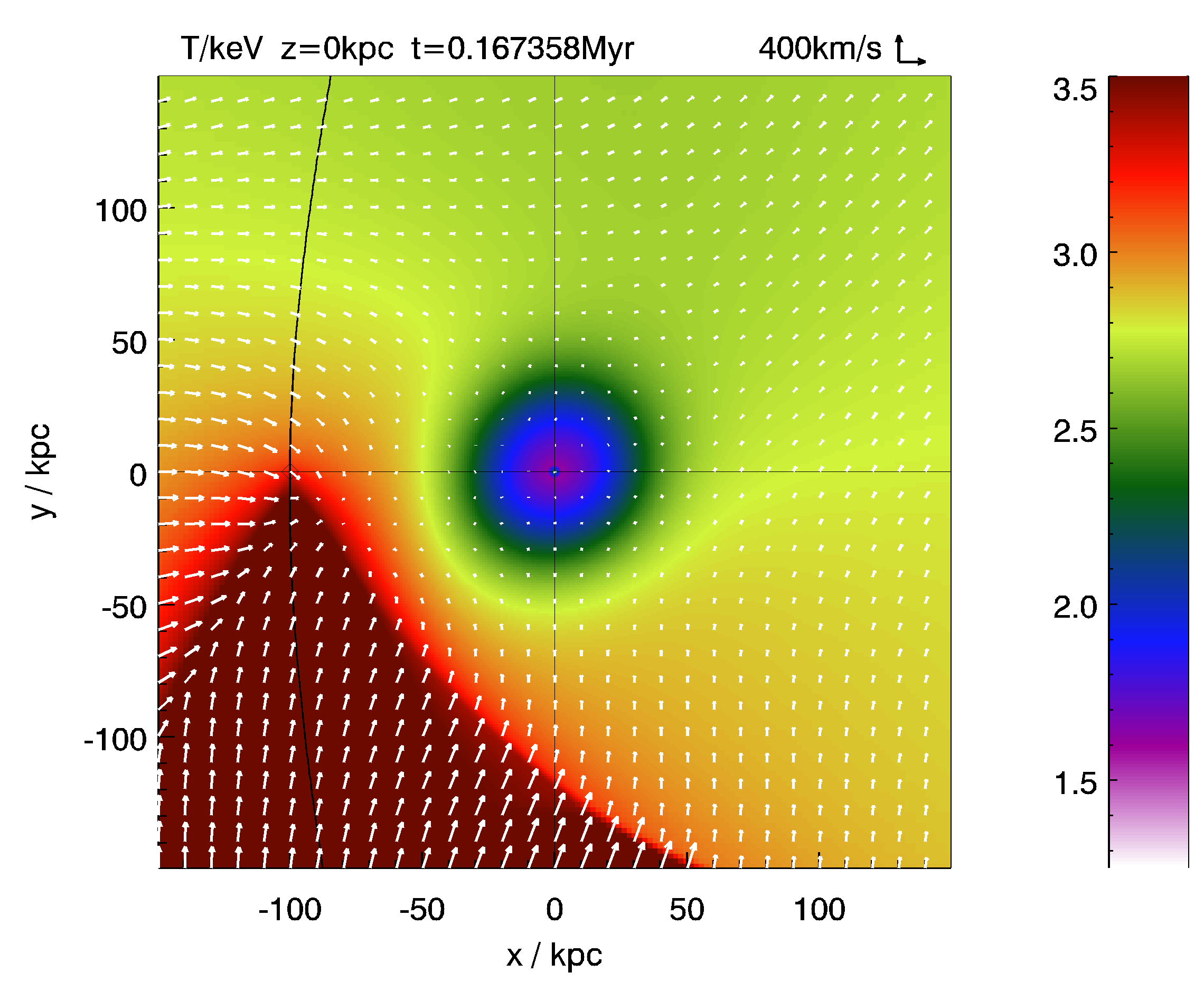}
\includegraphics[trim=300 80 330 0,clip,height=0.16\textheight]{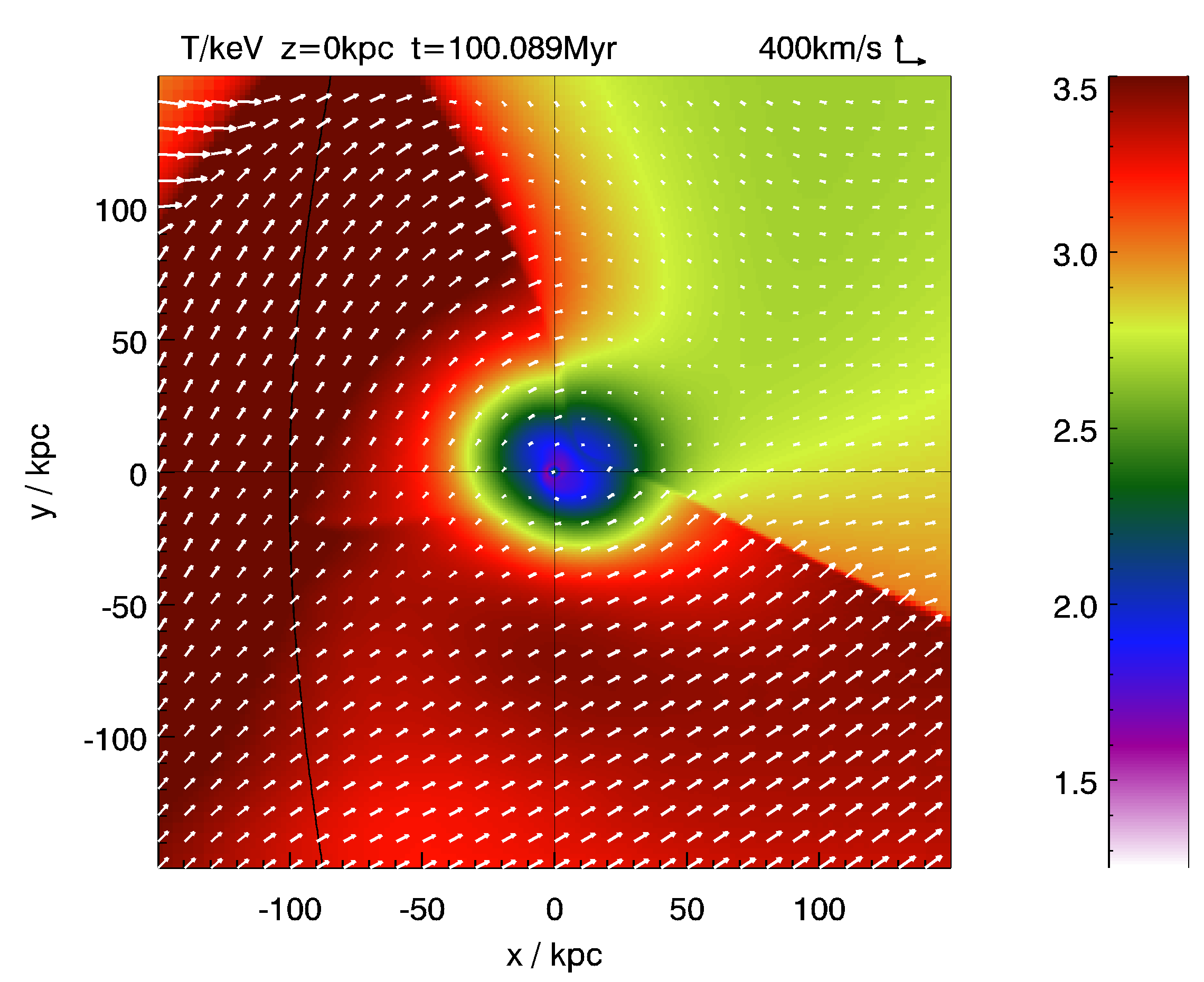}\newline
\includegraphics[trim=    0 80 330 0,clip,height=0.16\textheight]{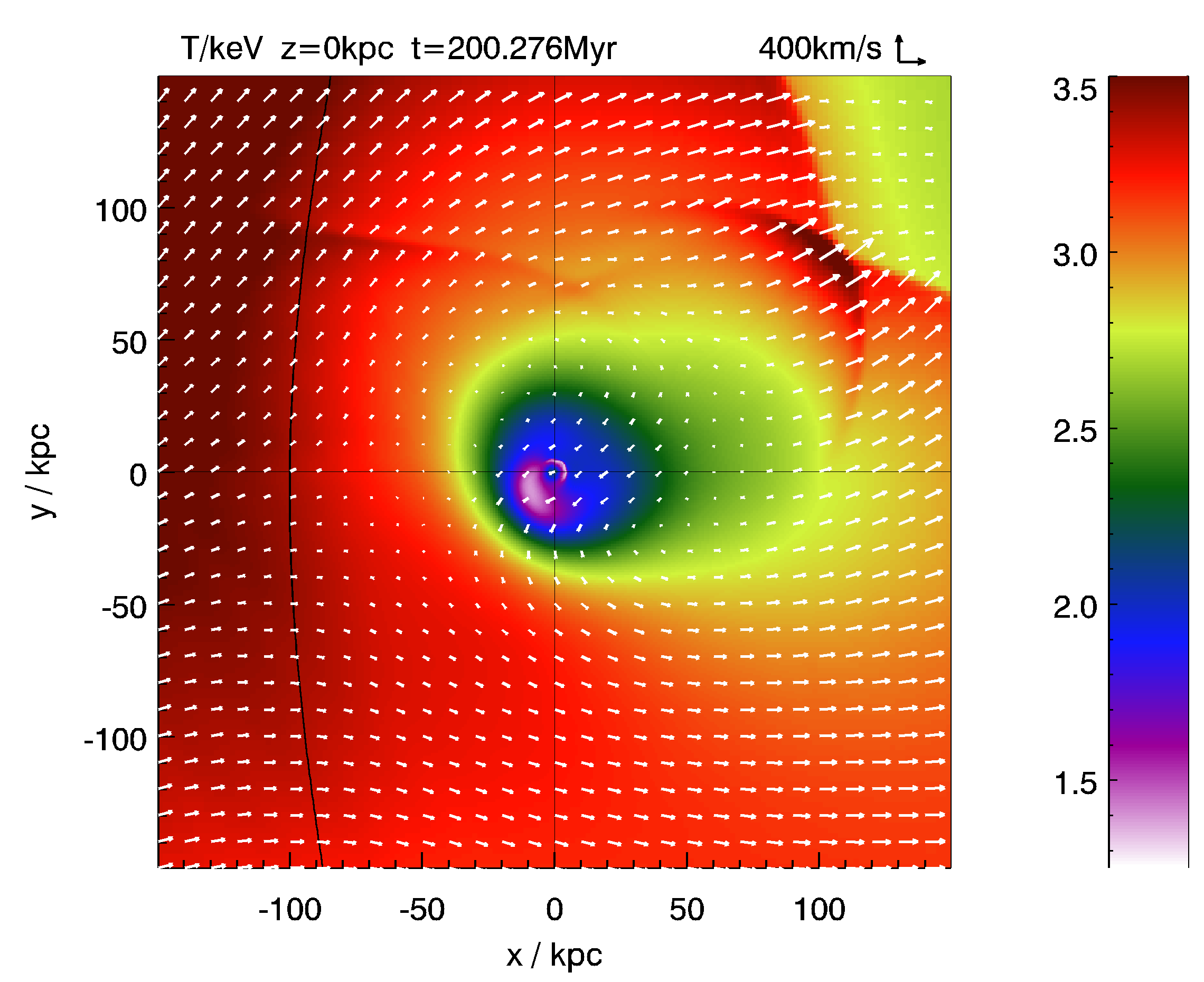}
\includegraphics[trim=300 80 330 0,clip,height=0.16\textheight]{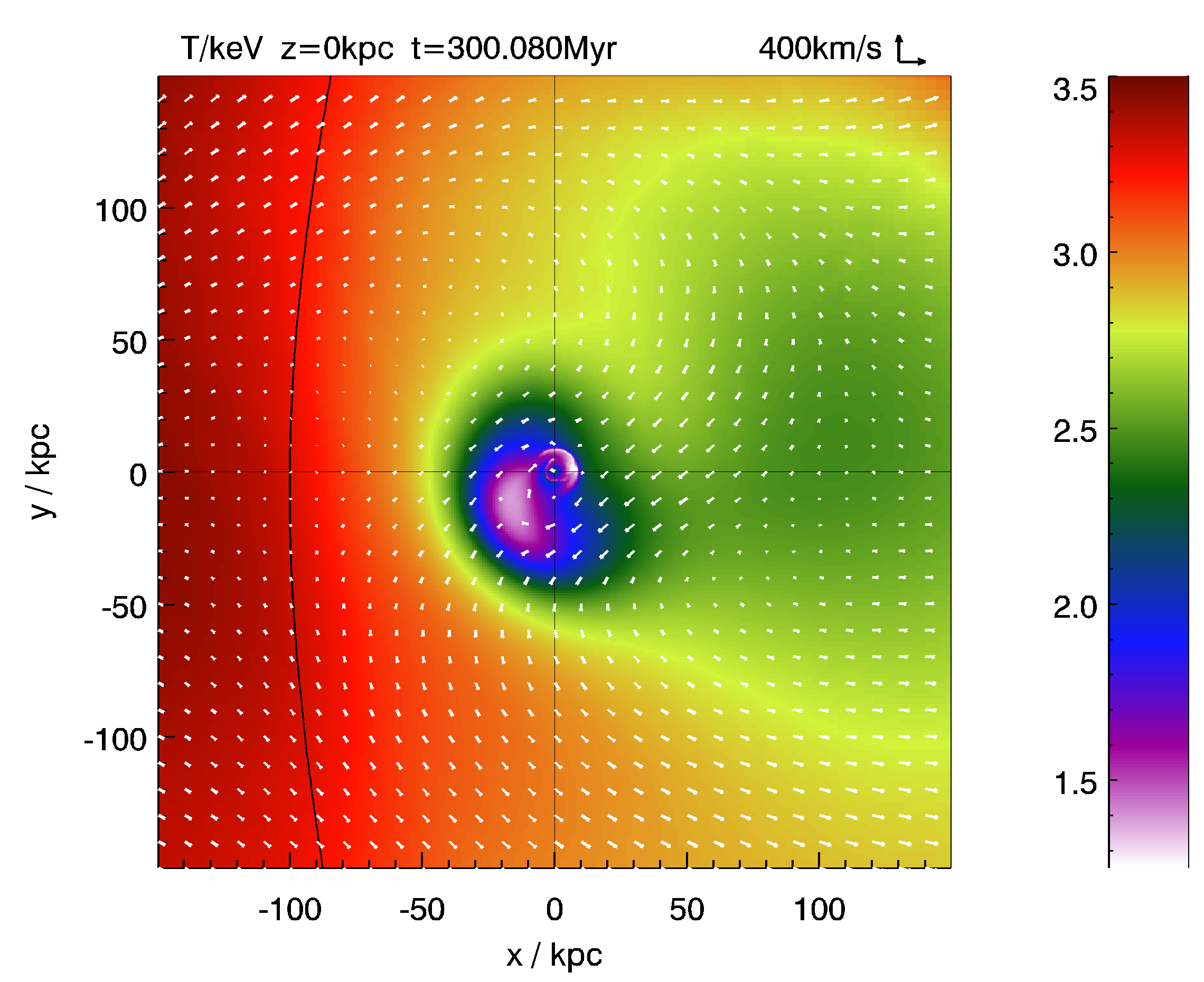}\newline
\includegraphics[trim=    0 80 330 0,clip,height=0.16\textheight]{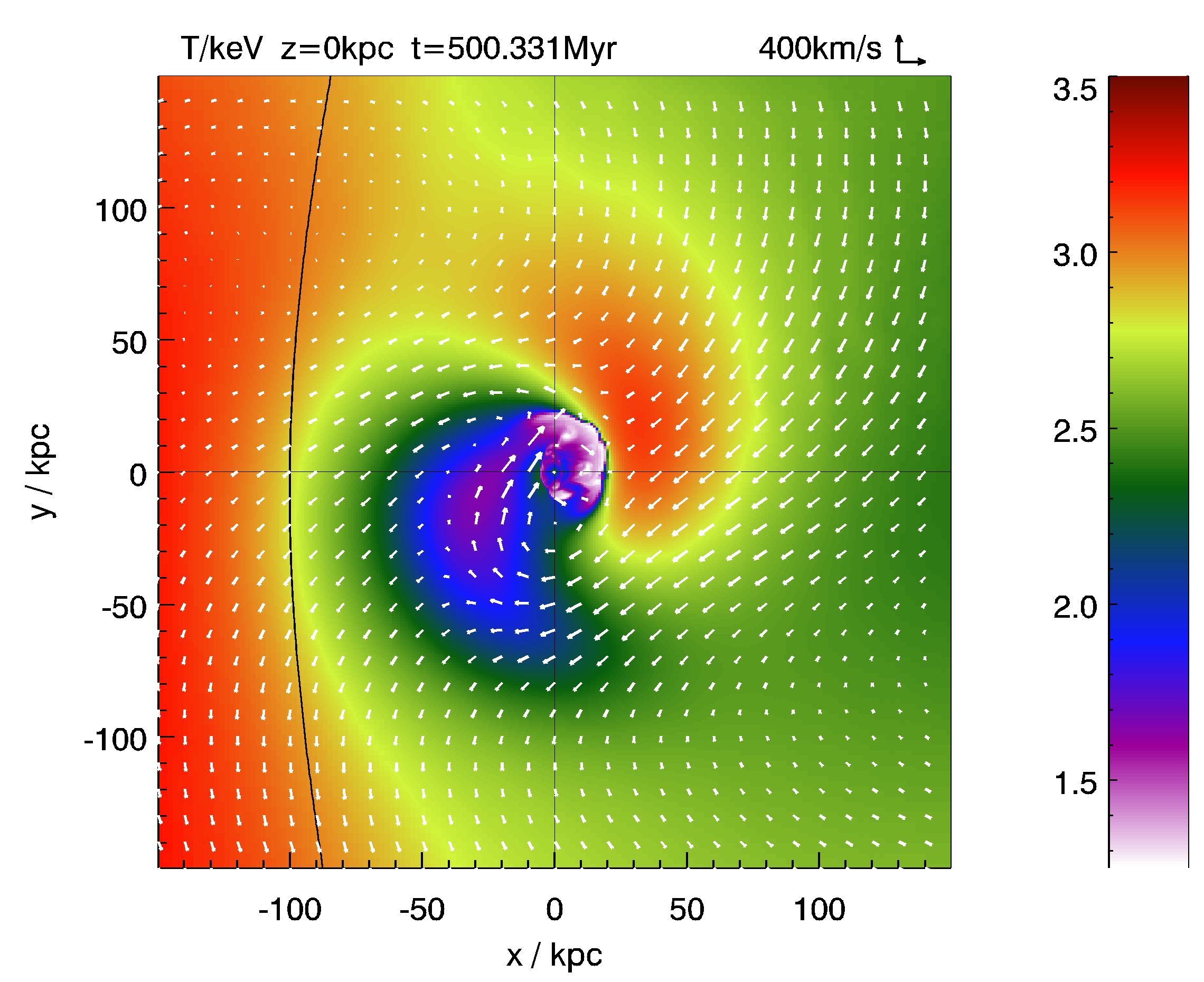}
\includegraphics[trim=300 80 330 0,clip,height=0.16\textheight]{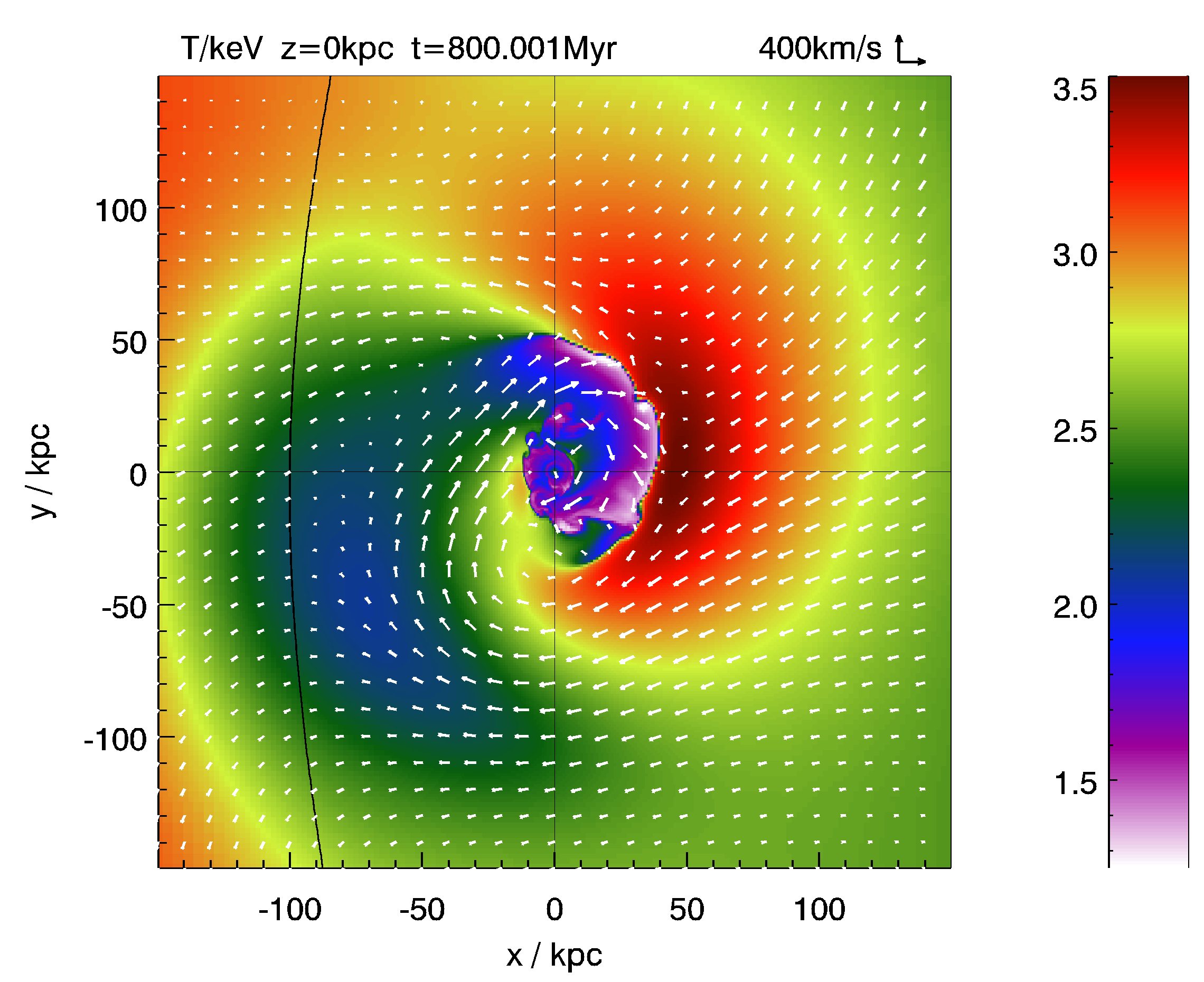}\newline
\includegraphics[trim=    0 80 330 0,clip,height=0.16\textheight]{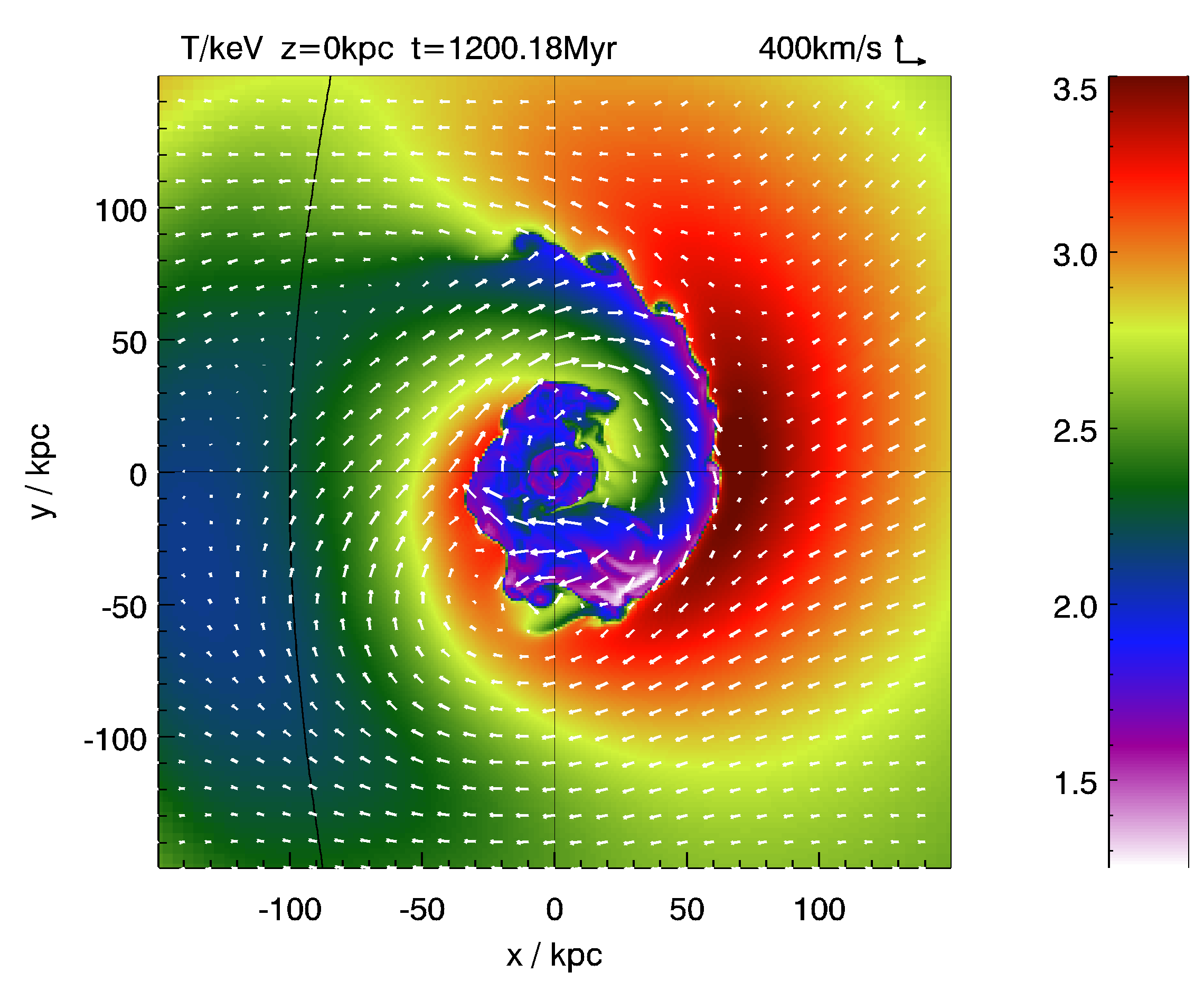}
\includegraphics[trim=300   0 0     0,clip,height=0.16\textheight]{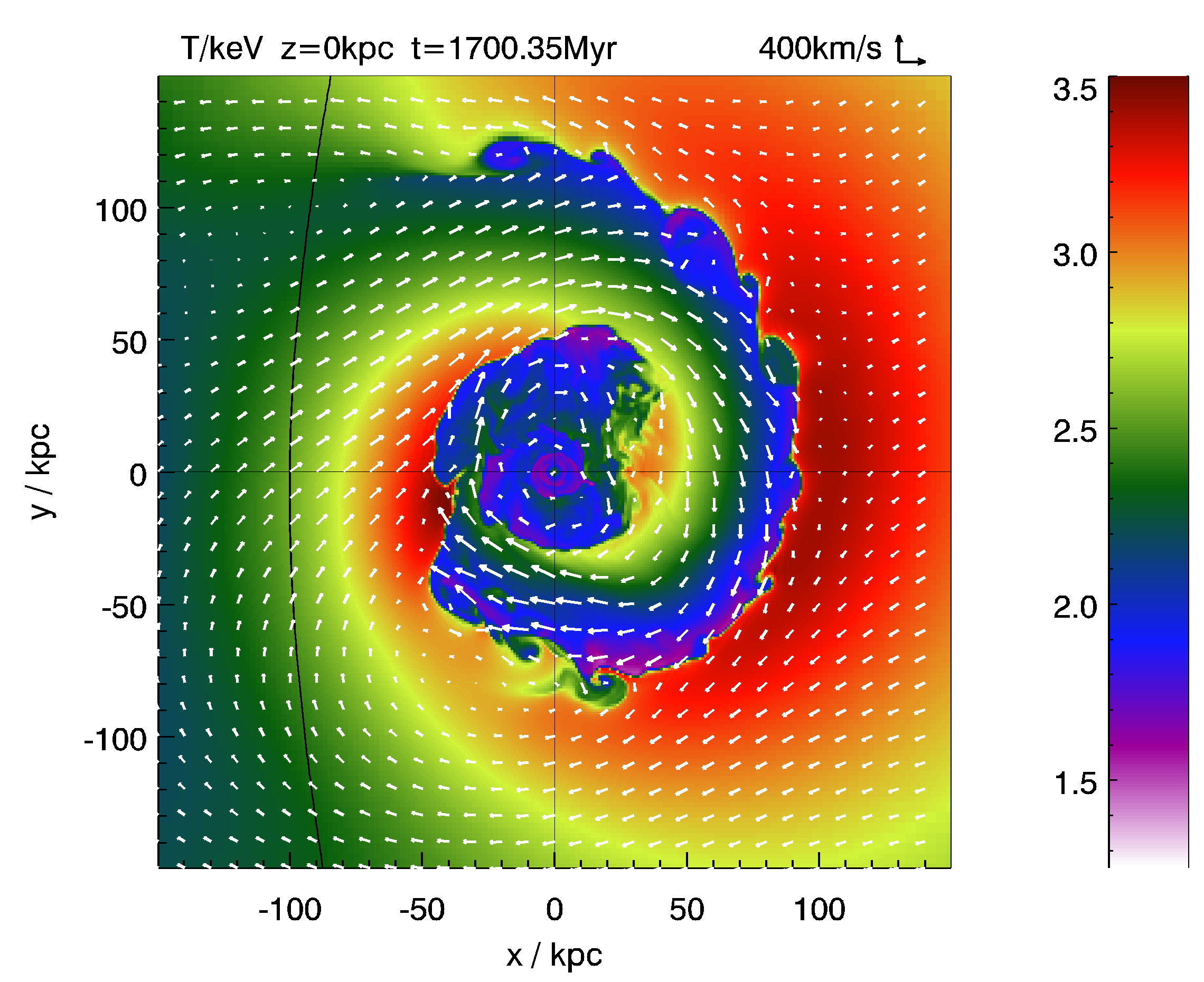}
\newline
\caption{Time series of temperature slices, taken in the orbital plane of the subcluster. White arrows indicate the velocity field (w.r.t. main cluster centre), colours code the ICM temperature in keV. The series is for the fiducial run M2a100dmin100dmax3 (high-resolution case). The time is noted in the title of each frame along with the scale bar for the velocity vectors. The top left panel shows a large-scale view at $t=-0.8\Gyr$. The remaining panels focus on the central $150\Kpc$. They share the temperature colour scale in the bottom left.}
\label{fig:slicetemp_massize}
\end{figure}
%FFFFFFFF

%
We demonstrate the evolution of the gas sloshing in Fig.~\ref{fig:slicetemp_massize}, which displays the ICM temperature (colour coded)  and velocity field (white arrows) in the orbital plane at different timesteps. The sloshing proceeds as described in Sect.~\ref{sec:intro_scenario}. The only effect not mentioned there is the adiabatic heating of the ICM near the subcluster due to infall of ICM into the subcluster potential (see top panel in Fig.~\ref{fig:slicetemp_massize}).

Qualitatively, the evolution is independent of the subcluster and orbit characteristics, which mainly influence the intensity of the contrast of density and temperature across the CFs (see Sect.~\ref{sec:compare}). 

\begin{figure}
%\begin{center}
\includegraphics[width=0.48\textwidth]{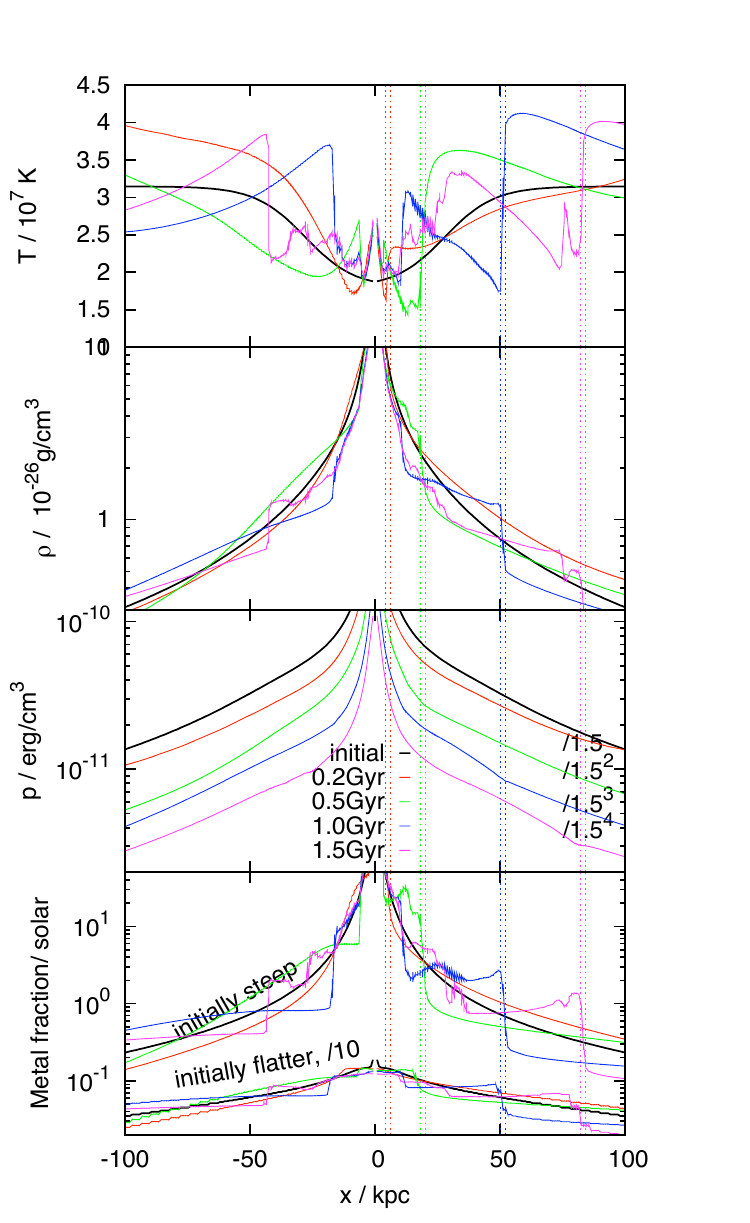}
\caption{Profiles of temperature, density, pressure,  and metallicity along the $x$-axis,  for the fiducial run (high-resolution case), at different timesteps (see legend). For clarity, the pressure profiles of different timesteps are divided by $1.5$, $1.5^2$, $1.5^3$, $1.5^4$. For the metallicity, we show two sets of profiles (see Sect.~\ref{sec:metals_ini}). For the upper set, the initial metal density is proportional to the light profile of M87. For the lower set, the initial metal density profile is flatter. The whole set of flatter profiles is divided by a factor of 10 to avoid confusion.  On the rhs side ($+x$-direction), we mark the outer CF by a thin vertical doublet line of the colour/style matching the legend.}
\label{fig:evolprofs}
%\end{center}
\end{figure}

In Fig.~\ref{fig:evolprofs} we take a closer look at the structure of the cool spiral and CFs by plotting temperature, density and pressure profiles along the $x$-axis. 
%
%FFFFFFFFFF
%FFFFFFFFFF
%
The CFs are indeed discontinuities in temperature and density within our resolution, but the pressure is continuous. However, the pressure gradient shows a discontinuity just inside of each CF.  Inside of each CF, there is a plateau in density accompanied by an increasing temperature towards smaller radii because the cool gas from the centre has been moved outwards. Going further inwards, in many profiles the next CF can be detected. The metal profiles will be discussed in detail in Sect.~\ref{sec:metals}.

The typical velocities induced by sloshing in our fiducial model are $300\Kms$, which is about half the circular velocity inside the inner 30 kpc of the Virgo cluster. This supports the idea of \citet{Keshet2009}, that the circular component of the sloshing motions contributes a significant radial acceleration working on the sloshing gas in the form of centripetal acceleration.

%% file: details.tex
%*******
\section{Preparation of synthetic observations} \label{sec:synthetic}
In order to highlight the simulation results, e.g., regarding the width of the CFs,  we do not convolve our maps with an instrument kernel, nor do we reduce the resolution of our maps to match current observations. 

%***
\subsection{Maps} 
\begin{itemize}
\item Synthetic X-ray images  are derived by projecting $n^2\sqrt{T}$ (where $n$ is the particle density and $T$ the gas temperature) along the LOS. This results in unnormalised brightness images.
\item We calculate  brightness residual maps by dividing each X-ray image by its azimuthal average. We favour this version over residuals w.r.t.~a fitted $\beta$-model because we want to highlight the deviation from the initial symmetry. 
\item Reducing the 3D temperature structure of a galaxy cluster to a 2D map requires (weighted) averaging along the LOS for each pixel of the temperature map.  A standard way is to calculate the emission-weighted projected temperature:
%-----------
\begin{equation}
T\Ew=\frac{\int W\, T dz}{\int W dz}\;\;\textrm{with}\;\;W=n^2\sqrt{T}.
\end{equation}
%======
\citet{Mazzotta2004} show that using the weights
%-----------
\begin{equation}
W=n^2/T^{3/4}
\end{equation}
%======
results in estimates for projected temperatures which are closer to the value derived by spectroscopical temperature fitting. Therefore we use this method. Emission-weighting results into very similar maps, except that the coolest regions appear slightly less cool, similar to what is found by \citet{Mazzotta2004}.
\item In analogy  to temperature maps, we derive metallicity maps by calculating the emission-weighted mean of the metallicity along each LOS.
\end{itemize}

%***
\subsection{Profiles of projected quantities} \label{sec:derive_profiles}
From our synthetic maps, we derive radial profiles  by averaging over an azimuthal range of $\pm 45\degree$ to both sides of  the desired direction. The same procedure was used to derive the observed profiles (S10).

Projected profiles are not sensitive to the direction of the LOS: e.g.~consider maps derived by projection along the $z$-axis and the $y$-axis. For both cases, profiles along the $x$-axis are very similar.

%***
\subsection{Determination of the CF radius} \label{sec:deriveCFradius}
For a given LOS and direction in a map, we derive the projected temperature profile as described in Sect.~\ref{sec:derive_profiles}. We define radial ranges with temperature slopes $\ge 0.02 \KeV / \Kpc$ as CFs. This sets an inner and an outer edge for each CF. We define the nominal radius of the CF to be centred between the front's inner and outer edge.  

The contrast of X-ray brightness or metallicity across the CF is the ratio of the this quantity at the inner and outer edge of the CF. 

%************************
\section{Details of comparison}
 %FFFFFFFFFF
\begin{figure}
\begin{center}
\includegraphics[width=0.45\textwidth]{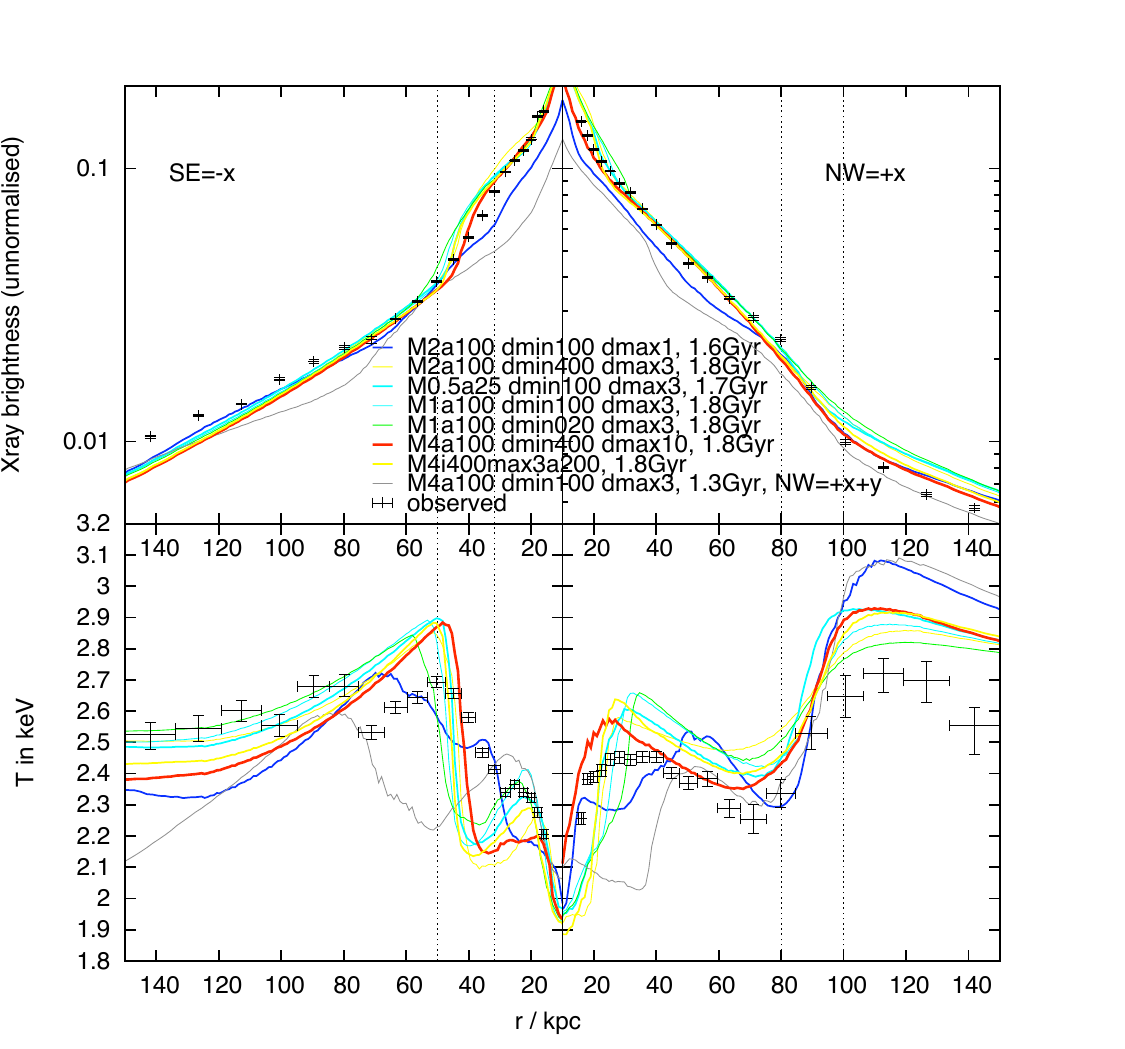}
\caption{
Comparison of azimuthally averaged profiles of X-ray brightness and projected temperature for runs not matching the observation. See Sect.~\ref{sec:compare_contrasts} and Fig.~\ref{fig:bestprof} for details.}
\label{fig:nomatchprof}
\end{center}
\end{figure}
%FFFFFFFFFF

For the sake of brevity, we have omitted some non-essential but still useful figures regarding the comparison of simulations and observations in the main text and collect them here.

Fig.~\ref{fig:nomatchprof} shows the projected profiles of X-ray brightness and temperature for the simulation runs that do not match the observations.

\begin{figure}
\phantom{x}\hfill M1a50dmin100dmax3 \hfill M2a100dmin100dmax10 \hfill\phantom{x}  \newline  
\includegraphics[trim=0 0 300 100,clip,height=4.2cm]{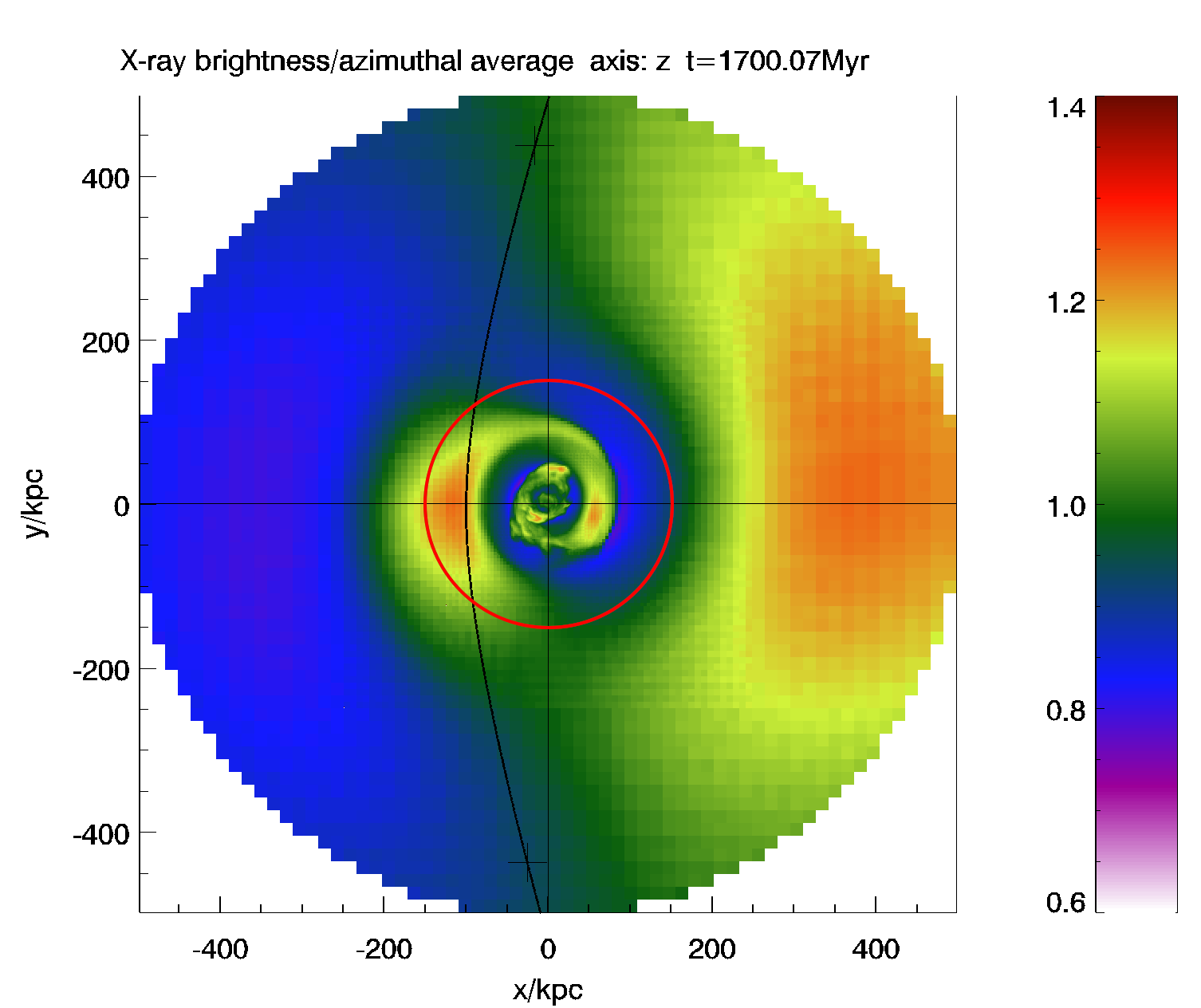}\hspace{-0.1cm}
\includegraphics[trim=220 0 300 100,clip,height=4.2cm]{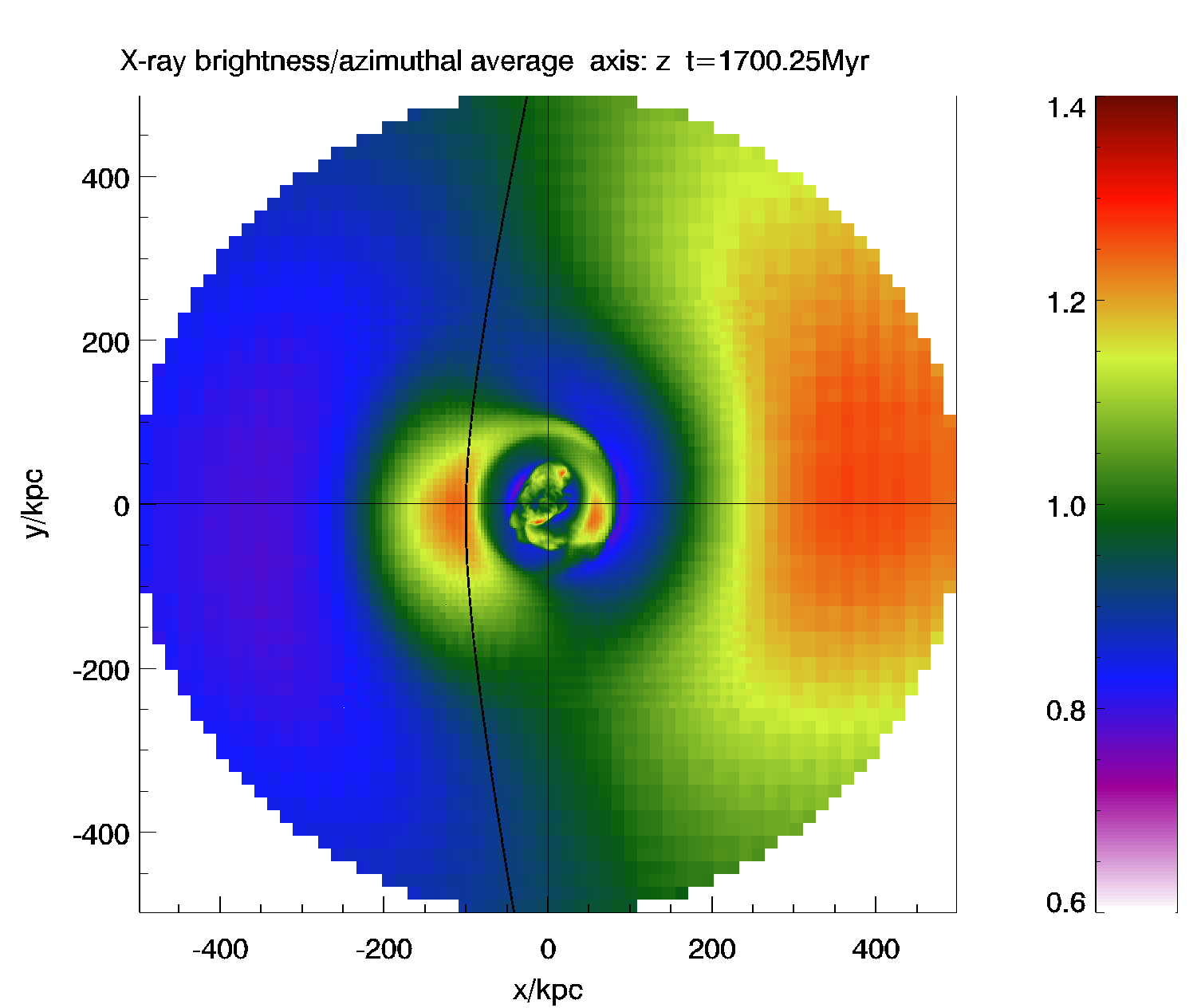}\hspace{-0.1cm}
\newline
\phantom{x}\hfill   M2a100dmin20dmax3 \hfill M4a100dmin400dmax3 \hfill\phantom{x}\hfill\phantom{x}  \newline
\includegraphics[trim=0 0 350 100,clip,height=4.2cm]{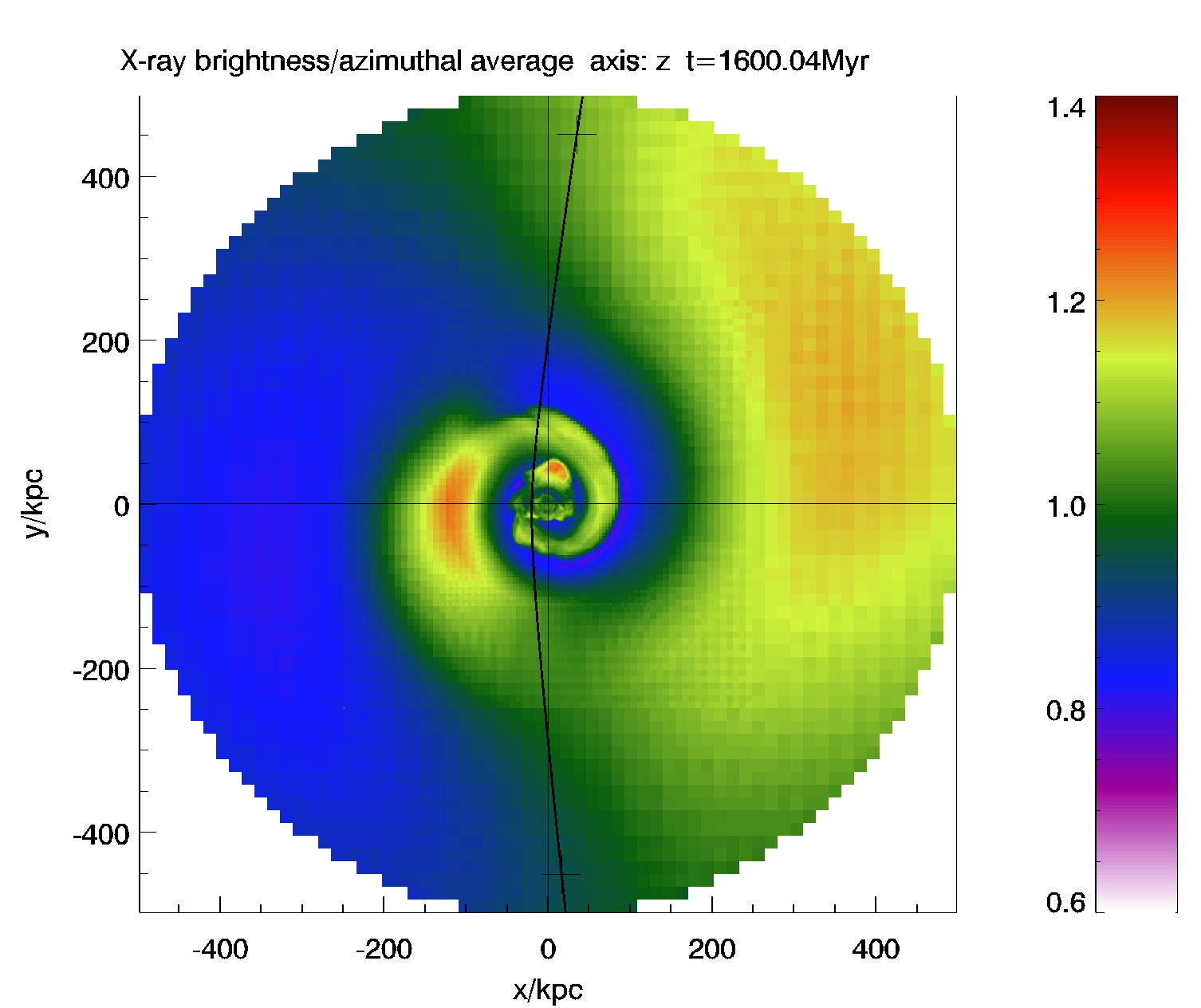}\hspace{-0.1cm}
\includegraphics[trim=230 0 0     100,clip,height=4.2cm]{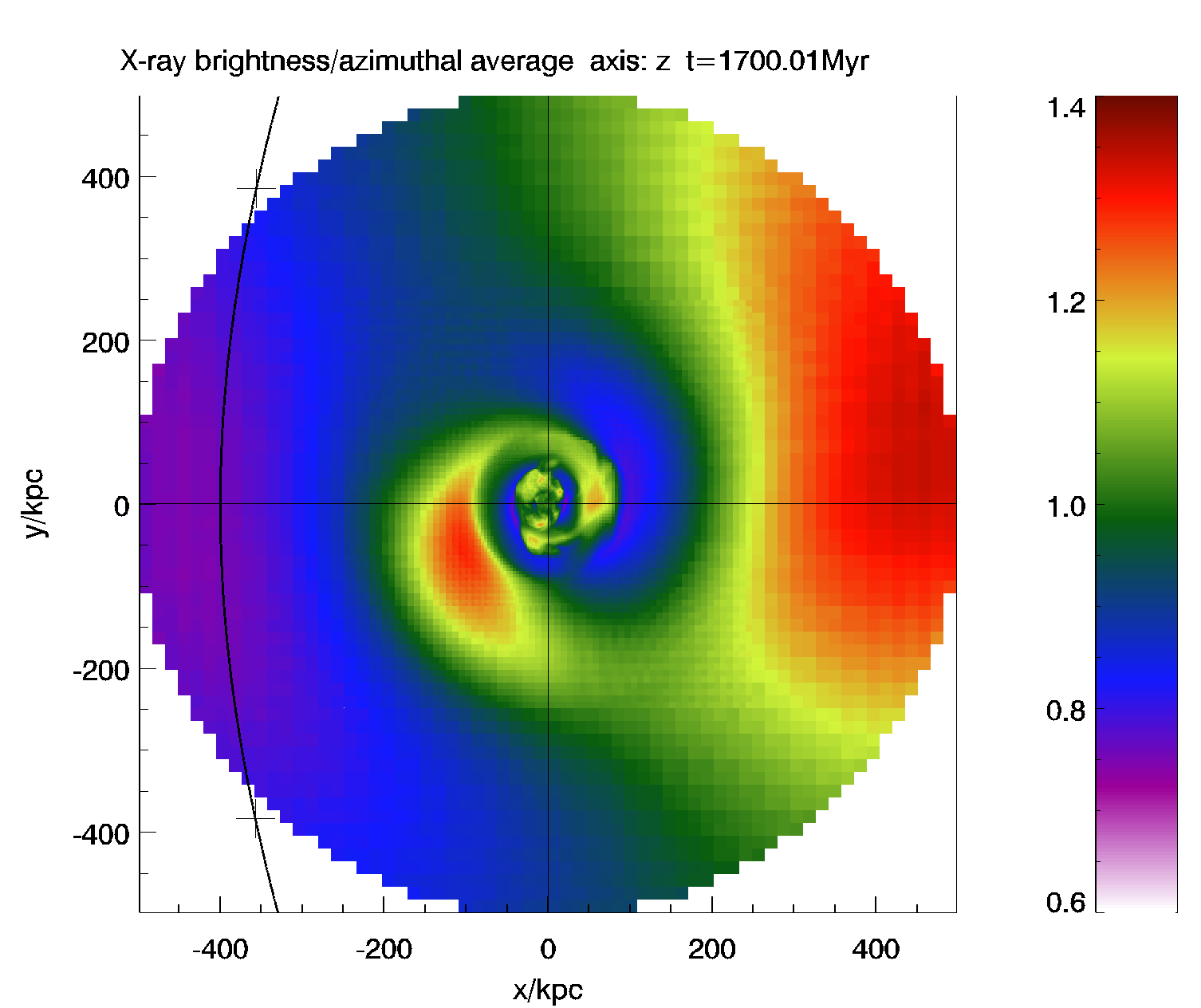}
\caption{
Brightness residual maps from  the four best-matching simulations (first group in Table~\ref{tab:exclude}, see labels above panels). The red circle in the first panel indicates the observed field of view.}
\label{fig:best1}
\end{figure}

\begin{figure}
M1a50dmin100dmax1\hfill M2a100dmin100dmax3 (high resolution)  \newline
\includegraphics[trim=0 0 300 100,clip,height=4.2cm]{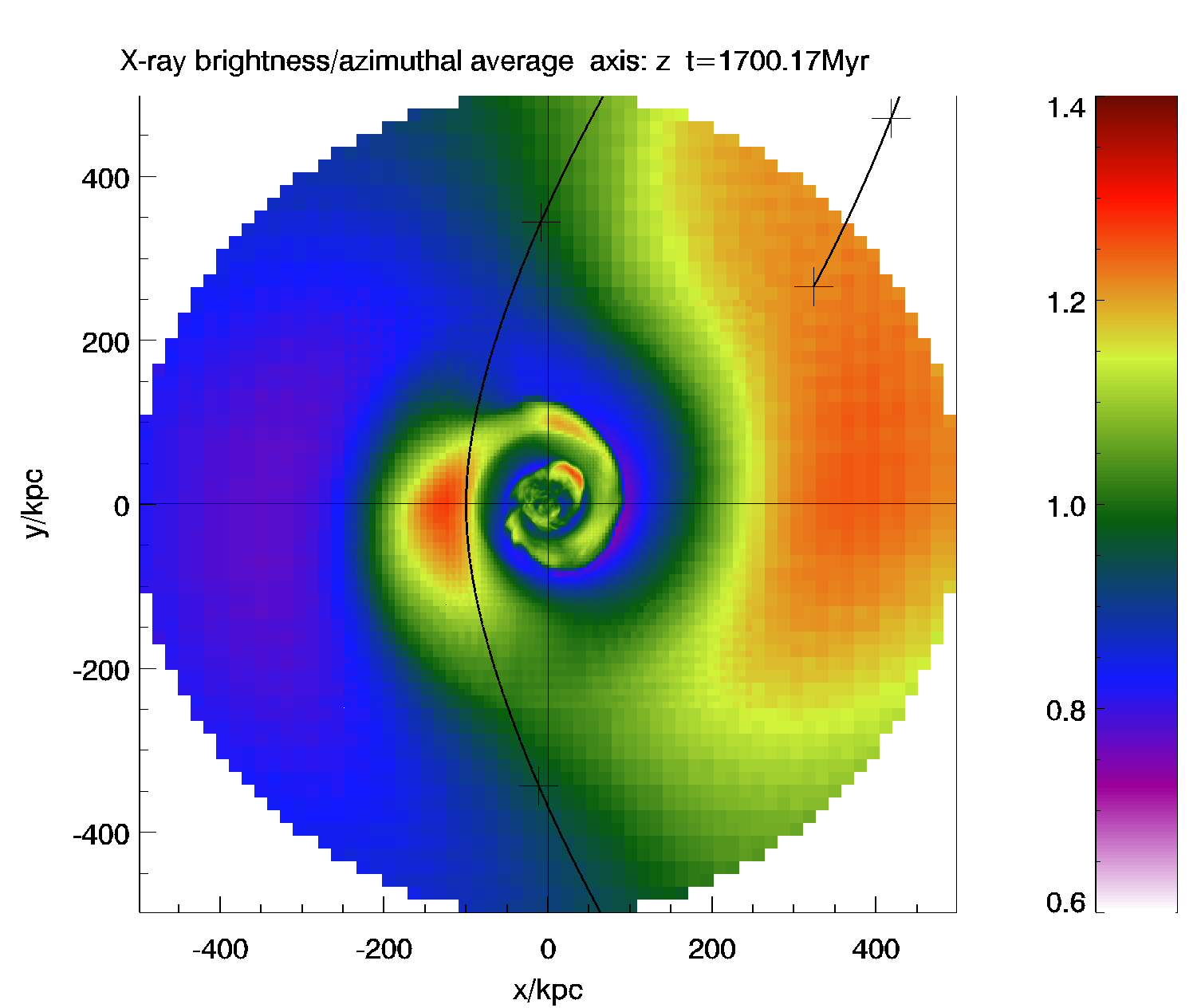}
\includegraphics[trim=220 0 300 100,clip,height=4.2cm]{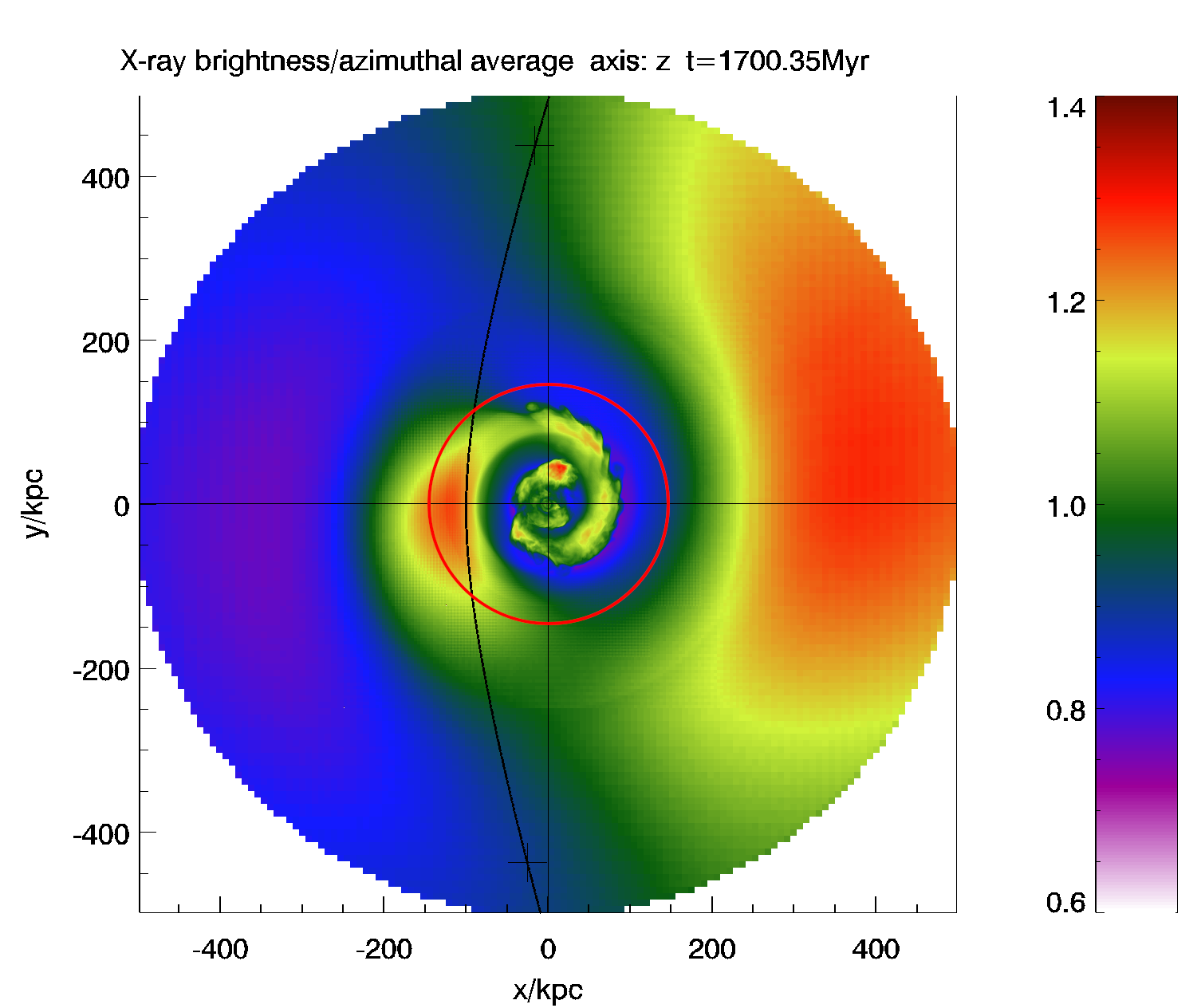}\newline
 M4a100dmin400dmax1 \newline
\includegraphics[trim=0 0 0     100,clip,height=4.2cm]{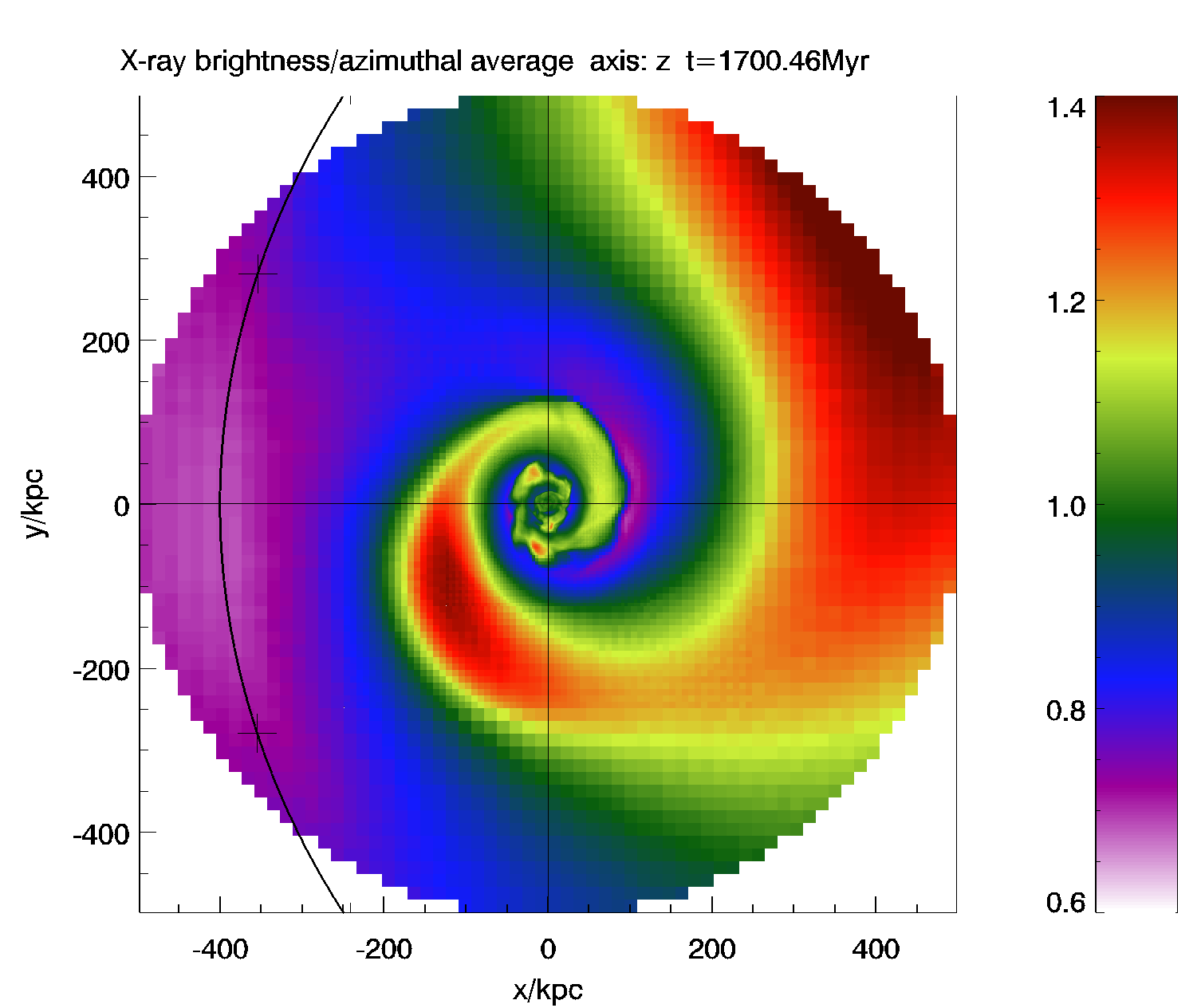}\newline
%\end{center}
\caption{
Same as Fig.~\ref{fig:best1}, but for second best simulations (second group in Table~\ref{tab:exclude}).}
\label{fig:best2}
\end{figure}

Figs.~\ref{fig:best1} and \ref{fig:best2} compare brightness residual maps for our best-match runs.

%
%FFFFFFFF
%FFFFFFFFF
%
%FFFFFFFF
%FFFFFFFFF